\documentclass[CRPHYS,Unicode,,manuscript,Volume]{cedram}
\usepackage{amsmath,amssymb}
\hypersetup{urlcolor=purple, linkcolor=blue, citecolor=blue}

\title{Geophysical flows over topography, a playground for laboratory experiments}
\alttitle{\'Ecoulements géophysiques en présence de topographie, un terrain de jeu pour la modélisation expérimentale}

\datereceived{03.05.2024}%
\daterevised{17.10.2024}%
\dateaccepted{17.10.2024}%

\author{\firstname{J\'er\'emie} \lastname{Vidal}\CDRorcid{0000-0002-3654-6633}}
\address{Universit\'e Grenoble Alpes, CNRS, ISTerre, 38000 Grenoble, France}
\email[J. Vidal]{jeremie.vidal@univ-grenoble-alpes.fr}

\author{\firstname{J\'er\^ome} \lastname{Noir}\CDRorcid{0000-0001-9977-0360}\IsCorresp}
\address{Institut f\"ur Geophysik, ETH Z\"urich, Sonneggstrasse 5, Z\"urich 8092, Switzerland}
\email[J. Noir]{jerome.noir@eaps.ethz.ch}

\author{\firstname{David} \lastname{Cébron}\CDRorcid{0000-0002-3579-8281}}
\addressSameAs{1}{Universit\'e Grenoble Alpes, CNRS, ISTerre, 38000 Grenoble, France}
\email[D. C\'ebron]{david.cebron@univ-grenoble-alpes.fr}

\author{\firstname{Fabian} \lastname{Burmann}\CDRorcid{0000-0001-8095-1081}}
\addressSameAs{2}{Institut f\"ur Geophysik, ETH Z\"urich, Sonneggstrasse 5, Z\"urich 8092, Switzerland}
\email[F. Burmann]{fabian.burmann@eaps.ethz.ch}

\author{\firstname{R\'emy} \lastname{Monville}\CDRorcid{0000-0002-9460-2293}}
\addressSameAs{1}{Universit\'e Grenoble Alpes, CNRS, ISTerre, 38000 Grenoble, France}
\email[R. Monville]{remy.monville@univ-grenoble-alpes.fr}

\author{\firstname{Vadim} \lastname{Giraud}\CDRorcid{0009-0003-6820-0148}}
\addressSameAs{2}{Institut f\"ur Geophysik, ETH Z\"urich, Sonneggstrasse 5, Z\"urich 8092, Switzerland}
\email[V. Giraud]{vadim.giraud@eaps.ethz.ch}

\author{\firstname{Yoann} \lastname{Charles}\CDRorcid{0000-0002-4861-8016}}
\addressSameAs{2}{Institut f\"ur Geophysik, ETH Z\"urich, Sonneggstrasse 5, Z\"urich 8092, Switzerland}
\address{Development of Advanced Engineering Solutions (DAES), Avenue des Grandes-Communes 8, Petit-Lancy 1213, Switzerland}

\thanks{JV, RM \& DC are supported by the European Research Council (ERC) under the European Union's Horizon 2020 research and innovation programme (grant agreement No 847433, \textsc{theia} project). JN and VG are supported by ETH grant \#ETH-04 22-1. FB received funding from the European Research Council (ERC) under the European Union’s Horizon 2020 research and innovation program (grant agreement No 833848, \textsc{uemhp} project). DC acknowledges the French Academy of Sciences, Electricit\'e de France, Labex \textsc{osug}@2020 (ANR10 LABX56).}

\altkeywords{topographie, rotation, stratification, \'ecoulements g\'eophysiques, noyaux plan\'etaires, oc\'eans de subsurface}

\begin{abstract}
Physicists face major challenges in modelling multi-scale phenomena that are observed in geophysical flows (e.g. in the Earth's oceans and atmosphere, or liquid planetary cores).
In particular, complexities arise because geophysical fluids are rotating and subject to density variations, but also because the fluid boundaries have complex geometries (e.g. the ocean floor) with wavelengths ranging from metres to thousands of kilometres.
Dynamical models of planetary fluid layers are thus often constrained by observations, whose interpretation necessitates a comprehensive understanding of the underlying physics.
To this end, geophysical studies often combine cutting-edge experiments across a wide range of parameters, together with theory and numerical simulations, to derive predictive scaling laws applicable for planetary settings. 
In this review, we discuss experimental efforts that have contributed to our understanding of geophysical flows with topography.
More specifically, we focus on (i) the flow response to mechanical (orbital) forcings in the presence of a large-scale (ellipsoidal) topography, (ii) some effects of small-scale topography onto bulk flows and boundary-layer dynamics, and (iii) the interaction between convection and roughness.
The geophysical context is briefly introduced for each case, and some experimental perspectives are drawn.
\end{abstract}
\begin{altabstract}
Les écoulements géophysiques, par exemple dans les océans ou les noyaux planétaires liquides, présentent souvent une turbulence caractérisée par une multitude d'échelles de temps et d'espace.
Celles-ci résultent notamment des effets de rotation globale et de stratification en densité, mais aussi des effets de paroi qui sont souvent irréguliers (e.g. la bathymétrie du plancher océanique).
Ainsi, peu d'études ont considéré la modélisation des écoulements géophysiques avec rotation, stratification en densité et effets topographiques.
En pratique, la meilleure approche consiste souvent à combiner des expériences de laboratoire sur une large gamme de paramètres, ainsi que des travaux théoriques et/ou numériques, pour ensuite extrapoler les résultats aux conditions géophysiques.
Dans cet article de revue, nous discutons les travaux principalement expérimentaux qui ont permis de mieux comprendre la dynamique des écoulements géophysiques avec des effets topographiques de paroi.
Premièrement, nous détaillons les écoulements engendrés par les forçages orbitaux (e.g. la marée ou la précession) pour un fluide contenu dans un ellipsoïde.
Ensuite, nous illustrons les effets d'une topographie de petite échelle sur certains écoulements en volume et de couche limite.
Enfin, nous discutons l'influence d'une paroi non lisse sur les écoulements engendrés par la convection. 
\end{altabstract}

\begin{document}

\maketitle

\section{Introduction}
Fluid dynamics enjoys great popularity among geophysicists and planetary scientists, which is not surprising considering the ubiquity of fluid phenomena in the universe. 
Many rocky planets possess shallow thin fluid layers (e.g. oceans or atmospheres), as well as thicker ones in depth (e.g. liquid cores or subsurface oceans in icy moons). 
From a physicist's perspective, geophysical fluid dynamics serves as both an applied and pure science. 
It is applied because we observe and predict the effects of geophysical flows in our daily life (such as weather forecasting, climate evolution, or Earth's magnetic field variations). 
Yet, it is also pure because most geophysical flows involve fundamental fluid dynamics processes (like waves, hydrodynamic instabilities, and turbulence) that are still not fully understood.

Indeed, several scientific questions currently challenge our geophysical flow models, such as the origin of the ancient magnetic fields on Earth \cite{landeau2022sustaining} and the Moon \cite{wieczorek2023lunar}. 
Similarly, controversies persist regarding estimates of heat transport by turbulent convection, which is key for understanding the long-term evolution of natural systems over billions of years. 
Actually, these questions arise because the internal dynamics of planetary bodies remains barely known.
In the middle of the nineteenth century, Hopkins \cite{hopkins1839nutations} proposed that one could probe planets' interiors from the variations of their rotation.
Indeed, if a planet or a moon has a hidden fluid layer, it could be inferred by looking at the planet's response to orbital perturbations, such as precession or nutations. 
This pioneering idea found many applications in the twentieth century, such as the confirmation of the liquidity of the Earth's outer core \cite{jeffreys1957theory}, or the existence of a molten core in Mercury \cite{margot2007large} and in the Earth's moon \cite{yoder1981free}.  
Nowadays, the increasing accuracy of the Earth's and moon's rotations provides a unique opportunity to go beyond the detection of their liquid cores to address the question of their internal dynamics. 
Meanwhile, space missions like the \textsc{esa} mission \textsc{juice} or the \textsc{nasa} mission  \textsc{clipper}, which are dedicated to the icy moons of Jovian planets, will provide unprecedented data to probe their subsurface oceans. 
However, global models based on Earth-like oceanic circulation are unlikely to be directly applicable to these envelopes \cite{soderlund2024physical}.
Specific models are thus still missing to properly model the fluid dynamics within such moons.

Therefore, understanding the dynamics of planets' and moons' fluid envelopes remains a central question in contemporary geophysics and planetary sciences. 
Before discussing recent progress on this topic, we shall first establish the fundamental physical components necessary for modelling these fluid envelopes.

\subsection{The key ingredients of geophysical fluid models}
Planetary fluid envelopes are unique by their large size, but also because of the physical mechanisms governing their dynamics.  
Notably, they often experience the effects of global rotation.
The Earth is currently rotating at an angular velocity $\Omega_s \simeq 7.3 \times 10^{-5}$~rad.s${}^{-1}$, and was even rotating faster in the past \cite{farhat2022resonant}. 
Global rotation is dynamically important if, first, the dynamics is in a low-viscosity regime (i.e. with a rotation time scale much smaller than the viscous time scale). 
This leads to the definition of the Ekman number $E$ given by
\begin{equation}
    E = \frac{\nu}{\Omega_s L^2},
\end{equation}
where $\nu$ is the kinematic viscosity of the fluid, and $L$ is the typical length scale of the system.
The low-viscosity regime is thus defined by $E \ll 1$. 
Moreover, geophysical turbulence also depends on the relative strength of the nonlinearities in the system. 
We introduce the Rossby number $Ro$ (comparing the rotation and nonlinear advection time scales), which is related to the classical Reynolds number $Re$ (which compares the viscous and nonlinear advection time scales) as
$Ro = Re E \, (L/\ell)^2$ with
\begin{subequations}
\begin{equation}
    Ro = \frac{u_\ell}{\Omega_s \ell}, \quad Re = \frac{\ell u_\ell}{\nu},
    \tag{\theequation a,b}
\end{equation}
\end{subequations}
where $u_\ell$ is the typical amplitude of the velocity at the flow length scale $\ell$.
The dynamics ignores rotation when $Ro \gg 1$, whereas it is strongly shaped by rotational effects when $Ro \ll 1$. 
Note that the length scale $\ell$ introduced above is still undefined. 
Actually, since geophysical systems exhibit many length scales in the range $0 < \ell \leq L$, different dynamical regimes can be expected as a function of the length scale (e.g. \cite{nataf2024dynamic} for liquid planetary cores).
In practice, the Rossby number can have values from $Ro \ll 1$ at large scales to $Ro \gtrsim 1$ at smaller scales.
The transition between these regimes is expected to be controlled by the Zeman length scale  defined as \cite{zeman1994note}
\begin{equation}
    \ell_z \sim \left ( \frac{\epsilon}{\Omega_s^3} \right )^{1/2},
\end{equation}
where $\epsilon$ is the mean energy dissipation rate per unit of mass. 
Turbulent structures with length scales $\ell \gg \ell_z$ are modified by global rotation and anisotropic, whereas the turbulence is unaffected by rotation and nearly isotropic when $\ell \ll \ell_z$ \cite{mininni2012isotropization,delache2014scale}.
If we consider the largest scales (assuming $\ell \sim L$), typical values are $Ro \sim 10^{-7}-10^{-6}$ for the Earth's liquid core with $E \sim 10^{-15}$ (assuming $u_\ell \sim 1-10$~km.y${}^{-1}$ for core flows driven by convection \cite{jones2015tog}, and the core radius $\sim 3480$~km).
This indicates that the Earth's liquid core is in a regime where viscous and advection effects are much smaller than rotational ones. 
Note also that a typical value for the Reynolds number is then $Re \sim 10^{8}$ for the largest scales.
The later example clearly illustrates a common difficulty in geophysical modelling, namely obtaining turbulent flows with $Re \gg 1$ in the regime $Ro \ll 1$, which is only accessible by considering low enough values of $E$.

\begin{figure}[tbp]
\centering
\begin{tabular}{cc}
\includegraphics[height=0.52\textwidth]{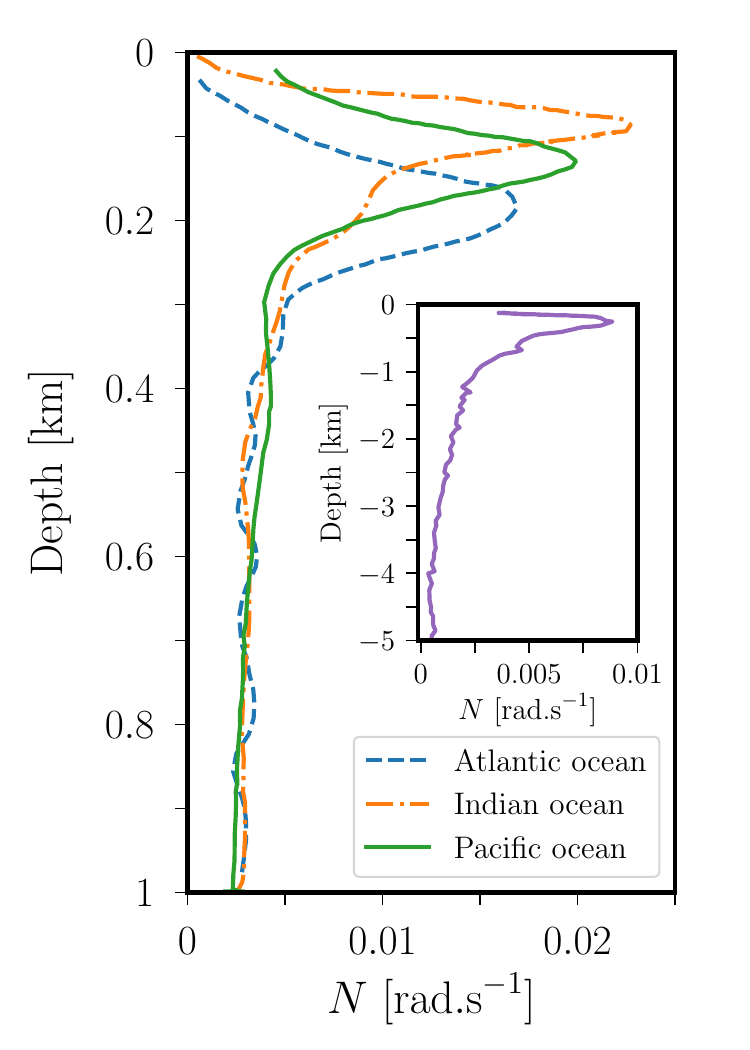} &
\includegraphics[height=0.52\textwidth]{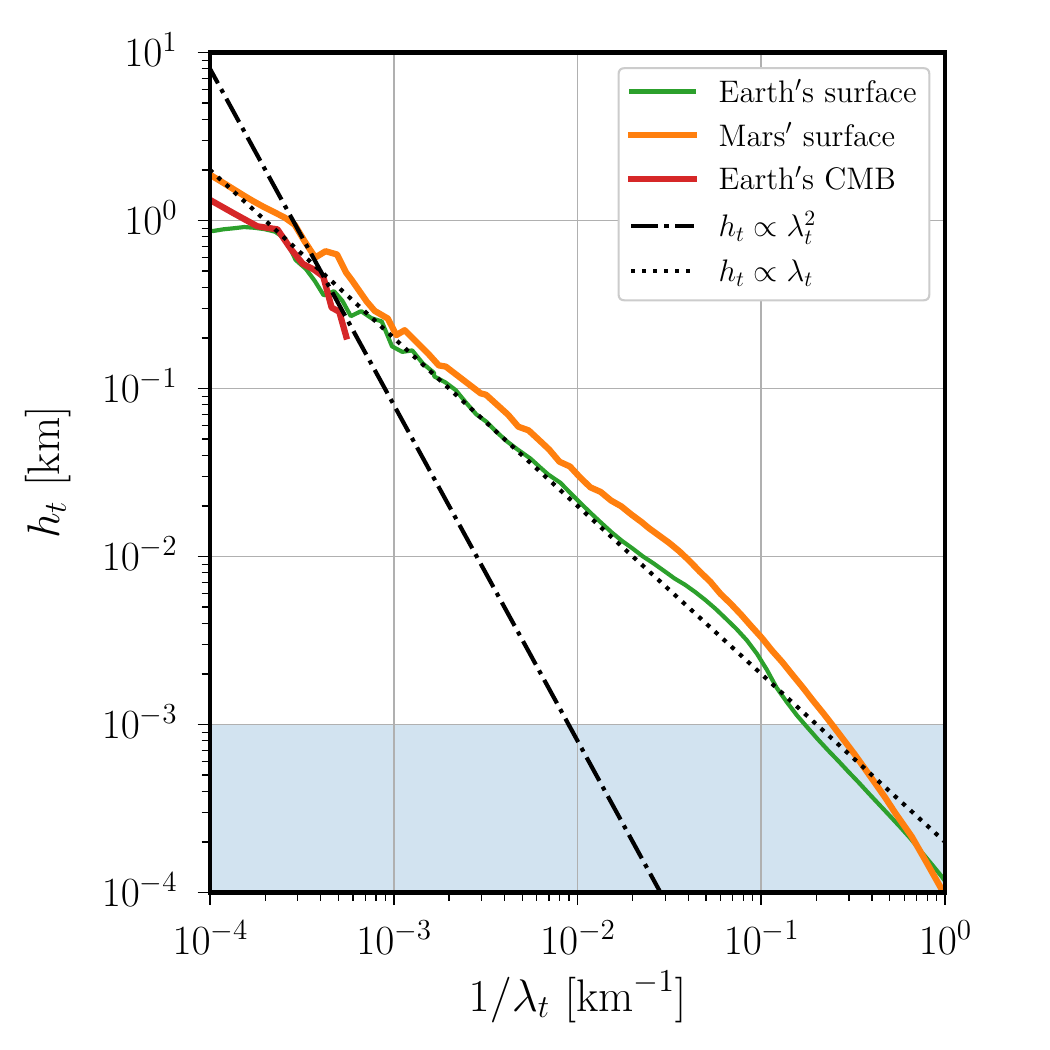} \\
(a) & (b) \\
\end{tabular}
\caption{\textbf{(a)} Illustrative profiles of the Brunt-V\"ais\"al\"a frequency $N$, as a function of depth, in the Earth's oceans (adapted from \cite{gent1985much}). Inset shows a typical profile in the deep ocean in the western North Pacific (adapted from figure 3.6 in \cite{talley2011descriptive}). \textbf{(b)} Spatial spectra of topography, where $h_t$ is the height of the topography at wavelength $\lambda_t$. Spectra deduced from the RMS of the observed power spectrum at the Earth's and Mars' surface\cite{rexer2015ultrahighdegree}, and a seismological model of the Earth's \textsc{cmb} (obtained from body waves) is also shown \cite{koelemeijer2021consistent}. Blue region shows the (laminar) Ekman boundary layer, of typical thickness $\mathcal{O}(\sqrt{\nu/\Omega_s})$, at the Earth's \textsc{cmb}  or at the bottom of the surface oceans (which have the same viscosity $\nu\approx 10^{-6}$~m$^2$.s$^{-1}$).
Empirical law $h_t \propto 2 \times 10^{-4} \, \lambda_t$ (dash-dot line) for the RMS height at the surface \cite{ermakov2018power}, and $h_t \propto 8 \times 10^{-8} \, \lambda_t^2$ (dotted line) for the \textsc{cmb} (from gravity RMS spectrum\cite{puica2023analytical}).}
 \label{fig:exmp}
 \end{figure}

Second, geophysical flows are often subject to buoyancy effects and density stratification.
The latter is usually characterised by the Brunt-V\"ais\"al\"a (BV) frequency, denoted below by $N$.
For a compressible fluid, the square of the BV frequency is given by \cite{chaljub2004spectral}
\begin{equation}
    N^2 = \frac{\boldsymbol{g}}{\rho} \boldsymbol{\cdot} \left ( \nabla \rho - \frac{\rho}{c_s^2} \boldsymbol{g} \right ),
    \label{eq:BVN}
\end{equation}
where $\boldsymbol{g}$ is the gravity field, $\rho$ is the fluid density, and $c_s$ is the speed of sound.
The BV frequency measures the departure of the actual density gradient to the adiabatic (isentropic) density gradient (given by $\rho \boldsymbol{g}/c_s^2$ in hydrostatic equilibrium).
As such, stably stratified fluids (e.g. a light fluid over a dense one in laboratory conditions) are characterised by $N^2>0$, whereas convection can occur for unstable stratification with $N^2<0$.
For example, the Earth's oceans are known to be stably stratified in density.
There is some variability as a function of the location and the seasons \cite{emery1984geographic}, but typical values are $N \sim 10^{-3}-10^{-2}$~rad.s${}^{-1}$ in the upper ocean and $N \sim 10^{-5}-10^{-4}$~rad.s${}^{-1}$ in the deep ocean (Figure \ref{fig:exmp}a).
On the contrary, the Earth's liquid core is convecting in bulk \cite{jones2015tog} (which implies $N^2 \sim 0$), but it might be either weakly stratified with $N/\Omega_s \leq \mathcal{O}(1)$ \cite{buffett2014geomagnetic,gastine2020dynamo} or strongly stratified with $N/\Omega_s \gg 1$ \cite{buffett2010chemical} in a thin layer near the core-mantle boundary\footnote{This is the interface between the liquid core made of molten iron, and the mantle made of silicate rocks.} (\textsc{cmb}).
Therefore, the two dynamical regimes should be investigated.

Both global rotation and stratification give rise to peculiar linear and turbulent flows, which make geophysical systems so unique and interesting. 
Yet, we can also draw some analogies between these two kinds of dynamics \cite{veronis1970analogy}.
For instance, they can both sustain wave motions in the form of inertial waves with angular frequencies bounded by $2 \Omega_s$ for rotating fluids \cite{greenspan1968theory}, or as internal (gravity) waves with angular frequencies bounded by $N$ for stably stratified fluids \cite{mowbray1967theoretical}.
In the Earth's atmosphere, internal gravity waves emitted over mountains carry energy from the surface to the upper atmosphere, and participate in mixing processes \cite{legg2021mixing} and angular momentum exchange between the atmosphere and the solid Earth  \cite{wurtele1996atmospheric}.
Similarly, the radiation of internal (gravity) waves over the sea floor bathymetry contributes to ocean mixing and tidal energy dissipation \cite{richet2018internal}.
For the dynamics, rotating and stratified flows are also often separated into bulk and thin boundary-layer regions. 
Finally, the associated turbulence is strongly anisotropic (e.g. with rotation \cite{godeferd2015structure} and stratification \cite{cortet2024turbulence,lefauve2024cras}), which is very different from the classical picture of isotropic Kolmogorov turbulence.

\subsection{The shape of planetary fluid envelopes}
\begin{figure}
    \centering
    \begin{tabular}{cc}
    \includegraphics[width=0.48\textwidth]{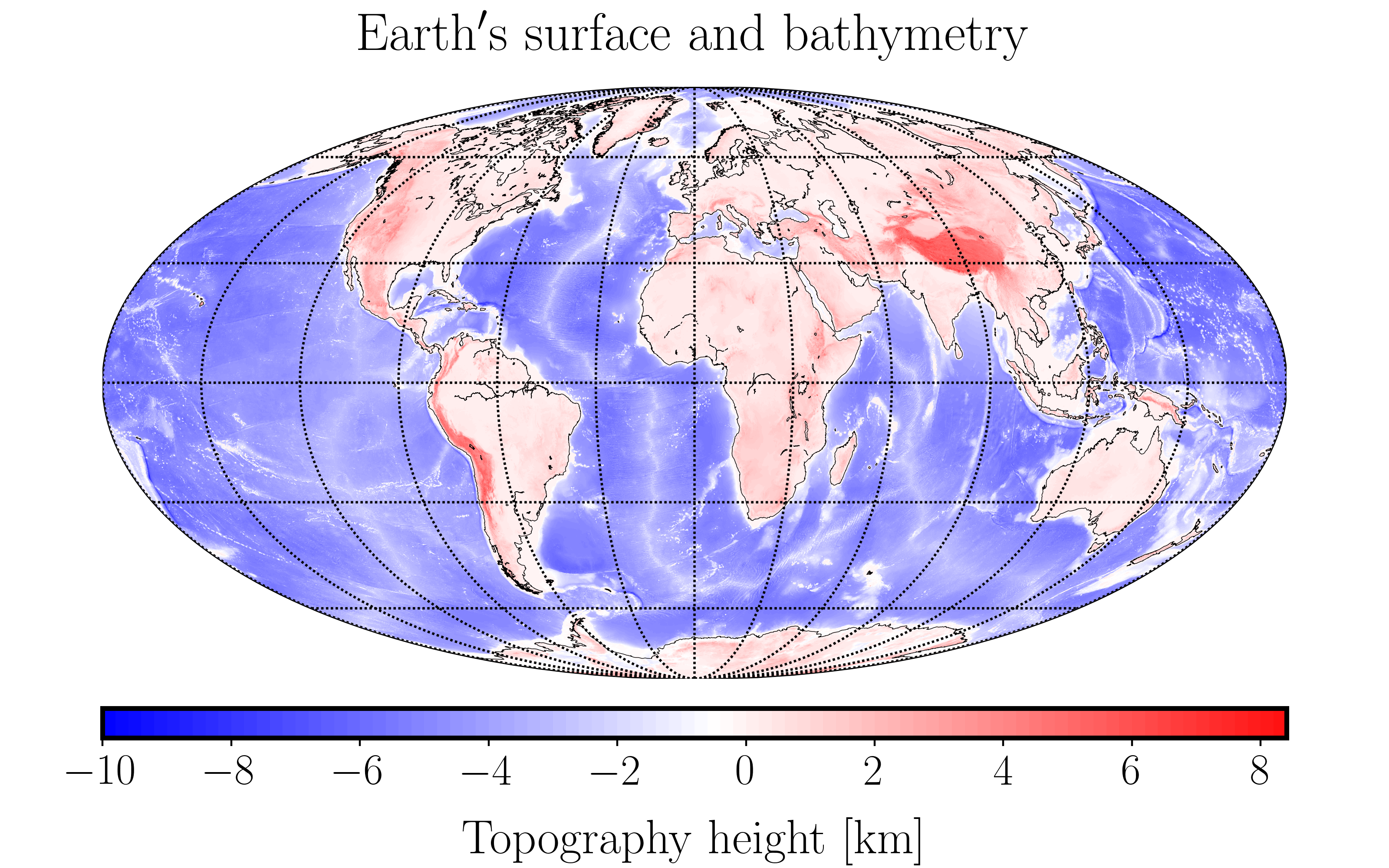} &
    \includegraphics[width=0.48\textwidth]{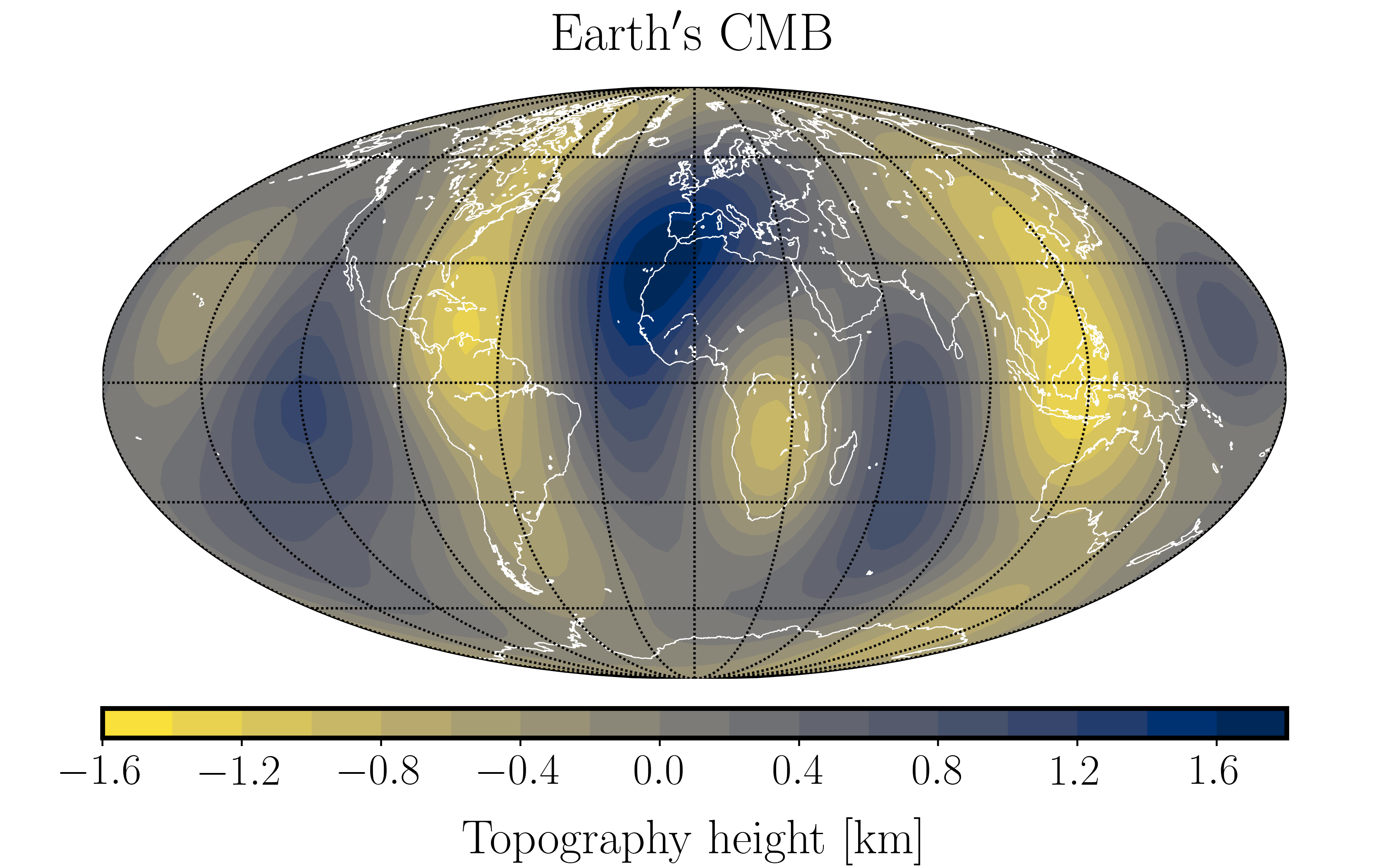} \\
    (a) & (b) \\
    \end{tabular}
    \caption{Landscape of topographies affecting geophysical flows. Colour bar shows the topography height (in km) on a map using a Mollweide projection. \textbf{(a)} \textsc{etopo1} global relief model of the Earth's surface, integrating continents' topography and oceans' bathymetry. Data provided in \cite{amante2009etopo1}. \textbf{(b)} Meso-scale topography at the Earth's \textsc{cmb}. Mean model obtained from seismological studies using body-wave and normal-mode inversions. Only the largest-scale components with spherical harmonics degree $\leq 6$ are shown. Data provided in \cite{koelemeijer2021consistent}.}
    \label{fig:topomap}
\end{figure}

To complete our picture of geophysical fluid envelopes, it is often needed to account for boundary effects. 
Planets' and moons' fluid envelopes are near-spherical bodies, on which a broad spectrum of topography wavelengths is superimposed (see Figure \ref{fig:exmp}b at the Earth's surface and \textsc{cmb}). 
At the leading order, the departure from the global spherical geometry is elliptical due to the combined action of centrifugal distortion and tides (as qualitatively expected from the theory of equilibrium figures for rotating fluid masses \cite{chandrasekhar1987ellipsoidal}). 
For the Earth's core, a typical value for polar ellipticity is $\sim 1/400$ (as estimated from \textsc{lod} variations \cite{davies2014strength} and the nutations \cite{dehant2022structure}).
Similarly, the equatorial ellipticity due to the lunar tides has the typical amplitude $\sim 10^{-7}-10^{-6}$ over geologic time scales, as estimated from hydrostatic tidal theory \cite{farhat2022resonant}. 
Contemplating the Earth's surface and oceans' floor (Figure \ref{fig:topomap}a), we observe that topographic features exist at all scales ranging from a few kilometres to a few hundred metres (i.e. meso-scales), but sub-metres structures are also present (i.e. roughness, not shown here). 
Meanwhile, mantle dynamics and global seismology suggest the presence of bumps and ridges at the Earth's \textsc{cmb}, with heights of a few kilometres and with horizontal extents of $1-10^3$~km (Figure \ref{fig:topomap}b), as well as small-scale roughness (e.g. due to thermo-chemical interactions between the liquid core and the solid mantle \cite{narteau2001smallscale,le2006dissipation}).
Fewer data are available for the ocean's floor or bottom ice layer of Jovian satellites.
Nonetheless, while they are insufficient to derive a precise picture, they all suggest that the interface is not smooth \cite{zebker2009size,iess2010gravity,nimmo2010shell}.

\begin{figure}
    \centering
    \includegraphics[width=0.8\textwidth]{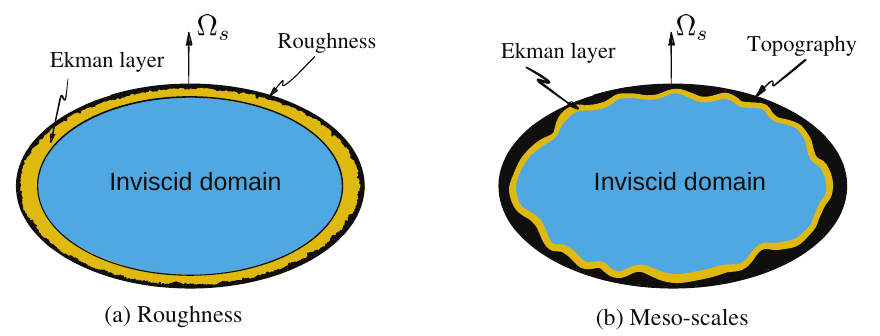}
    \caption{Illustration of the concept of \textbf{(a)} roughness and \textbf{(b)} meso-scale topography in a planetary context.}
    \label{fig:roughnessMesoscales}
\end{figure}

In the context of this review paper, we define thin-layer systems (e.g. the Earth's oceans and atmosphere) as fluid envelopes with a depth much smaller than the mean planetary radius, and thick-layer ones (e.g. planetary liquid cores and subsurface oceans) as envelopes whose depth is a large fraction of the radius. 
Let us finally introduce the concept of roughness and meso-scale topography (Figure \ref{fig:roughnessMesoscales}).
We refer below to roughness as small-wavelength topographies with height of the order of (or smaller than) the boundary-layer thickness.
The latter is typically $\sim 0.1-10$~m for a laminar viscous boundary layer in the Earth's oceans or in the Earth's core.
On the contrary, meso-scales are defined as topographies with small to intermediate wavelengths, whose height is large with respect to the Ekman-layer thickness but smaller than the large length scale $L$ (e.g. $100$~m to $100$~km). 

\subsection{Geophysical modelling, a challenge for numerics and experiments}
\begin{figure}
    \centering
    \begin{tabular}{cc}
        \includegraphics[width=0.45\textwidth]{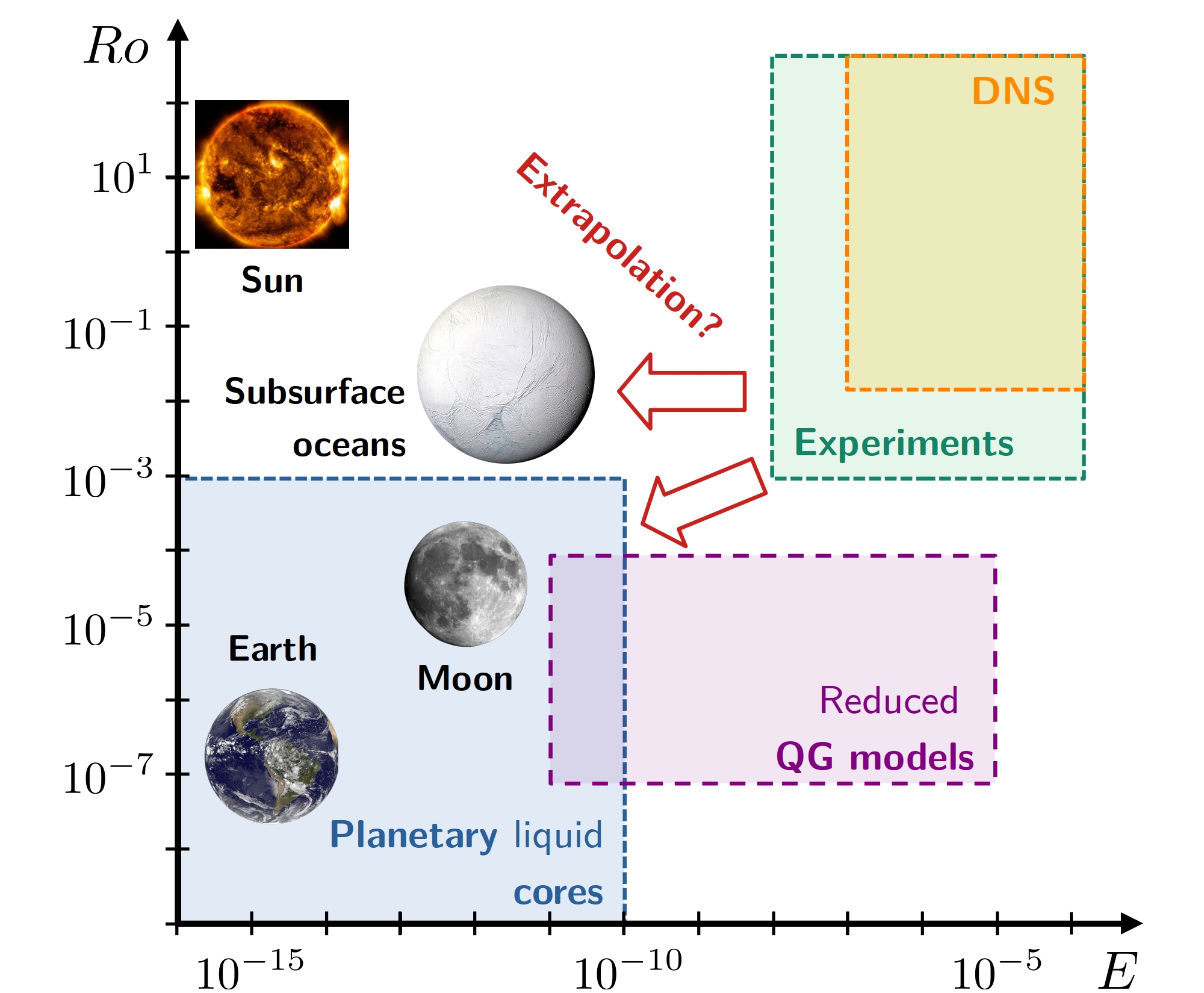} & 
        \includegraphics[width=0.45\textwidth]{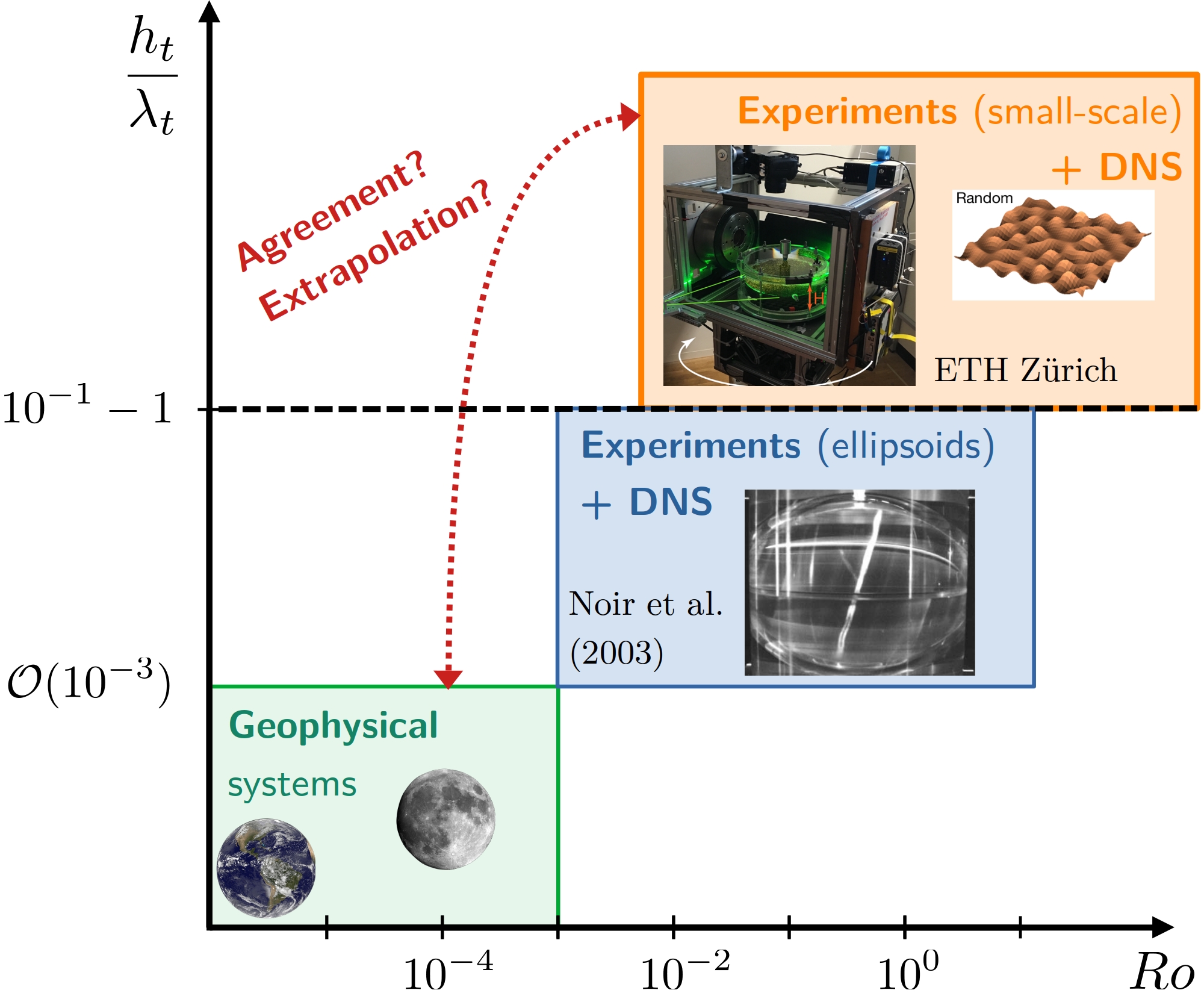} \\
        (a) & (b) \\
    \end{tabular}
    \caption{Schematic regime diagrams for the dynamics of thick-layer fluid systems. \textbf{(a)} Parameter space as a function of the Rossby number $Ro$ and the Ekman number $E$. It shows the huge gap between parameters characterising geophysical systems, and those that are accessible to experimental works and \textsc{dns} in global geometries. Reduced QG models refer to models of rapidly rotating convection in spherical geometries \cite{guervilly2019turbulent,barrois2022comparison}, in which the flow is assumed quasi-invariant along the rotation axis. \textbf{(b)} Parameter space depending on $h_t/\lambda_t$ (where $h_t$ is the topography amplitude at the wavelength $\lambda_t$, see Figure \ref{fig:exmp}b) and $Ro$.}
    \label{fig:parameters}
\end{figure}

Flows in geophysical fluid envelopes are mostly driven by two kinds of forcings, namely thermal-solutal convection \cite{jones2015tog} or mechanical forcings through coupling with a rigid boundary (e.g. at a \textsc{cmb}, or an icy crust in Jovian moons).
In geophysical systems, mechanical forcings typically result from orbital perturbations such as precession, librations, or tides \cite{le2015flows}. 
Convective flows in geophysical contexts are often characterised by slow time scales, while orbitally driven dynamics usually occur on time scales comparable to the planetary rotation time scale. 
This contrast gives rise to distinct regimes of geophysical turbulence. 
Rapidly rotating convection is expected to be quasi two-dimensional ($2-$D) in the planetary regime $Ro \ll 1$ \cite{nataf2024dynamic}.
In contrast, mechanical forcings can sustain anisotropic turbulence combining three-dimensional ($3-$D) flows (e.g. waves) and quasi $2-$D ones (e.g. mean zonal flows).
Since these phenomena are often simultaneously present in natural systems, geophysical flows often exhibit a wide range of time and length scales.
Moreover, geophysical systems are often characterised by rather extreme parameters (Figure \ref{fig:parameters}). 
One of the difficulties in geophysical studies is thus to account for such diverse phenomena, and to approach the appropriate range of parameters. 
In particular, modelling high-$Re$ turbulent flows in the low-$Ro$ regime is very challenging, as it requires lowering $E$ as much as possible. 
To this end, numerical and experimental approaches can be employed, and their pros and cons are summarised below. 
\vspace{0.5em}

\begin{paragraph}{Numerics}~
Direct numerical simulations (\textsc{dns}) usually offer a great flexibility to model flows in global geometries, or to account for phenomena that are difficult to control in laboratory conditions (e.g. dynamo action or thermal effects).
However, the main disadvantage is that probing the low-viscosity and rapidly rotating regime of geophysical flows is beyond the capabilities of any \textsc{dns}, even for the most advanced ones available today. 
For instance, the largest simulation of core convection struggled to consistently reach the value $E \sim 10^{-7}$ \cite{schaeffer2017turbulent}.
Moreover, it often takes months to only model a few turnover time scales, such that the long time scale of the flow remains currently out of reach in \textsc{dns}.
Fortunately, the expected almost $2-$D nature of rapidly rotating convection \cite{nataf2024dynamic} allows developing reduced models that can efficiently probe the low-$Ro$ turbulent regime \cite{guervilly2019turbulent,barrois2022comparison}.
Yet, we cannot easily adopt such an alternative route with topography. 
Indeed, the numerical challenge is more pronounced when incorporating topographic effects into the models. 
Despite efforts to develop efficient numerical methods capable of modelling topography (e.g. spectral-element methods), there remains a significant disparity of many orders of magnitude from geophysical values (e.g. with $E \sim 10^{-5}$ for the most turbulent flows in global ellipsoidal geometries \cite{favier2015generation,grannan2017tidally}). 
\end{paragraph}
\vspace{0.5em}
\begin{paragraph}{Experiments}~
As stated by K. S. Thorne and R. D. Blandford in their textbook \emph{Modern Classical Physics} \cite{thorne2017modern}, fluid dynamics is historically an experimental science.
Given the aforementioned limitations of \textsc{dns} for rapidly rotating flows with topography, experiments still appear as a valuable alternative nowadays. 
In short, experiments do not suffer from the difficulty of resolving a broad range of time and spatial scales, and they are not limited to weakly nonlinear regimes. 
Typically, a metre-size experiment using water and rotating up to $60$~ \textsc{rpm} corresponds to the values $E\sim10^{-7}$ and $Ro\sim 10^{-3}-10$. 
Another benefit of experiments is that very different shapes can be chosen for the fluid container. 
For that reason, they have proven invaluable to investigate ellipsoidal geometries or cavities with topographical features.
Briefly speaking, experiments must be designed with care in order to (i) ensure that the desired physical processes to investigate do control the dynamics, and (ii) provide sufficient quantitative information to test the models or derive scaling laws. 
Both aspects are paramount to successful experimental investigations, and they must be well-designed from the very beginning of the experimental development. 
A detailed discussion of experimental techniques is beyond the scope of the present review, but we can outline for the non-expert readers what is achievable by using non-intrusive measurements. 
There are different techniques to measure the velocity, pressure, temperature, or density.
They all provide partial views of the system, in contrast with \textsc{dns} giving access to all quantities in the fluid domain.
For instance, routinely used optical techniques in recent experiments \cite{sous2013friction,le2019experimental}, such as Particle Image Velocimetry (\textsc{piv}), typically allow recovering $2-$D maps of the velocity fields with certain spatial and temporal resolutions. 
With a typical mid-range setup, we can expect to measure velocities $u_\ell \sim 10^{-3} - 1$~m.s$^{-1}$, with a resolution of a few millimetres over a view field of tens of centimetres at a sampling frequency $\sim 10^2$~Hz. 
Dedicated highly customised \textsc{piv} systems, which are much more costly, can only marginally extend this range by a factor $\sim 10$. 
In some peculiar experimental setups, $3-$D reconstructions are also possible but difficult to perform (especially in rapidly rotating experiments). 
In opaque fluids (e.g. liquid metal), acoustic techniques must be employed, such as Ultrasonic Doppler Velocimetry (\textsc{udv}) \cite{noir2001experimental} or time of flight \cite{burmann2022low}, which only offer $1-$D profiles of a single velocity component (with roughly the same resolution range and slightly lower sampling frequency). 
Temperature, density, or pressure are often limited to point measurements.
More elaborated techniques, such as Laser Induced Fluorescence (\textsc{lif}) \cite{lherm2022rayleigh} or synthetic Schlieren \cite{sutherland1999visualization,dalziel2000whole}, can provide $2-$D maps but at the expense of a complex setup that is not always possible to mount on rotating devices. 
\end{paragraph}
\vspace{0.5em}

Contrary to non-rotating regimes (i.e. $Ro \to \infty$), for which laboratory experiments are miles ahead of \textsc{dns} for turbulent modelling, it appears that \textsc{dns} and experiments are rather close in terms of accessible parameters for rotating flows. 
Since the high-$Re$ and low-$Ro$ regimes are still largely unexplored, both numerical and experimental approaches are worth considering.
Moreover, since the extreme parameters relevant to planetary fluid envelopes are (and will remain) out of reach of both \textsc{dns} and experiments, we must resort to dimensional arguments and scaling laws for geophysical applications.
This requires a systematic exploration of the parameter space and, then, the results must be extrapolated using asymptotic laws constrained by the available data.
To this end, experiments are well suited to explore more complex small-wavelength geometries and, sometimes, extreme dynamical regimes than \textsc{dns}, but at the expense of limited quantitative information (in particular at low forcing). 
On the other hand, \textsc{dns} are suitable to study global geometries in laminar and less turbulent regimes, but with well-controlled physics and with access to all resolved quantities in the entire volume. 
Consequently, they often allow exploring different regimes in the parameter space (Figure \ref{fig:parameters}), which makes the two approaches complementary to infer robust scaling laws for geophysical applications.

\subsection{Outline of the paper}
Motivated by some applications discussed above, we will describe experimental works that have significantly contributed to our understanding of geophysical flows with topography.
We will only describe \textsc{dns} and theory when required for interpreting laboratory experiments.
Nonetheless, we point out that numerical and theoretical works remain essential to make progress in our understanding of these geophysical questions. As such, they would deserve an independent review. 
It is only through the combination of all three approaches that we shall gain insights into the geophysical dynamics of thick fluid envelopes.

The manuscript is divided as follows.
We first describe in \S\ref{sec:ellipsoid} the effects of large-scale ellipsoidal topographies on the flow dynamics driven by mechanical (orbital) forcings.
Next, we discuss in \S\ref{sec:smallscale} how a small-wavelength topography can modify rotating flows, with a particular focus on spin-up experiments with meso-scale topographies and the transition to turbulent (Ekman) boundary layers.
We then consider turbulent convection with small-wavelength topography in \S\ref{sec:convection}.
We briefly discuss in \S\ref{sec:stratification} why density stratification is challenging to account for with global rotation in experimental conditions, before ending the manuscript in \S\ref{sec:ccl}.

\section{Flow dynamics with a large-scale (ellipsoidal) topography}
\label{sec:ellipsoid}
We first consider the effects of large-wavelength topographies on the flow dynamics of thick-layer envelopes. 
Given the broad range of the subject, we have chosen to focus on flows driven by mechanical (orbital) forcings in liquid-filled ellipsoids without wall roughness, and we also discuss the case of spherical containers when appropriate (as the flow properties can be similar, e.g. for precession).
For geophysical applications, the mechanical forcings that have been mainly considered in the literature are 
\begin{itemize}
\item tidal forcing,
\item precession (i.e. a fluid-filled cavity whose axis of rotation is itself rotating around a secondary axis, which is fixed in the inertial frame),
\item librations (i.e. oscillations of the figure axes of an object with respect to a fixed, mean rotation axis), which can be either in longitude (i.e. an azimuthal harmonic modulation of the rotation rate of the cavity) or in latitude.
\end{itemize}
As discussed below, these forcings are capable of driving flow instabilities and (possibly) anisotropic turbulence in the bulk of the cavity, through either topographic or viscous couplings at the boundary.
Hence, they are worth considering for geophysical applications.

\begin{figure}
    \centering
    \includegraphics[width=0.93\textwidth]{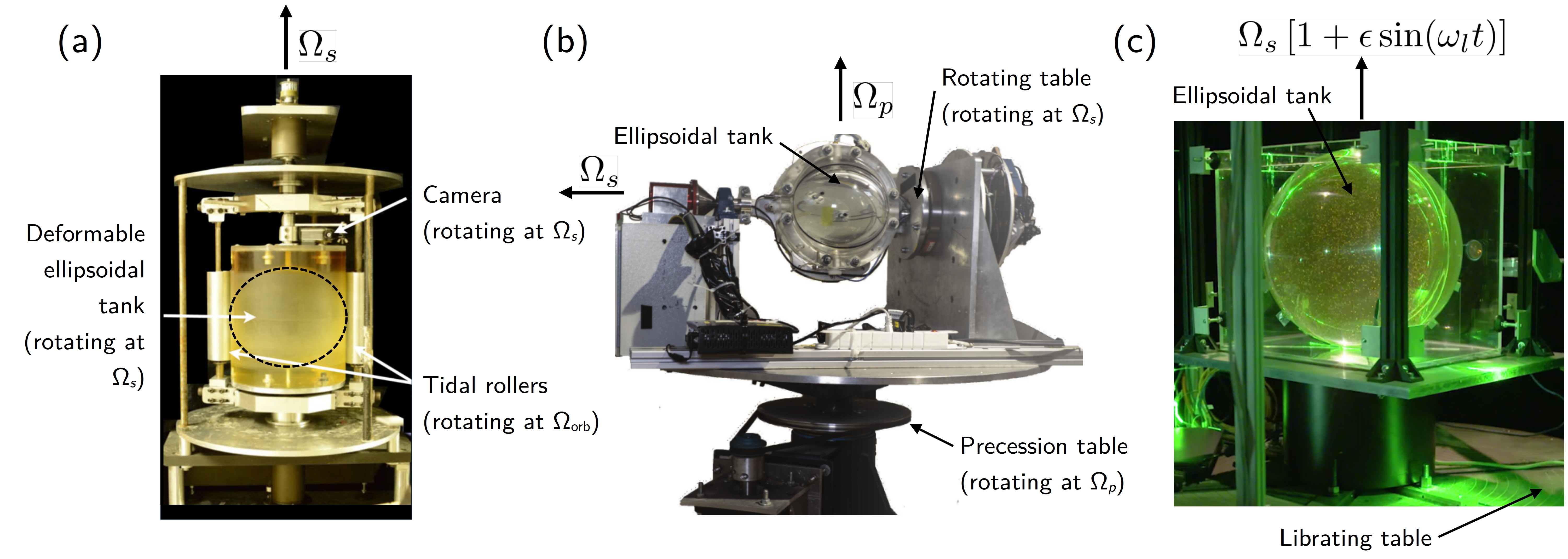}
    \caption{Illustrative experiments of mechanically driven flows in water-filled ellipsoids, which can go down to $E \sim 10^{-6}$ and $Ro \sim 10^{-2}$. \textbf{(a)} Tidal forcing. The fluid cavity, rotating at the angular velocity $\Omega_s$, is elliptically deformed by the action of two rollers rotating at $\Omega_\mathrm{orb}$. Adapted from \cite{grannan2017tidally}. \textbf{(b)} Precession. The ellipsoidal cavity is rotating around its $z-$axis at the angular velocity $\Omega_s \boldsymbol{1}_z$, which it is itself rotating at $\boldsymbol{\Omega}_p$ around another axis that is fixed in the inertial frame. The two axes are tilted by the precession angle $\alpha$ (here at $\pi/2$). Adapted from \cite{burmann2022experimental}. \textbf{(c)} Longitudinal librations. The cavity is rotating at the angular velocity $\Omega_s \left [1 + \epsilon \sin (\omega_l t) \right ]$, where $\Omega_s$ is the angular velocity of the fluid, $\epsilon$ is the libration amplitude, and $\omega_l$ is the libration (angular) frequency. Adapted from \cite{le2019experimental}.}
    \label{fig:manipellip}
\end{figure}

Investigating turbulent flows in ellipsoidal or spherical geometries is still very challenging, especially in the low-viscosity regime relevant to most geophysical systems (Figure \ref{fig:parameters}a). 
Starting with the seminal work of Malkus \cite{malkus1968precession}, 
flows driven in ellipsoids by orbital forcings have received increasing attention, especially using experimental and theoretical approaches.
Most studies have considered water-filled ellipsoids (Figure \ref{fig:manipellip}), but a few have also employed a liquid metal to explore hydromagnetic effects \cite{lacaze2006magnetic,herreman2009effects}.
A detailed discussion of all these studies is beyond the scope of the present paper.
We refer the readers to \cite{le2015flows} for an exhaustive review of the particularities of every forcing, and to \cite{le2022fluid} for a summary of the different experiments. 
Here, we first emphasise that experimental studies of mechanically driven flows in full ellipsoids (i.e. without an inner core) share some common points that can be understood in light of theoretical progress. 
Secondly, we discuss whether the experimental results can be extrapolated to planetary conditions, and outline some experimental perspectives to build more realistic models. 

\subsection{The full ellipsoid, a playground for joint experimental and theoretical works}
\label{subsec:fullellipsoid}
\subsubsection{Inertial modes, a cornerstone to understand rotating flows}
\begin{figure}
    \centering
    \begin{tabular}{cc}
        \includegraphics[width=0.45\textwidth]{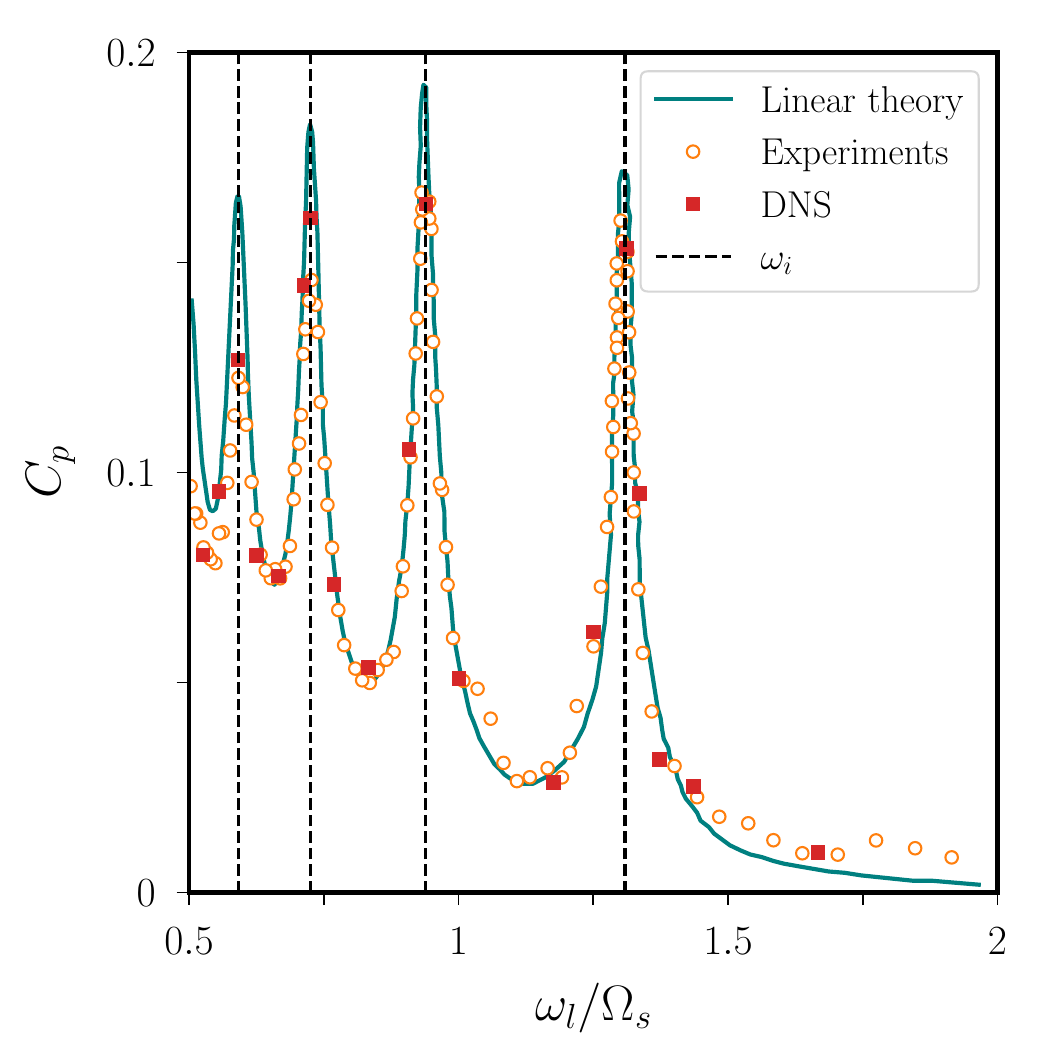} &
        \includegraphics[width=0.45\textwidth]{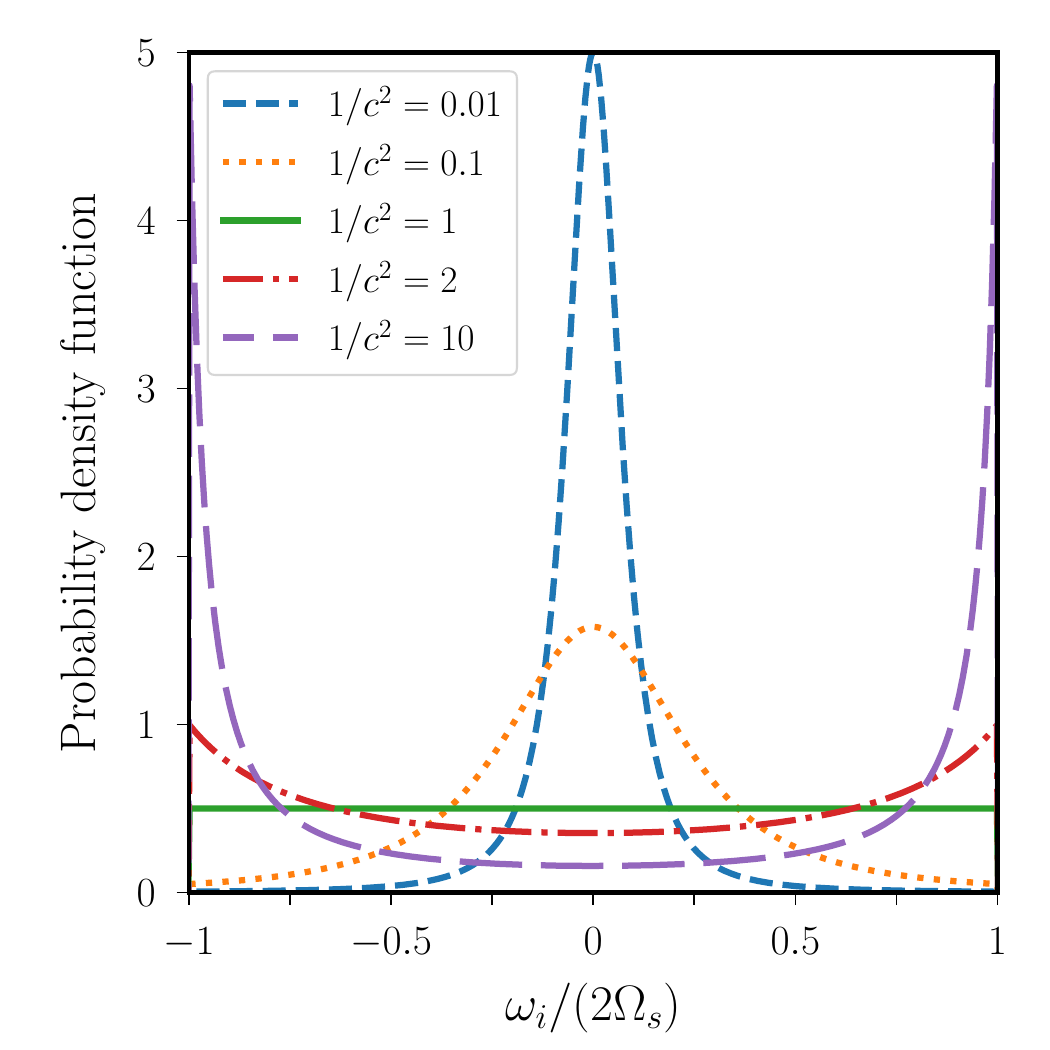} \\
        (a) & (b) \\
    \end{tabular}
    \caption{Inertial modes in a rapidly rotating sphere or ellipsoid. \textbf{(a)} Time average of the normalised pressure difference $C_p$ between the centre and the pole of a water-filled sphere subject to longitudinal librations (as in Figure \ref{fig:manipellip}c), where $\epsilon = \tilde{\epsilon} \omega_l/\Omega_s$ is the libration amplitude with $\tilde{\epsilon} = 8^\circ$, for the Reynolds number $Re_\omega = \omega_l a^2/\nu = 6.2 \times 10^{4}$ (with the radius $a = 10$~ cm of the sphere). Comparison between experiments, \textsc{dns}, and theoretical predictions. Adapted from figure 3 in \cite{aldridge1969axisymmetric} and figure 2 in \cite{sauret2010experimental}. \textbf{(b)} Asymptotic distribution of the normalised eigenvalues $\omega_i/(2\Omega_s)$ in the interval $[-1,1]$, in a spheroid with $a=b=1$ rotating along the $z-$axis, as a function of $1/c^2$. Adapted from figure 2 in \cite{CdV2023spectrum}.}
    \label{fig:inertialmodes}
\end{figure}

Rotating flows are often shaped by the action of inertial waves, whose restoring effect is global rotation \cite{greenspan1968theory}.
In a bounded geometry, these waves can admit counterparts known as inertial modes. 
We assume that the fluid is neutrally buoyant (i.e. $N=0$), and co-rotating with the rigid surrounding envelope at the steady angular velocity $\Omega_s \boldsymbol{1}_\Omega$ (where $\boldsymbol{1}_\Omega$ is a unit vector).
In the rotating frame, the fluid boundary is an ellipsoid given in Cartesian coordinates $(x,y,z)$ by $(x/a)^2 + (y/b)^2 + (z/c)^2 = 1$, where $(a,b,c)$ are the semi-major axes of the ellipsoid. 
In this rotating frame, an inertial mode is an inviscid solution of the eigenvalue problem given by
\begin{subequations}
\label{eq:inertialPB}
\begin{equation}
    \mathrm{i} \omega_i \boldsymbol{u}_i + 2\Omega_s (\boldsymbol{1}_\Omega \times \boldsymbol{u}_i) = - \nabla \Phi_i, \quad \nabla \boldsymbol{\cdot} \boldsymbol{u}_i = 0, \quad \left . \boldsymbol{u}_i \boldsymbol{\cdot} \boldsymbol{1}_n \right |_{\partial V} = 0,
    \tag{\theequation a--c}
\end{equation}
\end{subequations}
where $0 \leq |\omega_i| < 2 \Omega_s$ is the eigenvalue of the $i^\text{th}$ normal mode, $\boldsymbol{u}_i$ is the eigenvector for the velocity (respectively $\Phi_i$ for the pressure), and where $\boldsymbol{1}_n$ is the unit vector normal to the boundary $\partial V$.
Note that the degenerate solutions with $\omega_i=0$ are referred to geostrophic modes. 
They satisfy the Taylor-Proudman theorem 
\begin{equation}
  \boldsymbol{\Omega} \boldsymbol{\cdot} \nabla \boldsymbol{u}_i = \boldsymbol{0},
  \label{eq:TPtheo}
\end{equation}
showing that they are invariant along the rotation axis.
Inertial modes were first evidenced by the pioneering experiment of Aldridge \& Toomre \cite{aldridge1969axisymmetric}, who
considered a full sphere subject to longitudinal librations (as in Figure \ref{fig:manipellip}c).
Indeed, some inertial modes can be excited by librations when the libration frequency $\omega_l$ is close to some eigenvalues $\omega_i$ (Figure \ref{fig:inertialmodes}a).
Theoretical analysis later showed that such inertial modes can be excited because of viscous effects in the Ekman boundary layer \cite{zhang2013non,cebron2021mean,lin2023resonant}.
Other mechanisms can also sustain inertial waves or modes (after multiple reflections), such as boundary-layer instabilities \cite{sauret2013spontaneous} or bulk forcings \cite{morize2010experimental,zhang2017theory,lin2021triadic}.

Next, we may wonder how inertial modes are modified when the sphere is deformed into an ellipsoid.
Insight into this question can be gained by looking at the two simplest inertial modes, called the spin-over modes.
They are associated with a uniform vorticity along the equatorial $x-$ or $y-$axes. 
Their eigenvalues $\omega_\text{so}$ are given by \cite{vantieghem2014inertial}
\begin{equation}
    \omega_\text{so} (a,b,c) = \pm 2 \Omega_s \frac{ab}{\sqrt{a^2+c^2}\sqrt{b^2+c^2}}.
    \label{eq:spinover}
\end{equation}
The ellipsoidal deformation is thus responsible for a shift in frequency of the spherical eigenvalue $\omega_\text{so} (1,1,1) = \Omega_s$.
Similar effects exist for higher-order inertial modes, which are more or less pronounced depending on the modes' complexity.    
Therefore, detecting some inertial modes could be a way to probe the ellipticity of a fluid envelope (which is linked to Hopkins' original idea \cite{hopkins1839nutations}).
Figure \ref{fig:inertialmodes}~(b) illustrates a more subtle effect of the large-scale ellipsoidal topography.
As proved in \cite{CdV2023spectrum}, the frequency distribution of the eigenvalues is different in a sphere and an ellipsoid. 
The probability density function is uniform in a sphere, but non-uniform in an ellipsoid. 
For instance, low-frequency modes with $|\omega_i| \to 0$ are favoured when the ellipsoid is strongly elongated with $c/a\gg 1$, whereas high-frequency modes with $|\omega_i| \to 2 \Omega_s$ are enhanced in flattened ellipsoids with $c/a\ll 1$.
Consequently, this effect likely weakens the physical relevance of the oldest works, which employed strongly deformed ellipsoids to overcome diffusive effects, for modelling the wave-driven dynamics of geophysical flows in weakly deformed spheres.

Finally, a key property of inertial modes in an ellipsoid is that any divergenceless and square-integrable\footnote{i.e. any incompressible flow, either viscous or not, but with a finite value of the volume-averaged kinetic energy.} velocity $\boldsymbol{v}$ can be sought as a combination of the inviscid eigenvectors (including the geostrophic ones) as \cite{backus2017completeness,CdV2023spectrum}
\begin{equation}
\boldsymbol{v}(\boldsymbol{r},t) = \sum_{i=1}^{\infty} \gamma_i(t) \boldsymbol{u}_i(\boldsymbol{r}),
\label{eq:IMexpansion}
\end{equation}
where $\boldsymbol{r}$ is the position vector and $\{\gamma_i (t)\}_i$ are arbitrary time-dependent coefficients.
Therefore, many experimental results have been interpreted by using theories based upon inertial modes. 
So far, several phenomena have been experimentally observed, and the interplay with theory has allowed us to understand a somehow generic flow response in an ellipsoid (figures \ref{fig:poincare} to \ref{fig:turbellip}).

\subsubsection{Forced (laminar) flows, the first brick}
\begin{figure}
    \centering
    \begin{tabular}{cc}
        \includegraphics[width=0.44\textwidth]{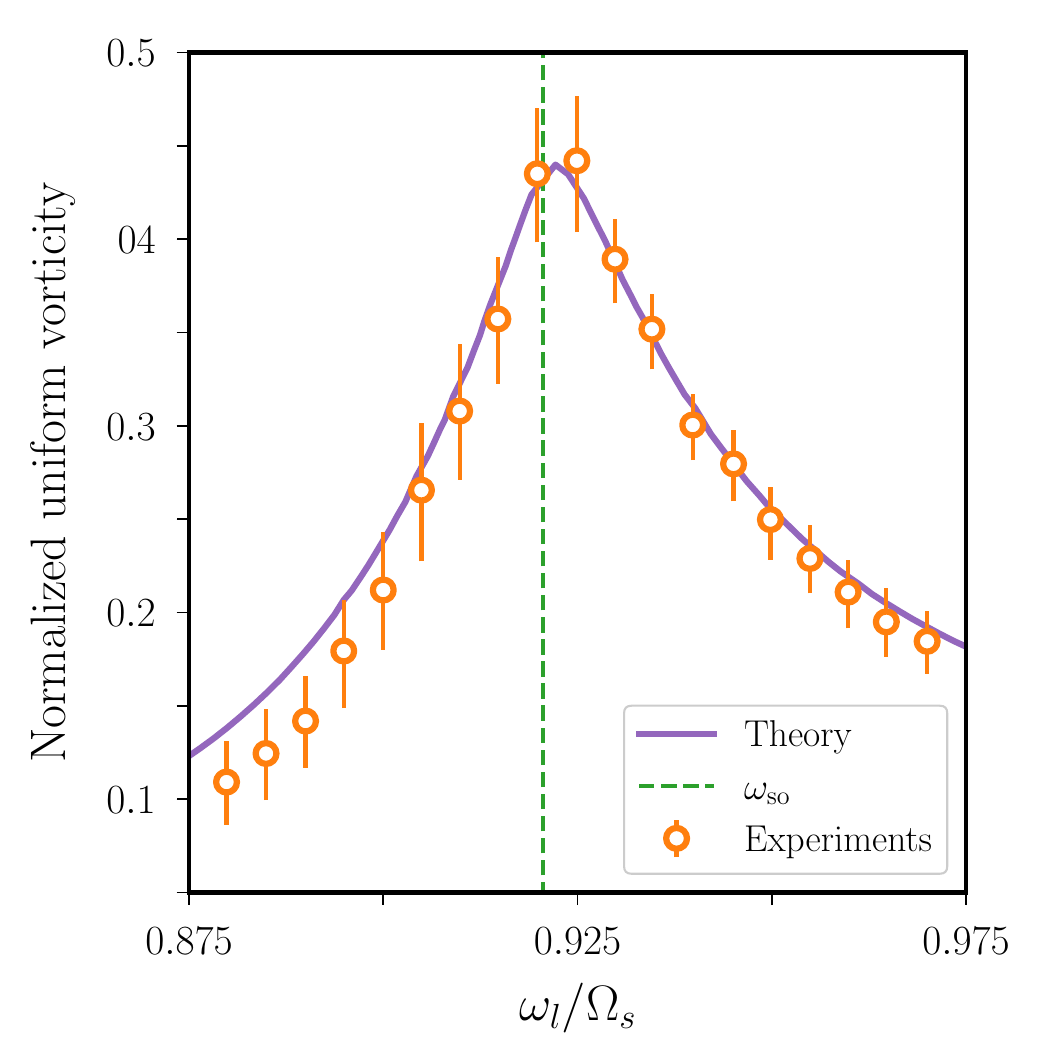} & 
        \includegraphics[width=0.44\textwidth]{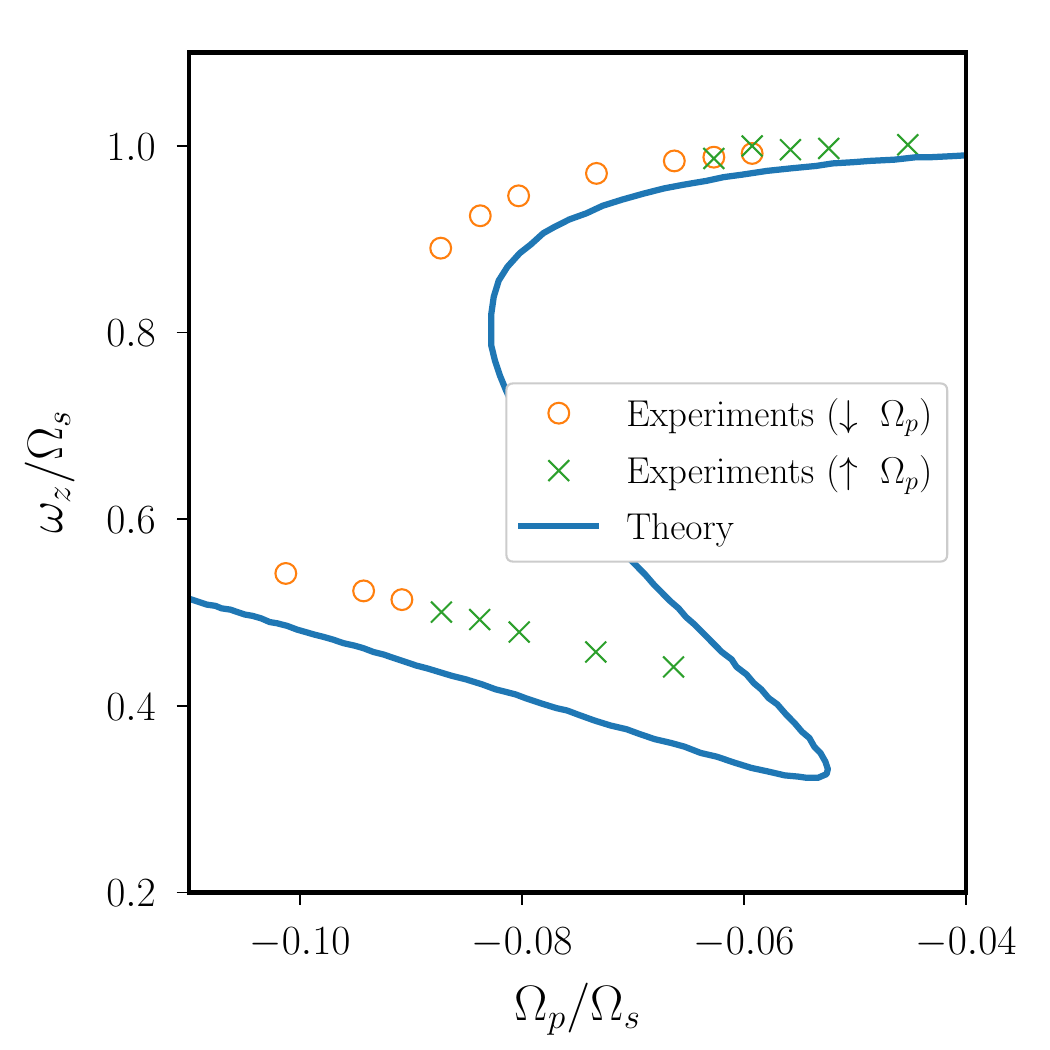} \\
        (a) & (b) \\
    \end{tabular}
    \caption{Laminar uniform vorticity flows, as obtained in experiments and predicted by theory. \textbf{(a)} Direct resonance for latitudinal librations. Triaxial ellipsoid with $a=0.078$~m, $b=0.125$~m and $c=0.1042$~m, subject to librations at $E=1.57 \times 10^{-5}$. Vertical dashed line shows the value of $\omega_\text{so}$ given by Equation (\ref{eq:spinover}) for this geometry. Experimental values obtained using Ultrasonic-Doppler-Velocimetry (\textsc{udv}) measurements, and the theory is given by a uniform-vorticity model \cite{vantieghem2015latitudinal}. See further details in \cite{charles2018flows}. 
    \textbf{(b)} Vertical component of the fluid rotation vector $\omega_z/\Omega_s$, as a function of $\Omega_p/\Omega_s$, in a spheroid with $c/a \simeq 0.847$. Precession forcing with $E = 2.3 \times 10^{-5}$ and a precession angle of $15^\circ$. Experimental values obtained using \textsc{piv} measurements. Hysteresis is found by either gradually decreasing ($\circ$) or increasing ($\times$) the value of $Po$. Theory obtained from \cite{busse1968steady}. Adapted from figure 4 in \cite{nobili2021hysteresis}.}
    \label{fig:poincare}
\end{figure}

Beyond inertial modes or waves, mechanical forcings often drive a (laminar) basic flow.
It can involve in Equation (\ref{eq:IMexpansion}) either a single modal component (in the case of a perfect resonance, as in Figure \ref{fig:inertialmodes}a) or a superposition of few large-scale modes. 
In the particular case of precession, the forced basic flow is close to a uniform-vorticity flow (even with viscosity \cite{busse1968steady,kida2020steady}), which is essentially a tilted rotation with an irrotational strain to satisfy the non-penetration at the boundary \cite{roberts2011flows,noir2013precession}. 
Such uniform-vorticity flows, which are often referred to as Poincar\'e flows, are the only ones to carry angular momentum in an ellipsoid \cite{ivers2017enumeration,vidal2023precession}.
As such, they are often expected to play a particular role in the rotational dynamics of the whole planet \cite{rekier2022earth}. 
Depending on the frequency of the forcing, the forced flow can exhibit direct resonances (Figure \ref{fig:poincare}a), whose amplitude is smoothed out by viscosity in experiments \cite{noir2013precession,vantieghem2015latitudinal,vidal2023precession}. 

\begin{figure}
    \centering
    \includegraphics[width=0.70\textwidth]{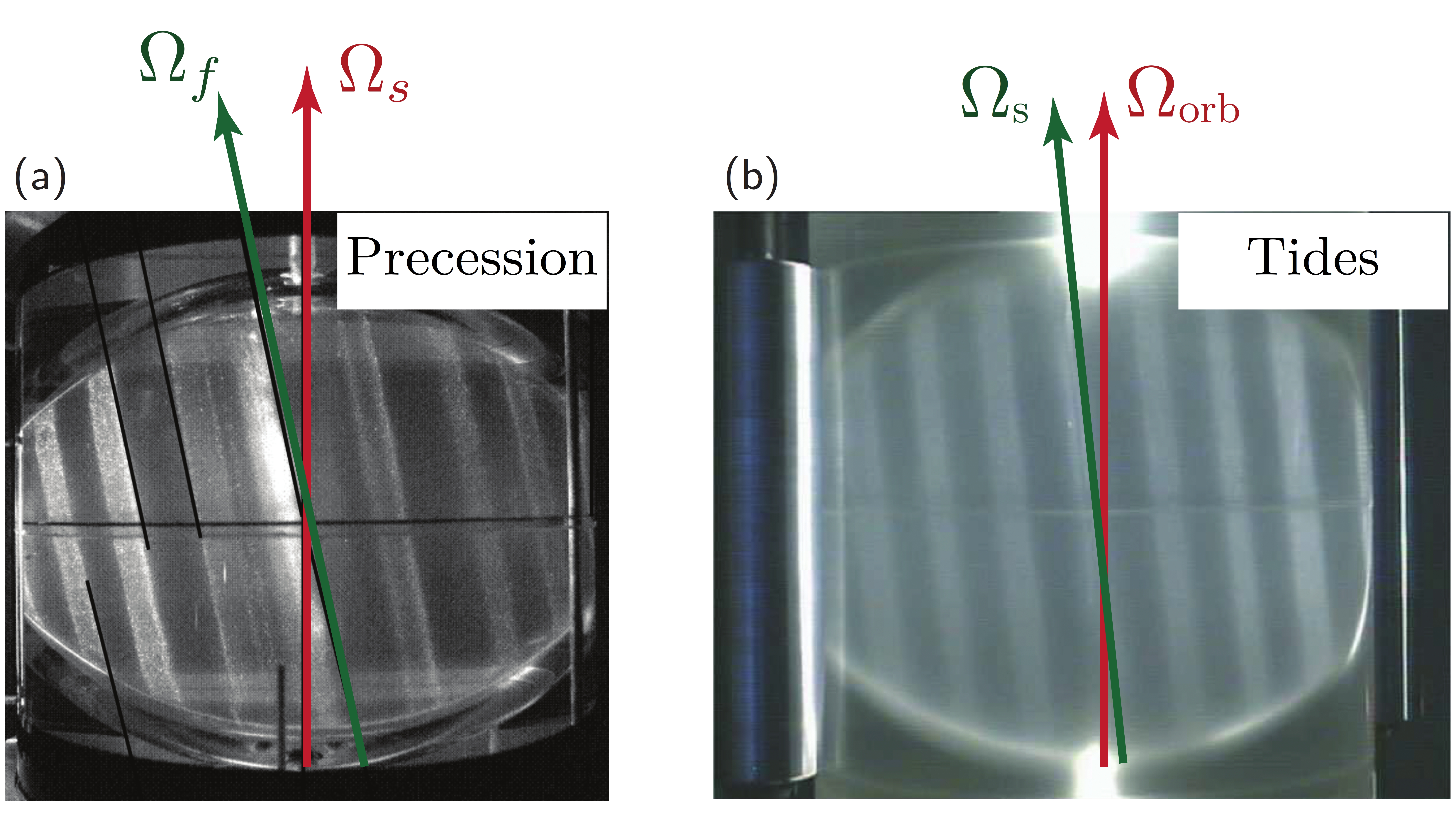}
    \caption{Similarities of mean (laminar) flows observed in ellipsoids subject to mechanical forcings. Visualisations in a meridional plane using Kalliroscope flakes. \textbf{(a)} Precession in a spheroid with $c/a = 0.96$. Experiment performed at $E=2.24 \times 10^{-6}$ and $\Omega_p/\Omega_s=-0.01$, for a precession angle of $20^\circ$. $\boldsymbol{\Omega}_f$ is the axis of rotation of the fluid, which is tilted from the mantle spin axis along $\Omega_s$. Adapted from \cite{noir2000ecoulements}. \textbf{(b)} Tidal flow in an ellipsoid. Experiment performed at $E = 10^{-5}$. Here, the orbital angular velocity along $\Omega_\text{orb}$ (enforced by the two rollers) is inclined by an angle of $5^\circ$ from the spin axis along $\Omega_s$ (which is aligned with $\boldsymbol{\Omega}_f$). Adapted from \cite{morize2010experimental}. }
    \label{fig:geosCylinderPrecTide}
\end{figure}

Despite their apparent simplicity, laminar basic flows in ellipsoids can have a rather complicated evolution in the parameter space. For instance, the uniform-vorticity flow in a precessing cavity can be non-unique for a broad range of precession rates (Figure \ref{fig:poincare}b), as found in experiments  \cite{nobili2021hysteresis,burmann2022experimental} and in agreement with prior theory \cite{busse1968steady,cebron2015bistable}.
Some viscously driven flows can also be important ingredients to build a complete model, such as the oblique shear layers generated by the breakdown of the oscillatory Ekman layer at the critical latitudes \cite{noir2001experimental,kida2011steady}, or laminar mean flows resulting from nonlinear interactions in the Ekman boundary layer (as reported in experiments for all three forcings \cite{suess1971viscous,sauret2010experimental,noir2012experimental,grannan2017tidally}) or from self-interactions of wave motions (as predicted \cite{tilgner2007zonal,le2020reflection,lin2021libration} and experimentally observed \cite{morize2010experimental}). 
It is interesting to note the similarity of the mean (zonal) flows obtained in both tidally-driven and precession-driven flows (Figure \ref{fig:geosCylinderPrecTide}). In both cases, the tilted rotation of the fluid (directly forced for precession, but resulting from a triadic instability for tidal forcing, see below) generates an almost identical pattern of zonal flows, which have yet to be fully understood.  

\subsubsection{The long route to rotating turbulence}
\begin{figure}
    \centering
    \includegraphics[width=0.97\textwidth]{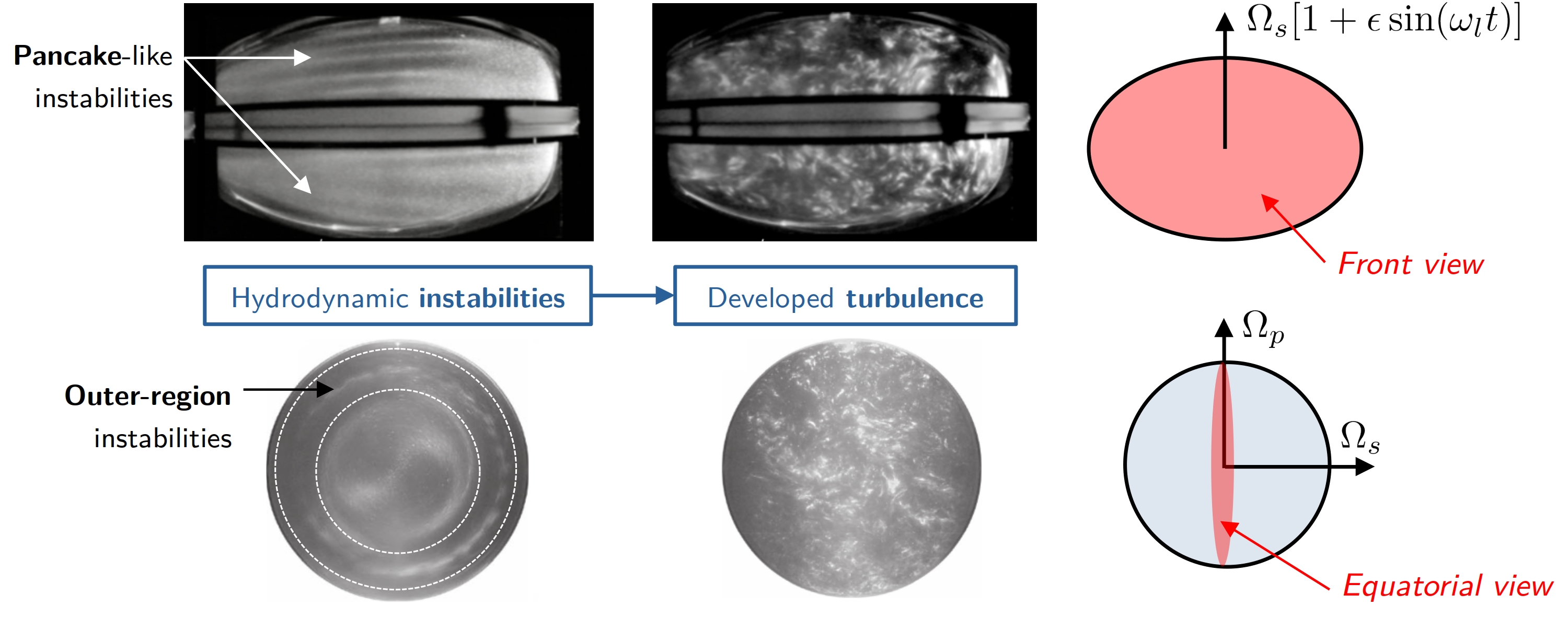}
    \caption{Schematic flow response in ellipsoids subject to mechanical forcings. \emph{Top}: Front visualisations of libration-driven flows in an ellipsoid at $E\simeq2.7 \times 10^{-5}$. Adapted from figure 3 in \cite{grannan2017tidally}. \emph{Bottom}: Equatorial visualisations (using reflective flakes) of precession-driven flows in a sphere (with orthogonal spin and precession axes) at $E \simeq 1.25 \times 10^{-5}$. Adapted from figures 5 and 13 in \cite{goto2014turbulence}.} 
    \label{fig:resellip}
\end{figure}

The forced laminar flow is usually linearly unstable to small-amplitude perturbations. 
As outlined in Figure \ref{fig:resellip}, we often observe the following route to turbulence in experiments, from the onset of hydrodynamic instabilities (when the forcing amplitude is strong enough to overcome diffusive effects) to anisotropic turbulence.
\vspace{0.5em}
\begin{paragraph}{(1) Instabilities}~
Linear stability theory suggests that triadic (parametric) instabilities are often triggered by mechanical forcings. 
In diffusionless fluids, these triadic (parametric) instabilities obey the generic resonance condition given by 
\begin{equation}
    \omega_0 = \omega_1 \pm \omega_2,
\end{equation}
where $\omega_0$ is the angular frequency of the forced Poincar\'e flow viewed in the frame rotating with the fluid (e.g. $\omega_l$ for longitudinal librations as in Figure \ref{fig:manipellip}c), and $\omega_{1,2}$ are the angular frequencies of two (free) inertial modes of the inviscid fluid. 
A triadic (parametric) instability can only grow upon the basic flow if its diffusionless growth rate $\sigma$ is larger than the diffusive damping effects in the system. 
For an ellipsoidal boundary without roughness, the leading-order dissipative mechanism in a hydrodynamic experiment is provided by the friction within the Ekman layer \cite{greenspan1968theory}. 
For a laminar Ekman layer, the onset criterion for triadic instabilities is thus given by
\begin{equation}
\sigma > \mathcal{O}(E^{1/2}).
\label{eq:growthvsdamping}
\end{equation}
For the mechanical forcings listed above, the diffusionless growth rate $\sigma$ of the flow instabilities is often proportional to the ellipsoidal deformation (at the leading order).
If so, criterion (\ref{eq:growthvsdamping}) means that the ellipsoidal deformation must be greater than the thickness of the Ekman layer (typically $\sim 0.1-1$~m in core conditions for a laminar Ekman layer) for the onset of instabilities. 
Other damping mechanisms, such as an enhanced spin-up because of meso-scale topographies or roughness (see \S\ref{sec:smallscale}), could thus naturally affect the viability of such instability mechanisms. 
The most famous example of such instabilities is the elliptical instability \cite{kerswell2002elliptical}, which can be driven by tidal forcing or librations in experiments \cite{lacaze2004elliptical,lacaze2006magnetic,herreman2009effects,le2010tidal,grannan2014experimental,grannan2017tidally,le2019experimental}. 
For precession, various triadic instabilities are expected in inviscid ellipsoids \cite{kerswell1993instability}, but their clear identification remains difficult given the accessible range of parameters in experiments \cite{nobili2021hysteresis,burmann2022experimental,burmann2024precessing}.
Indeed, current experiments using water-filled ellipsoids are usually limited to $E \sim \mathcal{O}(10^{-6})$ (see tables 1 \& 2 in \cite{le2022fluid}).
Another parametric instability of interest for precession is the so-called conical shear instability (CSI).
It results from triadic interactions between two inertial modes and the viscous conical shear layer, such that this instability can be observed in spherical geometries \cite{lin2015shear,cebron2019precessing}. 
The CSI was probably present in the seminal experiments of Malkus \cite{malkus1968precession} and Vanyo \cite{vanyo1995experiments}, but the unambiguous experimental observation only came later using a spheroid \cite{horimoto2020conical}. 
Note that the above parametric instabilities can also coexist with other hydrodynamic instabilities, such as the viscous shear-driven instability for tides \cite{sauret2014tide}, or boundary-layer instabilities for precession \cite{lorenzani2001fluid,cebron2019precessing} and librations \cite{noir2009experimental,sauret2013spontaneous}.
\end{paragraph}
\vspace{0.5em}

\begin{figure}
    \centering
    \begin{tabular}{cc}
    \centering
    \includegraphics[height=0.35\textwidth]{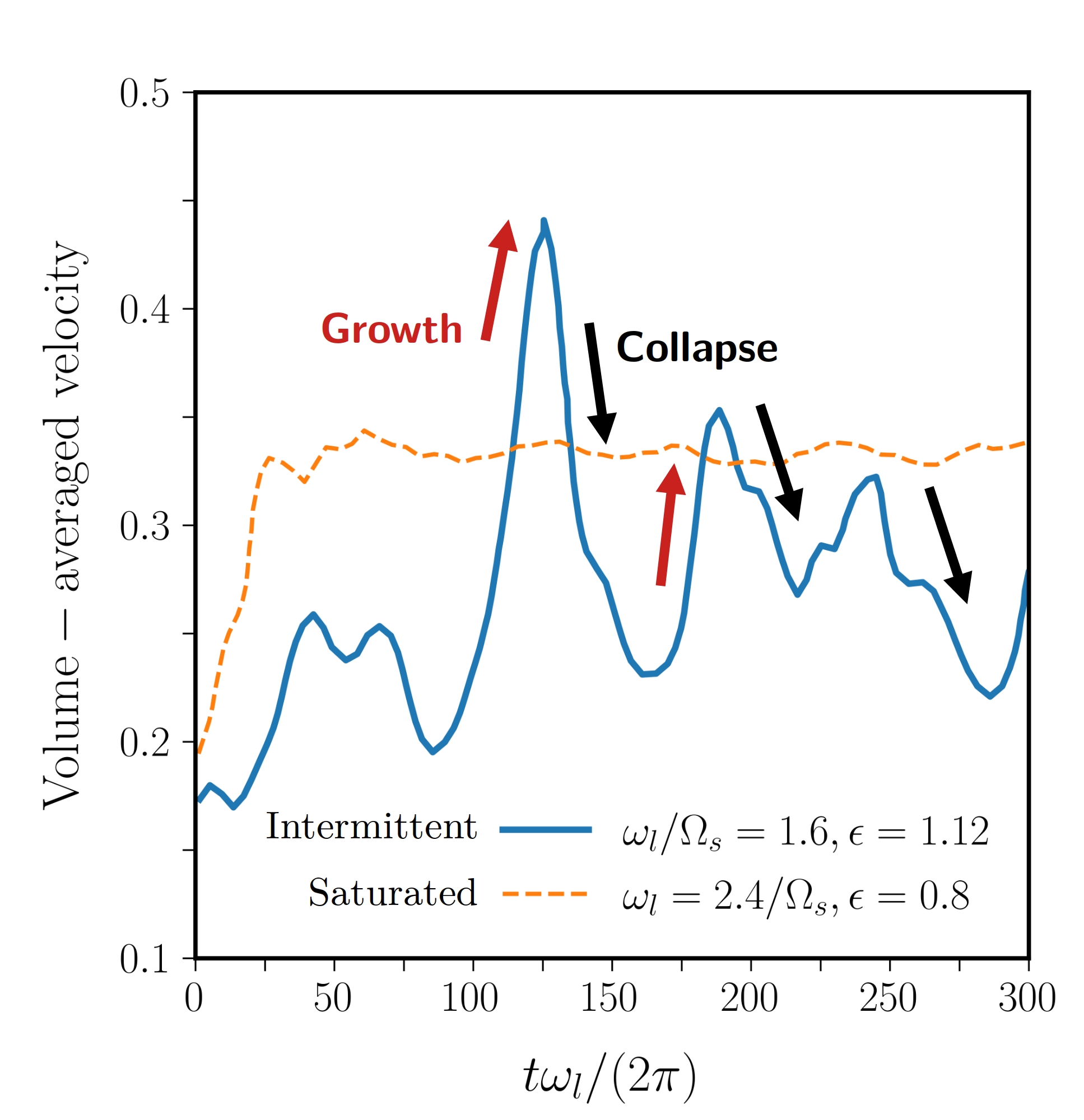} & 
    \includegraphics[height=0.35\textwidth]{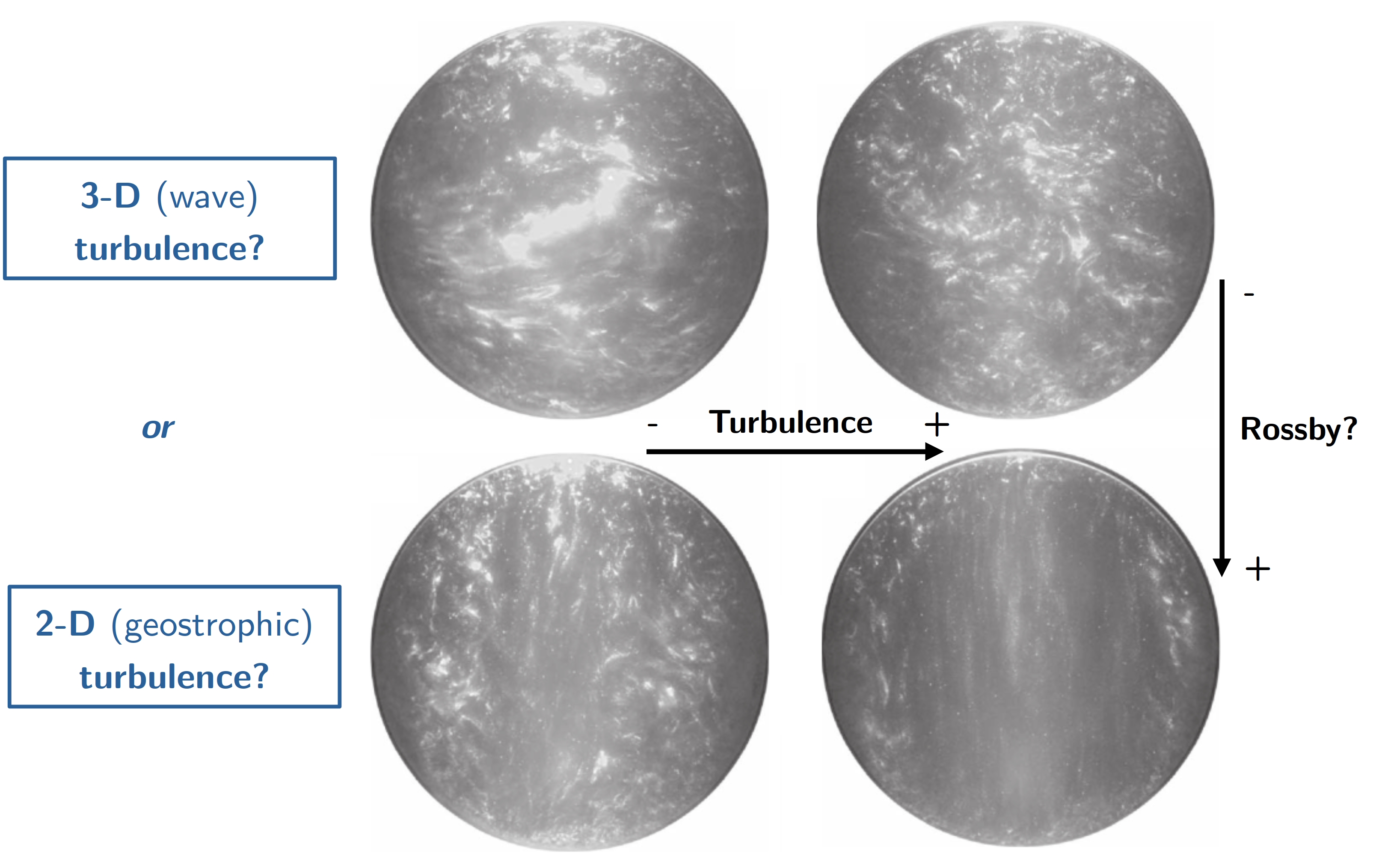} \\
    (a) & (b) \\
    \end{tabular}
    \caption{\textbf{(a)} Different turbulent regimes of libration-driven flows in an ellipsoid at $E\simeq2 \times 10^{-5}$. Vertical axis shows the volume-averaged velocity (measured from equatorial \textsc{piv} in the experiment), which is also averaged in time using four points per libration cycle to only retain the long-term time scales of the flow. Adapted from figure 6 in \cite{grannan2014experimental}. \textbf{(b)} $3-$D (wave) turbulence versus $2-$D (geostrophic) turbulence. Equatorial visualisations (using reflective flakes) of precession-driven turbulence in a sphere (with orthogonal spin and precession axes) at $E \simeq 1.25 \times 10^{-5}$. Adapted from figure 5 in \cite{goto2014turbulence}.} 
    \label{fig:turbellip}
\end{figure}

\begin{paragraph}{(2) Rotating turbulence}~Next, anisotropic turbulent flows are usually observed in experiments (Figure \ref{fig:resellip}).
The different mechanisms of linear hydrodynamic instability are rather well understood now. 
On the contrary, our understanding of the route to turbulence is still incomplete. 
Theoretical works suggest that such a transition could occur through nonlinear saturation of the primary flows, or because of various mechanisms involving inertial waves or modes \cite{kerswell1999secondary,le2017inertial,le2019experimental}. 
Determining the relevant turbulent regimes, which may be forcing-dependent, is still an open question.
As illustrated in Figure \ref{fig:turbellip}~(a), experiments have shown that the mechanically driven turbulence can be either intermittent \cite{noir2012experimental,grannan2014experimental,grannan2017tidally} (e.g. with successive chaotic cycles made of resonant growth, saturation, collapse and relaminarisation), or sustained \cite{goto2014turbulence,horimoto2017sustaining,horimoto2018impact,le2019experimental}. 
Moreover, the developed anisotropic turbulence could be either $3-$D or quasi $2-$D (Figure \ref{fig:turbellip}b). 
These two regimes have been observed in experiments, such as for precession \cite{goto2014turbulence} or librations \cite{le2019experimental}.
Finally, strong turbulence can be obtained if the forcing amplitude is large enough, which may be relevant for engineering applications \cite{meunier2020geoinspired,goto2023precessing}.
\end{paragraph}

\subsection{Unanswered questions for future experiments}
\label{subsec:extrapolation}
Experimental (and theoretical) works have allowed us to get satisfactory physical insight into the flow response of liquid-filled ellipsoids to mechanical forcings, from the generation of laminar flows to the transition to turbulence (figures \ref{fig:poincare} to \ref{fig:turbellip}). 
However, the story is not over and there is room for developing new experiments to attack some unsolved problems. 
As outlined below, obtaining a better fundamental understanding of the various turbulent states is crucial, in particular to extrapolate the results to geophysical conditions safely.

\subsubsection{Insight from turbulence studies}
\begin{figure}
    \centering
    \begin{tabular}{cc}
    \centering
    \includegraphics[width=0.48\textwidth]{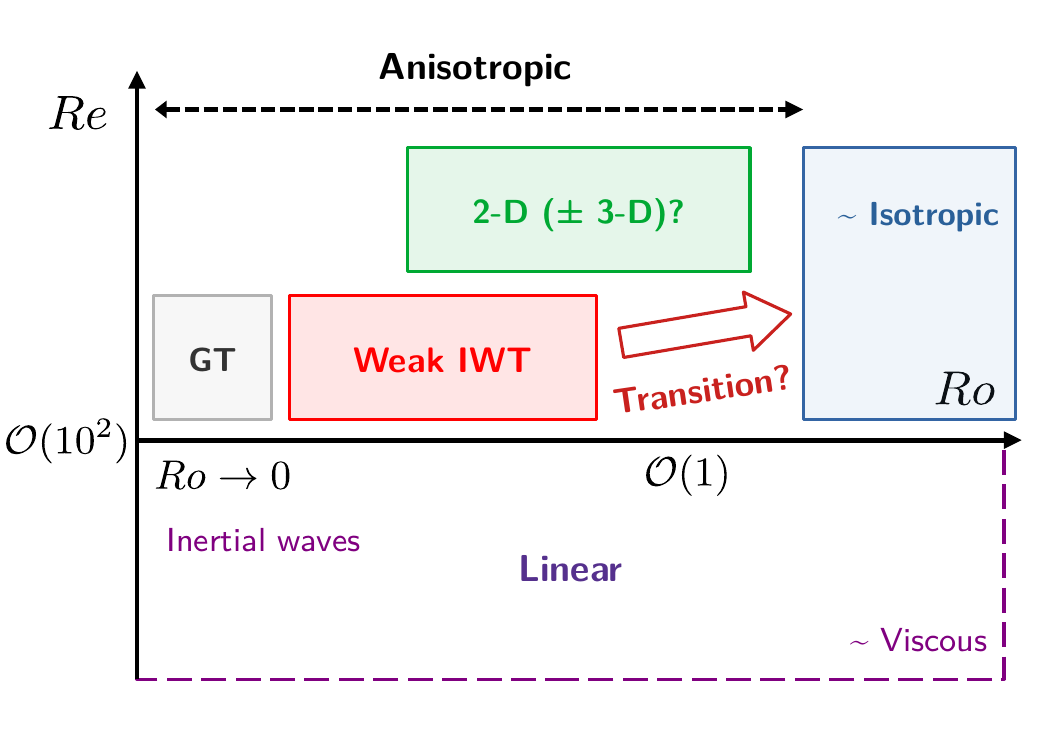} & 
    \includegraphics[width=0.48\textwidth]{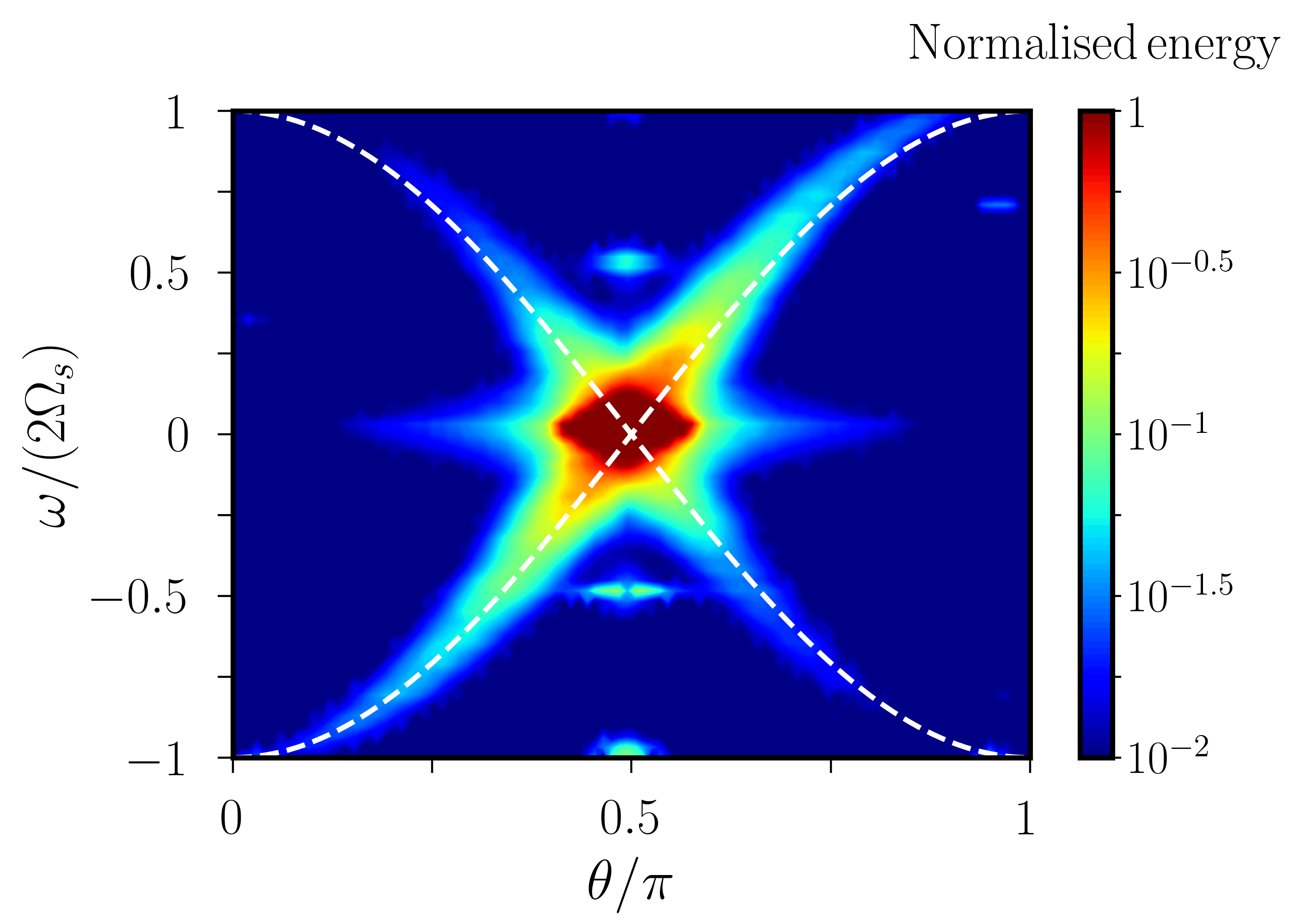} \\
    (a) & (b) \\
    \end{tabular}
    \caption{\textbf{(a)} Schematic regime diagram for rotating turbulence as a function of $Re$ and $Ro$. Adapted from figure 1 in \cite{godeferd2015structure}. GT: Geostrophic Turbulence. IWT: Inertial Wave Turbulence. \textbf{(b)} Experimental observation of inertial wave turbulence, adapted and modified from \cite{yarom2014experimental}. The experimental set-up is a rotating cylinder of $90$~cm in height and $80$~cm in diameter, which is rotating at $\Omega_s = 4 \pi$~rad.s${}^{-1}$. Colour bar shows the normalised density energy (in logarithmic scale), measured for the entire wave number range in the experiment, as a function of the frequency $\omega$ and the angle $\theta$ between the rotation axis and the wave vector (measured from the rotation axis). White dashes curves show dispersion relation (\ref{eq:dispIW}) of plane inertial waves, along which the energy is concentrated. The strongest energy is near $\theta \approx \pi/2$, which evidences a strong geostrophic component.}
    \label{fig:WT}
\end{figure}

The ability of mechanically driven flows to exhibit different states of rotating anisotropic turbulence (Figure \ref{fig:turbellip}b) is really fascinating. 
Indeed, it echoes the fundamental problem of rotating turbulence in low-$Ro$ flows \cite{godeferd2015structure}.
As sketched in Figure \ref{fig:WT}(a), different regimes of rotating turbulence can be obtained as a function of $Ro$ and $Re$.
For instance, rotating turbulence can be in a quasi $2-$D regime in which energy is concentrated in the form of quasi-geostrophic flows (i.e. with flows almost invariant along the rotation axis). 
Since rapidly rotating convection is known to sustain low-frequency and quasi $2-$D flows when $Ro \ll 1$ (e.g. \cite{guervilly2019turbulent,barrois2022comparison} in spherical geometries), the regime of quasi-geostrophic turbulence is widely studied in the geophysical community \cite{nataf2024dynamic}. 
However, when $Ro$ is not vanishingly small, $3-$D inertial waves can also take part in the anisotropic turbulence and transfer energy to smaller scales through a direct energy cascade. 
Some experiments have indeed shown the key role played by inertial waves \cite{yarom2014experimental,campagne2015disentangling,yarom2017experimental,salhov2019measurements,monsalve2020quantitative}, which can still coexist with quasi $2-$D columnar motions and an inverse cascade \cite{duran2013turbulence,campagne2014direct,shaltiel2024direct}.
Note that different theories or scaling laws have been proposed for rotating turbulence, for instance using phenomenological arguments \cite{zhou1995phenomenological,nazarenko2011critical,baqui2015phenomenological}, various closure models \cite{galtier2003weak,cambon2004advances,bellet2006wave} or from \textsc{dns} \cite{smith1999transfer,thiele2009structure}. 
They often yield different predictions, against which experimental results can be tested.
Currently, most rotating experiments are designed in the light of weak Inertial Wave Turbulence (IWT).
The latter is indeed of great interest for turbulence modelling because it provides, in the asymptotic regime $Ro \ll 1$, a natural closure of the turbulent equations \cite{galtier2023multiple}. 

Weak IWT considers weakly nonlinear interactions (when $Ro \ll 1$) of plane inertial waves, which satisfy the plane-wave dispersion relation given by
\begin{equation}
    \omega(\boldsymbol{k}) = \pm 2 \Omega_s \frac{k_\parallel}{|\boldsymbol{k}|}
\label{eq:dispIW}
\end{equation}
from equations (\ref{eq:inertialPB}a,b), where $\boldsymbol{k}$ is the wave vector with $k_\parallel = \boldsymbol{k} \boldsymbol{\cdot} \boldsymbol{1}_\Omega$ and the perpendicular wave vector $\boldsymbol{k}_\perp = \boldsymbol{k} \times \boldsymbol{1}_\Omega$ of norm $k_\perp$.
In the theory, the weak nonlinearities take the form of three-wave interactions. 
For wave-vector triads $[\boldsymbol{k},\boldsymbol{p},\boldsymbol{q}]$ satisfying the resonance conditions given by
\begin{subequations}
\label{eq:disprelWKE}
\begin{equation}
    \boldsymbol{k} + \boldsymbol{p} + \boldsymbol{q} = \boldsymbol{0}, \quad \omega(\boldsymbol{k}) + \omega(\boldsymbol{p}) + \omega(\boldsymbol{q}) = 0,
    \tag{\theequation a,b}
\end{equation}
\end{subequations}
(weak) IWT is governed by a Wave Kinetic Equation (WKE) written in symbolic form as \cite{galtier2022physics}
\begin{equation}
    \partial_t \mathcal{N}(\boldsymbol{k}) = Ro^2 \mathcal{T}(\boldsymbol{k}),
    \label{eq:WKE}
\end{equation}
where $\mathcal{N}(\boldsymbol{k}) = \mathcal{E}_a(k_\perp, k_\parallel)/\omega(\boldsymbol{k})$ is the wave action with the axisymmetric kinetic energy spectrum\footnote{It is defined as $\mathcal{E}_a(k_\perp,k_\parallel) = 2 \pi k_\perp \mathcal{E}(\boldsymbol{k})$ with the kinetic energy spectrum $\mathcal{E}(\boldsymbol{k})$, such that the total kinetic energy is $\propto \int^\infty \mathcal{E}_a (k_\perp, k_\parallel) \, \mathrm{d}k_\perp \mathrm{d} k_\parallel$.} $\mathcal{E}_a(k_\perp,k_\parallel)$, and where $\mathcal{T}(\boldsymbol{k})$ is a collision integral (defined over the wave vectors satisfying the above triadic conditions). 
The solution of the WKE with a steady non-zero $\mathcal{N}(\boldsymbol{k})$, known as the Kolmogorov-Zakharov (KZ) spectrum, has received much interest in the turbulence community. 
Finding an explicit KZ solution of the general WKE is a difficult task but, here, the problem is considerably simplified in the anisotropic limit $k_\parallel \ll k_\perp$.
In this regime, the WKE admits the anisotropic KZ solution given by \cite{galtier2003weak,gelash2017complete}
\begin{equation}
    \mathcal{E}_a(k_\perp, k_\parallel) \sim \sqrt{\epsilon \Omega_s} k_\perp^{-5/2} k_\parallel^{-1/2}.
    \label{eq:KZspectrum}
\end{equation}
The KZ solution predicts a net direct energy cascade within an inertial range, where the rate of spectral energy transfer is assumed to be equal to the mean energy dissipation rate per unit of mass $\epsilon$. 
First experimental observations \cite{yarom2014experimental} did find that rotating turbulence can exhibit energy along the inertial-wave dispersion relation (Figure \ref{fig:WT}b), which was then obtained in some \textsc{dns} \cite{clark2015spatio,le2017inertial}.
Later on, dedicated experiments \cite{monsalve2020quantitative} and simulations \cite{le2020evidence,yokoyama2021energy} confirmed the main spatial properties of anisotropic KZ spectrum (\ref{eq:KZspectrum}).
Moreover, formula (\ref{eq:KZspectrum}) predicts that the one-dimensional spectrum should scale as $|\boldsymbol{k}|^{-2}$ at large scales (for which we might assume $k_\perp \sim k_\parallel$), which is consistent with early experimental \cite{baroud2002anomalous} and numerical findings \cite{bellet2006wave,thiele2009structure}.

It has been proposed that mechanically driven flows in ellipsoids could sustain IWT in geophysical conditions (e.g. for tidal flows \cite{le2017inertial,le2019experimental}). 
Despite IWT is appealing from a fundamental viewpoint, its relevance for mechanically driven turbulence in ellipsoids is currently not obvious.
Indeed, it is unclear if the underlying assumptions of IWT are satisfied or not in geophysical conditions. 
In particular, we remind the reader that IWT is a local turbulence theory, which is  valid for homogeneous flows (i.e. unbounded) when $Ro \ll 1$. 
First, mechanically driven flows in ellipsoids likely occur in the low-$Ro$ regime, but this regime could only be valid at large scales (e.g. $Ro \sim 10^{-6}$ for planetary liquid cores with $l\sim L$). 
Indeed, at smaller scales, the local Rossby number is expected to increase as $\ell \to \ell_z$ (assuming that $u_\ell/\ell$ increases as $\ell$ decreases, e.g. $u_\ell/\ell \sim \epsilon^{1/3} \ell^{-2/3}$ for a Kolmogorov cascade \cite{nataf2024dynamic}). 
Second, rotating turbulence is often shaped by spatial inhomogeneities. 
The latter notably manifest in the form of inertial modes at large scales, as discussed for ellipsoids in \S\ref{subsec:fullellipsoid}, but which are also found in other geometries \cite{bewley2007inertial,lamriben2011excitation,boisson2012inertial}.
Therefore, it would be necessary to extend the WT theory to account for finite-size geometrical effects to assess the validity of the WT predictions for the largest scales characterised by $Ro \ll 1$. 
The continuous dispersion relation given by Equation (\ref{eq:dispIW}) should then be replaced by a discrete one in bounded geometries, which often prevents the occurrence of many triadic interactions.
For IWT, \textsc{dns} in periodic domains  \cite{bourouiba2008discreteness} showed that discretisation effects become non-negligible when $Ro \lesssim 10^{-3}$. 
Consequently, it is still unclear whether the KZ spectrum can properly describe the low-$Ro$ regime at rather large scales in bounded geophysical systems. 
Note that a first extension of the IWT theory has been done by considering a rotating channel that is infinite in the horizontal directions perpendicular to the rotation axis \cite{scott2014wave}. 
Yet, the fully bounded case remains to be considered theoretically and, then, compared to new experiments.

\subsubsection{$2-$D or not $2-$D? That is another question}
Another weakness of IWT is that it does not account for geostrophic flows,which are associated with $k_\parallel = 0$ in Equation (\ref{eq:dispIW}).
Indeed, there is no direct resonant triad satisfying Equations (\ref{eq:disprelWKE}a,b) that can transfer energy from $3-$D waves to the geostrophic component at the leading asymptotic order in $Ro \ll 1$ \cite{smith1999transfer}.
Actually, this results is also valid for any inviscid flow in a bounded geometry in the low-$Ro$ regime \cite{greenspan1969non}. 
If energy were only supplied to the $3-$D waves, then geostrophic structures would not be excited and no inverse cascade of energy could occur in the dynamics. 
However, the preferred regime of rapidly rotating turbulence could be that of quasi-geostrophic turbulence when $Ro \to 0$. 
This may result from the anisotropic properties of the inertial-wave dispersion relation  \cite{davidson2006evolution,staplehurst2008structure}, but also because the geostrophic manifold could be a global attractor of the rotating flows when $Ro \to 0$ \cite{gallet2015exact}. 
If true, this could thus challenge the validity of the WT theory for geophysical flows with $Ro \ll 1$.
Moreover, most experiments and \textsc{dns} of rotating turbulence are only performed for moderately small values of $Ro$ (Figure \ref{fig:parameters}).
For such values, nearly geostrophic components are often observed (e.g. \cite{le2017inertial,le2019experimental} for tidal flows).
Therefore, it is essential to understand the mechanisms feeding and sustaining geostrophic flows from a fundamental viewpoint, but also for geophysical applications.

So far, the transition between these $2-$D and $3-$D regimes is not fully understood. 
Indeed, the dominant mechanisms sustaining geostrophic flows may be forcing-dependent and $Ro$-dependent.
Moreover, they may also be distinct in forced and decaying turbulence.
Nonetheless, two canonical nonlinear mechanisms are usually put forward to explain the growth of geostrophic components \cite{smith1999transfer}.
The first one involve near-resonant triadic interactions \cite{bretherton1964resonant}, for which triads $[\boldsymbol{k},\boldsymbol{p},\boldsymbol{q}]$ satisfy the detuned resonance conditions given by
\begin{equation}
    \omega(\boldsymbol{k}) + \omega(\boldsymbol{p}) + \omega(\boldsymbol{q}) = \mathcal{O}(Ro),
    \tag{\theequation a,b}
\end{equation}
together with $k_\parallel = 0$ and $p_\parallel = - q_\parallel$.
This is the scenario investigated in \cite{smith2005near,clarkdileoni2016quantifying,le2020near}, which can lead to the onset of geostrophic flows with a growth rate $\propto (k Ro)^2$ when $k Ro \ll 1$  or $\propto k Ro$ when $k Ro \gtrsim 1$ \cite{le2020near}.
The second mechanism involves quartetic interactions (or four-wave interactions) \cite{smith1999transfer}.
For primary triads $[\boldsymbol{k},\boldsymbol{p},\boldsymbol{q}]$ satisfying conditions (\ref{eq:disprelWKE}), we can find quaterts $[\boldsymbol{p},\boldsymbol{q},\boldsymbol{r},\boldsymbol{s}]$ satisfying the quatertic resonance conditions given by
\begin{subequations}
\label{eq:dispquatertic}
\begin{equation}
    \underbrace{\boldsymbol{p} + \boldsymbol{q}}_{=-\boldsymbol{k}} + \boldsymbol{r} + \boldsymbol{s}  = \boldsymbol{0}, \quad \underbrace{\omega(\boldsymbol{p}) + \omega(\boldsymbol{q})}_{=-\omega(\boldsymbol{k})} + \omega(\boldsymbol{r}) + \omega(\boldsymbol{s}) = 0,
    \tag{\theequation a,b}
\end{equation}
\end{subequations}
with $s_\parallel = 0$. 
We refer the reader to Appendix B in \cite{smith1999transfer} for a detailed example.
This mechanism, which also predicts a growth rate $\propto (k Ro)^2$, was observed numerically \cite{kerswell1999secondary} and experimentally \cite{brunet2020shortcut}.
Note that other mechanisms have also been invoked such as various interactions with eddies \cite{mininni2009scale,bourouiba2012non,buzzicotti2018energy,lam2023supply}, and the shape of the fluid domain could also play a role \cite{van2020critical}.

Back to geophysical flows, nearly $2-$D turbulence is unambiguously the preferred regime for rapidly rotating convective flows \cite{nataf2024dynamic}.
On the contrary, it is unclear which regime is preferred for mechanically-driven turbulence in geophysical (i.e. $Ro \ll 1$ and $E \to 0$) or experimental conditions (with finite values of $Ro$ and $E$).
Fortunately, future experiments could shed new light on this controversial question.
For a new-generation experiment with a metre-size ellipsoid filled with water at room temperature, it seems very difficult to lower the Ekman number below $E \sim 10^{-7}$ with a typical value $\Omega_s\sim 10$~rad.s${}^{-1}$ for the angular velocity of currently used rotating tables. 
Nonetheless, this may be sufficient to start disentangling the various mechanisms and turbulent regimes in an ellipsoid. 
Finally, it would also be worth extending the theory to account for ellipsoidal (or spherical) boundaries for a better quantitative comparison with the experiments.
Indeed, since geostrophic flows have very different properties in unbounded and bounded systems (e.g. due to the beta effect \cite{greenspan1968theory}), the finite-size geometry could quantitatively affect the predictions previously obtained in either unbounded or cylindrical geometries.  

\subsubsection{Towards geophysical extrapolation}
\begin{table}
\caption{Proposed scaling laws for the diffusionless growth rate $\sigma$ and the flow amplitude $u_\ell$ for precession and tidal forcings in ellipsoids. The proposed numerical prefactors can be found in the original works. $\eta_e = |a^2-b^2|/(a^2+b^2)$ is the equatorial ellipticity of the ellipsoid with semi-major axes $(a,b,c)$, and $\eta_p$ its polar flattening. $\Delta \omega$ is the differential rotation between the fluid and the mantle due to precession. EI: elliptical instability, VSDI: Viscous shear-driven instability, CSI: conical shear instability, SI: Shear instability.}
\begin{tabular}{lcccc}
    \hline
    {Forcing} & \multicolumn{2}{c}{Tides} & \multicolumn{2}{c}{Precession} \\
    {Mechanism} & EI & VSDI & CSI & KSI
    \\
    \hline
    Growth rate $\sigma$ & $\mathcal{O}(\eta_e |\Omega_s-\Omega_{orb}|)$ & - & $\mathcal{O}(|\Delta \omega| E^{1/5})$ & $\mathcal{O}(|\Delta \omega| \eta_p)$  \\
    Turbulence amplitude $u_\ell$ & $\mathcal{O}(\eta_e |\Omega_s-\Omega_{orb}| a)$ & $\mathcal{O}(\eta_e^2 \Omega_s a E^{-\chi})$ & $\mathcal{O}(|\Delta \omega| E^{2/5} a)$ & $\mathcal{O}(|\Delta \omega| \eta_p a)$ \\
    References & \cite{barker2013non,grannan2017tidally,vidal2019fossil} & \cite{sauret2014tide} & \cite{lin2015shear,vidal2023precession} & \cite{kerswell1993instability,barker2016turbulence,vidal2023precession}\\
    \hline
\end{tabular}
\label{table:instab}
\end{table}

In addition to the fundamental and enthralling question of the properties of rotating turbulence, we also have to obtain scaling laws to correctly extrapolate the experimental results to geophysical conditions (which are still out of reach). 
Some scaling laws, which have been proposed for linear and nonlinear features of mechanically driven flows, are gathered in Table \ref{table:instab}.
For instance, scaling laws are crucial if we want to estimate the contribution of mechanically driven flows to tidal dissipation in the subsurface oceans of some icy moons \cite{lemasquerier2017libration,wilson2018can}, or to assess their viability to sustain planetary magnetic fields (to go beyond prior numerical studies \cite{reddy2018turbulent,vidal2018magnetic,cebron2019precessing}).
In the latter case, dimensional analysis tells us that the typical magnetic field strength $B$ driven by dynamo action could be given at the fluid surface by \cite{davidson2013scaling}
\begin{equation}
    B \propto \sqrt{\mu \rho_m} \, (L \mathcal{P})^{1/3},
    \label{eq:scalingB}
\end{equation}
where $\mu$ is the fluid magnetic permeability, $\rho_m$ is the mean fluid density, and $\mathcal{P}$ is the mean energy production rate per unit of mass.
Moreover, on the long time scales characterising dynamo action, we may also expect the dynamics to be close to a steady state yielding $\mathcal{P} \sim \epsilon$ from the energy balance, where $\epsilon$ is the mean energy dissipation rate per unit of mass.
Hence, finding a quantitative estimate of $\epsilon$ is often a cornerstone in turbulence theory and for geophysical extrapolation. 
Direct measurements of $\epsilon$ are difficult because they require a well-resolved velocity field. 
Since we have $\epsilon = \epsilon (u_\ell,\ell)$ by dimensional analysis, a first and accessible step towards geophysical extrapolation is therefore to estimate the typical strength $u_\ell$ of mechanically driven turbulent flows  (e.g. Table \ref{table:instab}).
Yet, not all the predictions have been quantitatively validated with experiments.
Moreover, another pitfall is that prior laws were obtained without a clear understanding of the underlying rotating turbulence regime.
As an example, the energy dissipation law is expected to be given from formula (\ref{eq:KZspectrum}) by \cite{galtier2003weak}
\begin{equation}
    \epsilon \sim \frac{u_\ell^4 l_\parallel}{\Omega_s l_\perp^3}
\end{equation}
with $\ell_\perp \sim 1/k_\perp$ and $\ell_\parallel \sim 1/k_\parallel$, which differs from the usual scaling $\epsilon \sim u_\ell^3/\ell$ in isotropic Kolmogorov turbulence.
Since these dissipation scalings would yield different estimates from Equation (\ref{eq:scalingB}), new experiments are required to validate the scaling laws for the different regimes.

Additional physical ingredients would also be worth including in new experiments, such as a shell geometry (i.e. with an inner core). 
Prior experimental works \cite{lacaze2005elliptical,lemasquerier2017libration} showed that the flow response to mechanical forcings in a shell is qualitatively similar to that in a full ellipsoid, even if the mathematical theory of inertial modes is very different in a shell geometry \cite{rieutord2000wave}. 
This similarity should thus be further investigated using future experiments. 
Including density stratification (with $N^2 > 0$) would also be of great importance for other geophysical systems. 
For example, there are (nearly) isolated and lenticular vortices in the oceans \cite{bashmachnikov2015properties} or Jovian planets \cite{marcus1993jupiter}, whose dynamics (from birth to death) could affect mixing processes.
Dedicated experiments could be useful to model the dynamics inside \cite{labarbe2021diffusive,vidal2024igw} and outside \cite{le2021numerical} such geophysical eddies, which can be modelled by ellipsoidal vortices \cite{aubert2012universal,de2017laboratory,lemasquerier2020remote}. 
Investigating the flow response in an ellipsoid with coupled global rotation and density stratification is currently very challenging (see \S\ref{sec:stratification}), but it may be amenable to experimental work in the future. 
Finally, the ongoing (and future) space missions will likely shed new light on turbulent flows in Jovian planets. 
Therefore, the flow compressibility may be worth considering in next-generation experiments to model the dynamics of gas giants in a global geometry (e.g. following a pioneering experimental work \cite{su2020acoustic}).

\section{Slowly varying flows over meso-scale and rough topographies}
\label{sec:smallscale}
We now consider how small-wavelength topographies (i.e. meso-scale and roughness) can affect rotating flows. 
Several phenomena well studied in non-rotating but possibly stratified fluids, such as the drag due to an obstacle or wall-driven turbulence, remain poorly quantified for rotating flows.
Moreover, small-wavelength topography could also inhibit the growth of bulk (turbulent) flows driven by orbital mechanical forcings (as discussed in \S\ref{sec:ellipsoid}), by enhancing the dissipation in Equation (\ref{eq:growthvsdamping}).
For all these reasons, it is key to assess the effects of small-wavelength topographies on rotating flows. 
We discuss in \S\ref{subsec:meso} how meso-scale topographies and roughness affect bulk rotating flows in thick-layer systems, and we investigate how a turbulent boundary layer (BL) is modified by roughness in \S\ref{subsec:turbulentEkman}.

\subsection{Bulk flows over meso-scale topographies and roughness}
\label{subsec:meso}
\subsubsection{Meso-scale topography}
Flows interacting with topographical features have garnered interest within the oceanographic and atmospheric research communities, benefiting from the abundance of data available to study both the dynamics of the flows and the characteristics of the topography. 
For atmospheric and oceanic applications (i.e. thin-layer systems), it is customary to solve the dynamics using the $f-$plane approximation, that is to only retain the component of the planet's rotation vector $\boldsymbol{\Omega}$ onto the local direction of gravity by introducing the Coriolis parameter $f= 2 \Omega_s \cos \theta$, where $\theta$ is the colatitude in spherical coordinates.
In the context of stratified fluids with $0<|f|<N$, where $N$ is the BV frequency as defined above in Equation (\ref{eq:BVN}), energy transfer and dissipation is well captured by the propagation and breaking of Lee waves (i.e. internal gravity waves emitted over mountains), thereby promoting efficient vertical mixing processes \cite{legg2021mixing}. 
Experimental investigations have revealed the existence of various flow regimes, ranging from linear wave radiation to blocked flows. 
They depend on the relationship between the potential energy acquired by a fluid parcel as it traverses from the base of a valley to the summit of a topographic feature, and the kinetic energy available within the background flow (as quantified by a Froude number). 
In scenarios characterised by low values of the Froude number, linear and weakly nonlinear Lee waves propagate in the ocean, radiating energy away from the topography. 
Conversely, when the Froude number is large, the available kinetic energy is not sufficient to transport parcels over the topography.
This leads to the emergence of blocked flows and regimes of wave emission, but with an effective wavelength larger than that of the topographical features. A more detailed description of flows over topography in stratified fluids can be found in \cite{legg2021mixing}.

\begin{figure}
    \centering
    \includegraphics[width=0.9\textwidth]{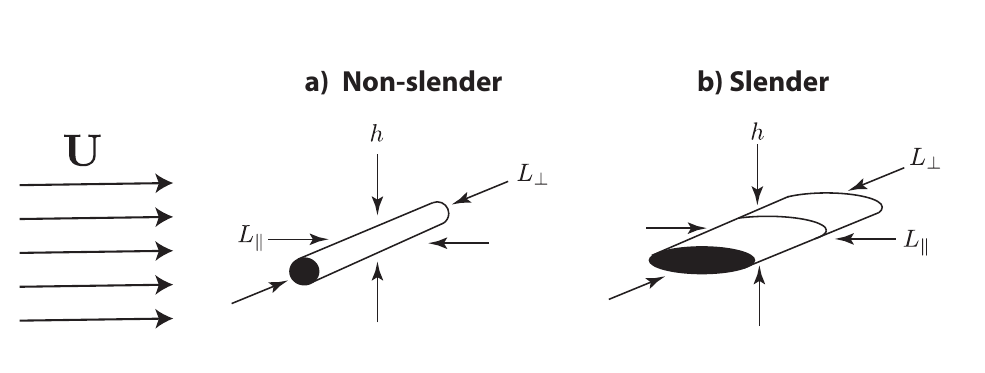}
    \caption{Illustration of the concept of \textbf{(a)} a non-slender obstacle with $h \gtrsim L_\parallel$ and \textbf{(b)} a slender obstacle with $h \ll L_\parallel$, where $h$ is the obstacle's thickness, $L_\parallel$ is its stream-wise dimension (i.e. along the direction of the velocity $\boldsymbol{U}$), and $L_\perp$ is its cross-stream dimension. In the two cases, the obstacle is elongated with $L_\perp \gg L_\parallel$.}
    \label{fig:slendernonslender}
\end{figure}

\begin{figure}
    \centering
    \includegraphics[width=0.87\textwidth]{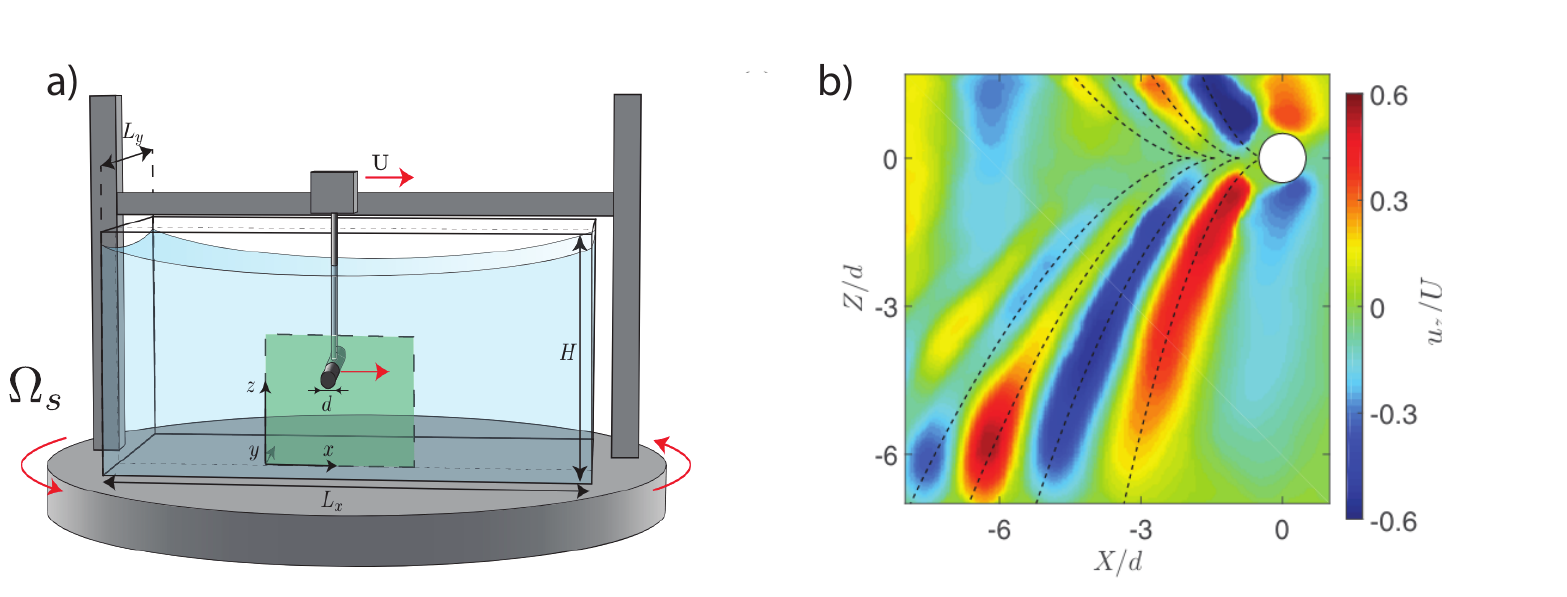}
    \caption{Inertial waves radiated by an elongated (non-slender) cylinder steadily towed in a rotating tank. Adapted from \cite{machicoane2018wake}. \textbf{(a)} Sketch of the experimental setup. \textbf{(b)} Cross-section showing the vertical velocity from \textsc{piv} normalised by the towing speed. In the experiment, the Rossby number is small and the obstacle Reynolds number is less than $10^3$.}
    \label{fig:machiacoane}
\end{figure}

In thick-layer fluid systems, the departures from a strictly adiabatic density gradient (i.e. when $N^2 \neq 0$) are expected but undocumented.
Hence, the flow dynamics with weak or without stratification (i.e. either weakly stable or neutral) also deserves consideration for such geophysical applications. 
However, such dynamical regimes have gained much less attention relative to strongly stratified regimes.
Let us delve into the simplest scenario of an incompressible, neutrally buoyant fluid in rapid rotation steadily passing a single obstacle. 
As described below, the observed flows are contingent upon the obstacle's aspect ratio and the fluid velocity. 
\vspace{0.5em}
\begin{paragraph}{Single bump}~For isolated bumps, defined as objects with a comparable horizontal extent $L$ in both the stream-wise and cross-stream directions, Taylor columns (i.e. flow structures that are almost invariant along the rotation axis) dominate the flow when $H Ro/L \ll 1$ (where $H$ is the total fluid column height). 
Physically, it means that the vertical travel time of inertial waves over the whole fluid column above the bump, which is given by $H/(\Omega_s L)$ for inertial waves with the same wavelength as the topography, must be much shorter than the typical time it takes to horizontally translate the obstacle by a distance $L$. 
Conversely, a wake of inertial waves predominates when $H Ro / L \gg 1$. 
These phenomena have been validated in experiments employing various shapes of obstacles, such as elevated straight cylinders/bumps \cite{taylor1923experiments,hide1966experimental,hide1968slow} or cylindrical depressions \cite{boyer1984rotating}.
\end{paragraph}
\vspace{0.5em}
\begin{paragraph}{Single ridge}~
For elongated obstacles like ridges, the dynamics additionally hinges on the cross-sectional shape. 
It has been demonstrated that the formation of Taylor columns is inhibited for elongated and non-slender objects of vertical ($h$) and horizontal stream-wise ($L_{\parallel}$) dimensions with $h\gtrsim L_{\parallel}$, and with significant cross-stream dimensions (i.e. $L_{\perp} \gg {L_{\parallel}}$, see Figure \ref{fig:slendernonslender}a).
In such cases, the flow manifests purely as a wake of inertial waves \cite{heikes1982observations,johnson1982effects} (even at vanishingly small values of $Ro$).
This was recently confirmed by laboratory experiments \cite{machicoane2018wake}, where an elongated cylinder is steadily towed in a rotating square tank for small to moderate values of $Ro$.
As shown in Figure \ref{fig:machiacoane}, bulk flows dominated by $3-$D inertial waves are observed in such experiments. 
Conversely, experiments conducted with slender elongated obstacles (i.e. $h \ll L_{\parallel} \ll L_{\perp}$, see Figure \ref{fig:slendernonslender}b) showed geostrophic dominated dynamics of blocked flows at low values of $Ro$, and distorted zonal circulations at larger values of $Ro$ \cite{weeks1997transitions}. 
\end{paragraph}

\vspace{0.5em}
\begin{paragraph}{2-D topography}~
Considerably less attention has been devoted to spatially more intricate topographies in neutrally buoyant fluid (e.g. arrangements of multiple hills), particularly concerning angular momentum transfer and energy dissipation within the fluid.
This issue was recently tackled experimentally \cite{burmann2018effects}, by investigating the impact of meso-scale bottom topography on the spin-up time, that is the time required for the fluid's rotation to realign with that of the container after a sudden increase. This time is a proxy for the rate of angular momentum exchange. 
In the experiment, the bottom of a straight circular cylinder was paved with a chessboard-like array of square blocks with varying horizontal extents and fixed height (Figure \ref{fig:spin_up}).
The decay rate of the mean kinetic energy was monitored in horizontal cross-sections, using \textsc{piv} velocity fields. 
The evolution of the mean energy is exponential, and the spin-up time is defined as the $e$-folding time. 
Topographic effects yield a significantly shorter spin-up time, up to 10 times for a specific horizontal length scale of the blocks, meaning that in this experiment the axial torque can be up to 10 times larger with a topography than for a flat cylinder.
In the canonical flat end walls case, the spin-up time follows a scaling law $\tau \propto E^{-1/2}$\cite{greenspan1968theory}. 
With topography, this scaling is altered with an exponent decreasing with the wavelength. 
At the smallest wavelength considered in this study, it is almost independent of $E$.
In addition, the authors observed a vertical propagation of inertial waves generated near the bottom topography, with a typical wavelength comparable to (or smaller than) the topographic blocks. 
These observations suggest that the mechanism by which the BL communicates with the bulk has evolved from a large-scale Ekman circulation for a flat cylinder to a more efficient inertial wave radiation-dominated regime with topography. 
This process is similar to the dissipation of tidal energy by the emission of Lee waves in the bottom ocean \cite{legg2021mixing}. 
A spin-up time independent of Ekman at the smaller wavelength suggests that the initial transfer of angular momentum from the surface to the fluid is governed by strong nonlinear dynamics near the bottom topography, which ultimately radiates energy in the bulk via inertial waves. 
This scenario for small-wavelength topography at small (but finite) values of $Ro$ requires further investigation, including a more detailed flow visualisation near the topography. 
\end{paragraph}
\vspace{0.5em}

While such experiments represented an initial step to get insights into the effects of complex topography patterns on fluid dynamics (and its interaction with the surrounding solid), several unresolved questions remain.
In particular, a detailed characterisation of the flow near the topography, and the development of a model to forecast torque and energy dissipation, deserve further consideration.

\subsubsection{Rough surfaces, a challenge for experimental approach}
\begin{figure}
    \centering
    \includegraphics[width=0.85\textwidth]{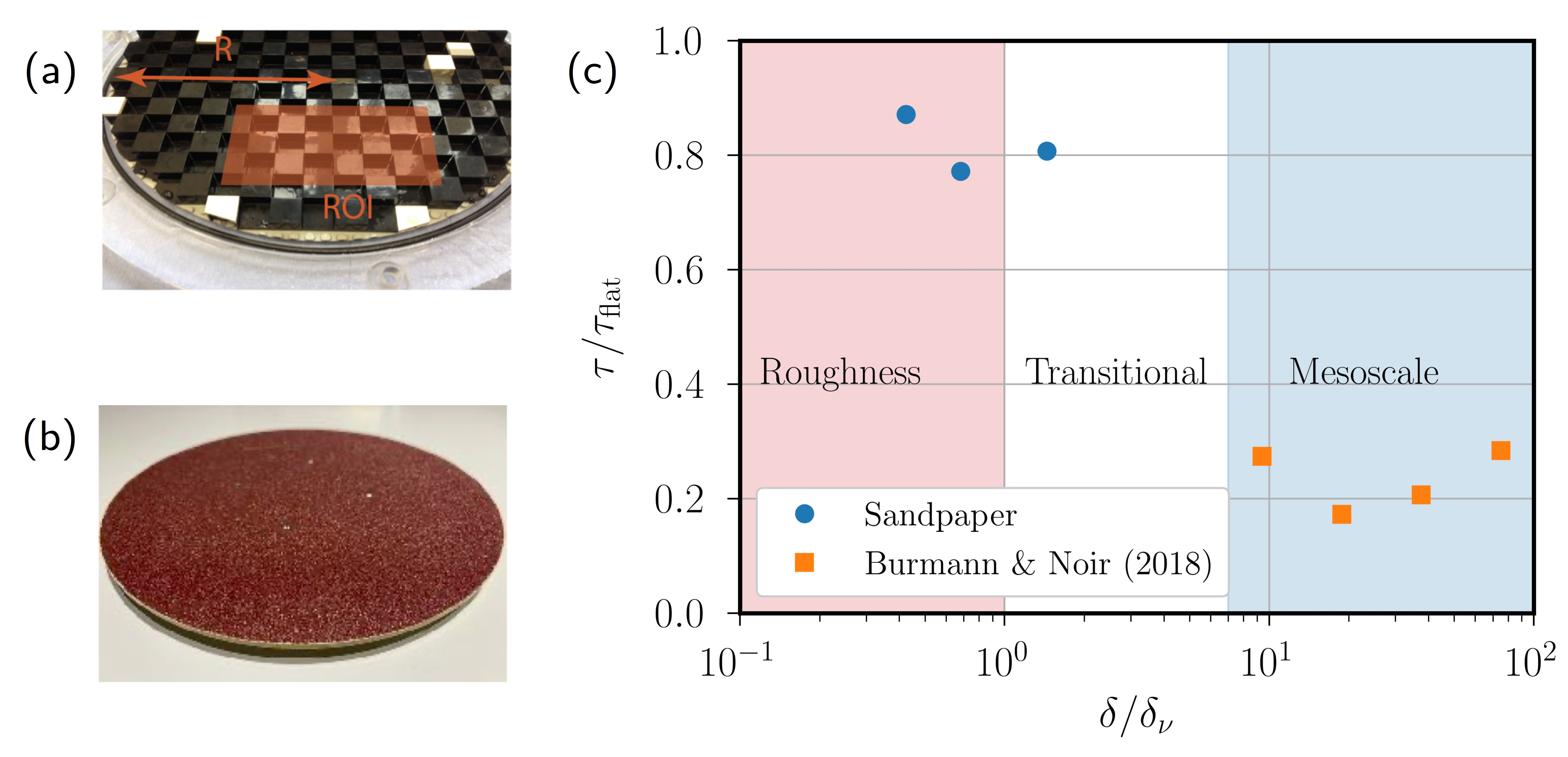}
    \caption{\textbf{(a)} Meso-scale topography at the bottom of a cylinder (of radius $R$) in spin-up experiments \cite{burmann2018effects}. ROI: Region of Interest considered in the original paper. \textbf{(b)} Roughness in spin-up experiments with rough end walls conducted at \textsc{eth} Z\"urich. \textbf{(c)} Spin-up time $\tau$ (scaled by that of a flat end wall cylinder $\tau_\text{flat}$) as a function of the horizontal length of topography $\delta$ (scaled by the thickness $\delta_\nu$ of the Ekman layer). Red shaded and blue shaded areas indicate roughness and meso-scale topography, respectively.}
    \label{fig:spin_up}
\end{figure}

Roughness, that is topography embedded in the BL, is also likely to be present in geophysical systems (e.g. at the Earth's \textsc{cmb} \cite{narteau2001smallscale,le2006dissipation}, Earth's surface and ocean floor). 
Its presence will affect the dissipation processes and transfer of angular momentum.
Accounting for roughness in numerical models is also difficult, and it often entails solving the large-scale flows using \textsc{dns} in which the effects of the roughness are parameterised using models of boundary layers dynamics (see for instance \S\ref{subsec:turbulentEkman} below for turbulent boundary layers). Controlled experiments do not suffer from the limitations of \textsc{dns} and can be used to validate the idealised models of rough end-walls. 
In a typical laboratory experiment, employing water at room temperature in a tank of typical size $L\sim50$~cm rotating at the angular velocity $\Omega_s\sim 2\pi $~rad.s${}^{-1}$, the Ekman number is of order $E \sim 10^{-6}$.
This results in a viscous BL thickness of approximately $\delta_{\nu} \sim 50-100$~$\mu$m. 
This entails employing textured walls, with a typical height and wavelength on the order of $\delta\sim 1-100$~$\mu$m. 
The primary challenge in such experiments lies in controlling the roughness. 
Indeed, at such sub-millimetre scales, constructing consistently rough walls that remain flat and statistically homogeneous on the small scales of the roughness ($\delta\sim 1-100$~$\mu$m) is non-trivial.
We present in Figure \ref{fig:spin_up} previously unpublished experimental data from a preliminary experiment on spin-up with rough end walls, conducted at \textsc{eth} Z\"urich\footnote{Conducted during the graduate internship of Y. Charles.}. Here, sandpaper of varying grain sizes is affixed to the end walls of a cylinder, and the azimuthal velocity within the fluid bulk is measured using Laser Doppler Velocimetry (\textsc{ldv}). 
The spin-up time is determined as the $e$-folding time of the decay in the amplitude of the azimuthal circulation. 
Not surprisingly, these initial findings suggest that the spin-up time is shortened when the roughness is embedded within the Ekman layer.
Once the grain size exceeds the BL thickness, a shift in dynamics occurs, still associated with an enhanced coupling but not as significant. In all cases, the increase of the torque by the roughness with respect to a flat wall is merely of the order of 20$\%$, much less than that of a meso-scale topography in the same range of Rossby and Ekman numbers.
A speculative scenario at low values of $Ro$ posits that the roughness enhances angular momentum exchange and energy dissipation within the BL itself, while the spin-up is still communicated to the bulk through large-scale Ekman pumping, less efficient than the wave radiation with meso-scale topography. This preliminary data suggest that further experiments are worth considering, in particular to characterise the large scale circulation.

\subsubsection{Experimental perspectives}
Extensive research has been devoted to exploring the impact of meso-scale topography on the dynamics of geophysical fluids within the oceanic and atmospheric communities. 
However, these findings often lack direct applicability to thick-layer fluid systems. 
Due to the inherent challenges of flows over topography for \textsc{dns} (particularly in nonlinear regimes), experiments are indispensable to fill the gap between the numerically accessible regimes and the inaccessible regimes of planets and moons. 
Meanwhile, the observed pivotal role of inertial waves suggests the potential development of linear and weakly nonlinear wave drag models to estimate the torque and energy dissipation (alike that of Lee waves in the Earth's oceans and atmosphere \cite{legg2021mixing}), or reduced quasi-geostrophic models as in the recent works on flows over rough seafloors \cite{radko2023generalized,radko2023sandpaper}.  
Experiments over a broad range of parameters and topography shapes will be necessary to test those models, and extend them beyond the weakly nonlinear limit.
This would entail using challenging velocity measurements near the topography, together with measures of global quantities such as the torque exerted by the fluid on the solid boundaries. 

Finally, the interaction between the large-scale flows driven by an oscillatory mechanical forcing (as the ones discussed in \S\ref{sec:ellipsoid}) and small-wavelength topography remains unexplored.
Ever since it was proposed by Hopkins \cite{hopkins1839nutations} in the middle of the nineteenth century that one could probe the internal structure of planets from their rotation, the role of the topographic torque between cores or subsurface oceans and their surrounding shells has been debated(e.g. at the Earth's \textsc{cmb} \cite{dehant2017understanding,rekier2022earth}). 
In contrast to the steady flows discussed above, the mechanically driven flows are oscillatory with periods close to the rotation period. 
The interaction of oscillating flows with topography in neutrally buoyant fluids remains inadequately documented, especially within thick-layer fluid systems.
While canonical problems in ellipsoidal geometries have received extensive attention, the modification of inertial waves or modes, and their nonlinear couplings with topography, remains largely unexplored even in the simplest scenarios involving single bumps or ridges. 
Further experimental work is thus needed to elucidate the mechanisms of energy dissipation and angular momentum transfer for such geophysical applications.

\subsection{Turbulent boundary layers with flat and rough end-walls}
\label{subsec:turbulentEkman}
\begin{figure}
\centering
\includegraphics[width=0.9\textwidth]{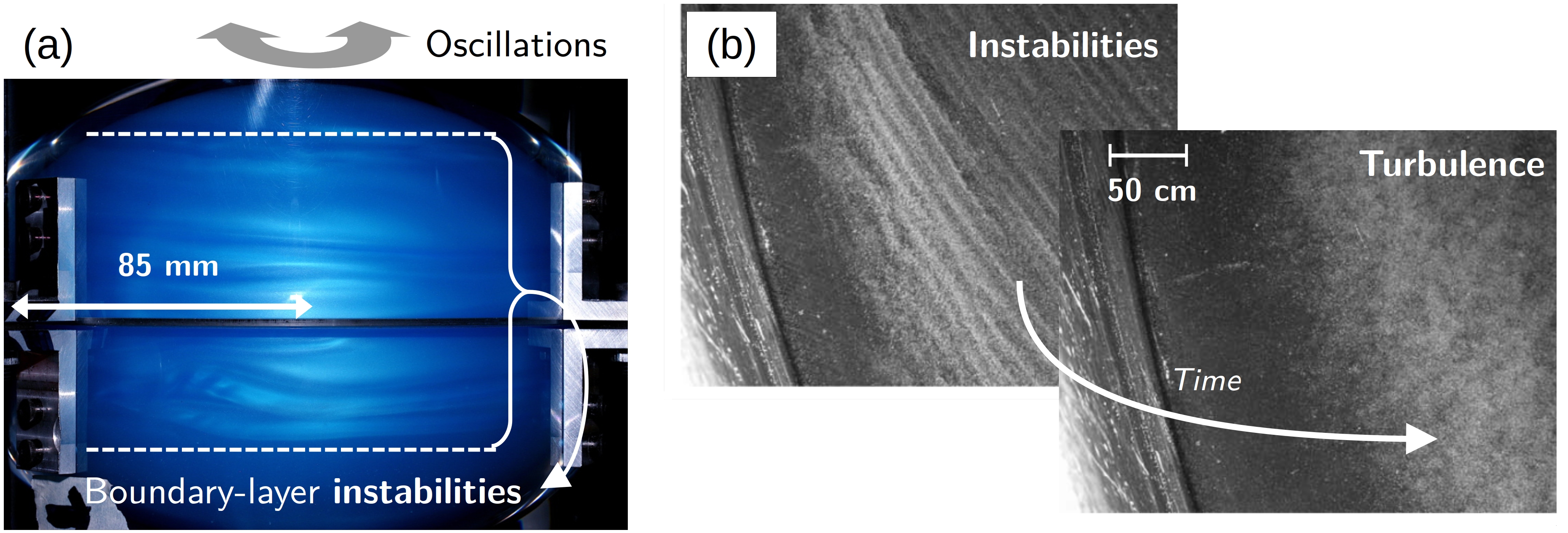}
\caption{Transition from laminar to turbulent Ekman boundary layers (BLs). \textbf{(a)} BL instabilities driven by longitudinal librations in a sphere at $Re_\delta \sim 100$ and $E \simeq 5 \times 10^{-5}$. Flow visualisation using Kalliroscope flakes (which make the volume of fluid opaque, except for the first mm below the boundary). Adapted from \cite{noir2009experimental}. \textbf{(b)} Ekman BL instabilities (in the form of rolling waves \cite{caldwell1970characteristics}, here oriented with an angle of about $20{}^\circ$ from the geostrophic flow), and turbulence triggered during the spin-down of a rotating tank. Visualisation using particle-streak photographs. Adapted from figure 3 in \cite{sous2013friction}.}
\label{fig:BLinstab}
\end{figure}

Like in the Earth's atmosphere \cite{coleman1999similarity}, the Ekman boundary layer (BL) could be turbulent in thick fluid envelopes.
For that, the local Reynolds number must be much larger than the critical values $55-150$ for the onset of boundary-layer instabilities \cite{lilly1966instability,caldwell1970characteristics,aelbrecht1999experimental}. 
For instance, the mechanical (orbital) forcings discussed in \S\ref{sec:ellipsoid} can sustain some boundary-layer instabilities in experimental conditions (e.g. for librations \cite{noir2009experimental}, see Figure \ref{fig:BLinstab}a), but also in some geophysical conditions (e.g. at the Moon's \textsc{cmb}, where the local Reynolds number is $\sim 10^4-10^5$ because of precession-driven flows \cite{cebron2019precessing}). 
Similarly, a transition to turbulent BLs is also commonly observed in laboratory experiments during spin-down of a rotating tank (Figure \ref{fig:BLinstab}b).
A description of turbulent Ekman BLs is interesting for geophysical modelling.
For instance, wall friction is enhanced by a turbulent BL (compared to the classical $E^{1/2}$ scaling for a laminar Ekman BL), such that wall-driven turbulence could weaken or inhibit the growth of the mechanically driven bulk flows discussed in \S\ref{sec:ellipsoid}.
However, obtaining a turbulent BL dynamics is challenging using global \textsc{dns}, in which the boundary layers are often laminar because of the numerical difficulty to resolve for both the largest scales of the flows and the smaller ones associated with turbulent BLs.
Therefore, it is worth obtaining mathematical parametrisations for turbulent Ekman BLs, which could be later implemented in global numerical models (or used for geophysical extrapolation).

In practice, the ultimate goal is often to obtain an estimate of the drag coefficient ${\tau_\star}/{(\rho U^2)}$ in geophysical conditions, where $\tau_\star$ is the surface shear stress on the wall, $\rho$ is the density (which is constant for an incompressible fluid), and $U$ is the amplitude of the geostrophic velocity far enough from the wall.
Initially motivated by atmospheric applications, the turbulent Ekman BL forced by a steady geostrophic flow on a plane has been studied using experiments, \textsc{dns} and theory, complemented by more than a hundred years of meteorological data  \cite{hess2002evaluating,hess2002evaluating2}. 
Classical similarity theory provides asymptotic laws for the velocity in a turbulent Ekman BL \cite{csanady1967resistance,blackadar1968asymptotic,spalart1989theoretical}, which is often in good agreement with measurements. 
First, we remind the readers the basics of the similarity theory for a highly turbulent Ekman BL with roughness, and we discuss its usual simplified version for smooth boundaries. 
Constraints obtained from experimental works are later discussed and, finally, we outline some remaining open questions for geophysical modelling and future experimental works.

\subsubsection{Theoretical considerations for turbulent Ekman BLs}
\begin{figure}
\centering
\begin{tabular}{cc}
    \includegraphics[width=0.43\textwidth]{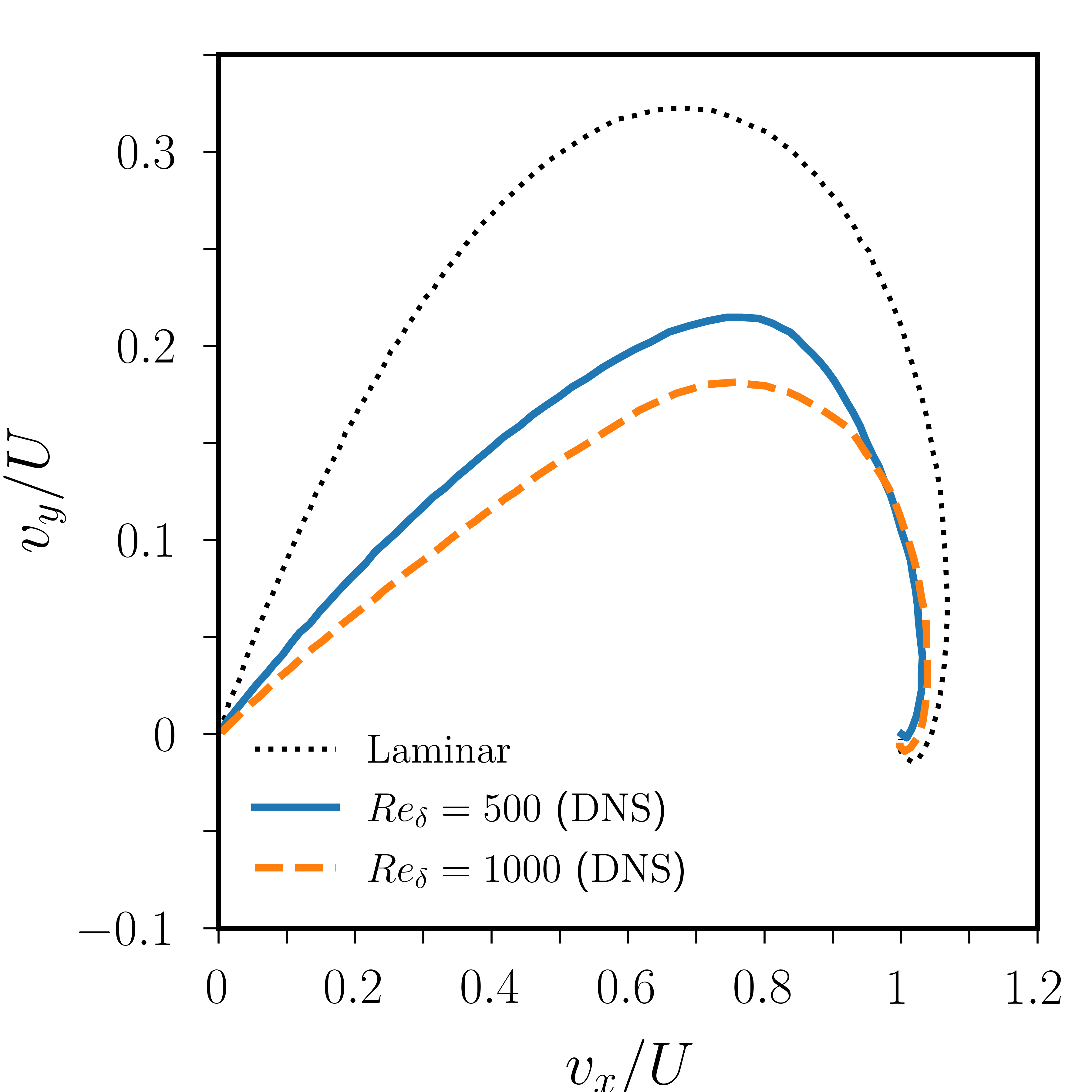} &
    \includegraphics[width=0.43\textwidth]{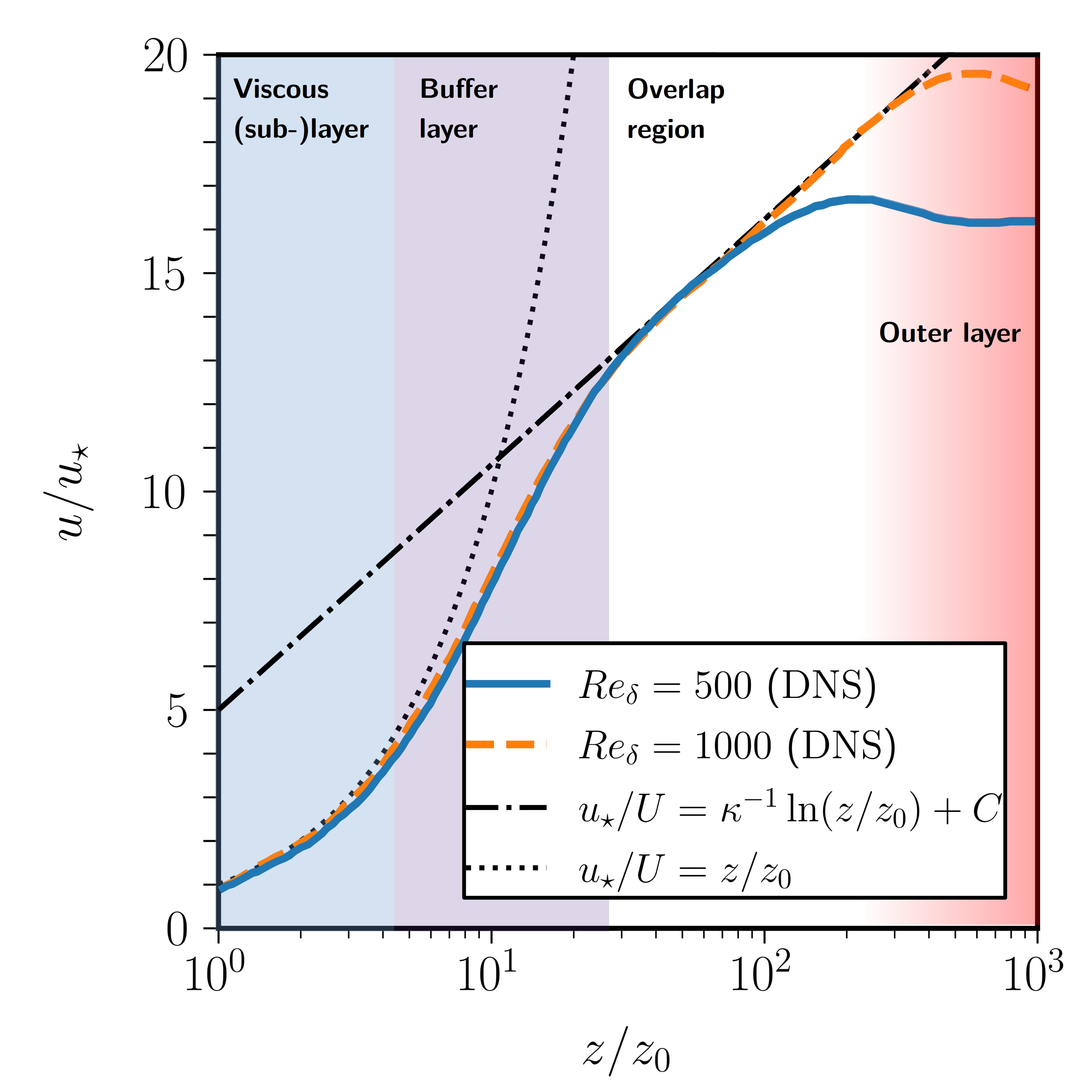} \\
    (a) & (b) \\
\end{tabular}
\caption{Neutral turbulent Ekman BL over a smooth plane boundary, for a geostrophic (bulk) flow $U$ in the $x$ direction. Adapted from figure 2 in \cite{ansorge2014global}. \textbf{(a)} Hodograph showing the modification of the Ekman spiral (obtained for a laminar BL) as a function of $Re_\delta$ for a turbulent Ekman BL. $(v_x,v_y)$ are the Cartesian components of the mean velocity in the coordinate system aligned with the geostrophic flow $\boldsymbol{U}$. The surface veering angle $\beta$ is given by the slope near the origin. \textbf{(b)} Normalised mean velocity $u/u_\star$ as a function of the normalised distance $z/z_0$ to the wall. In the \textsc{dns}, $u_\star$ is estimated from $u_\star^2 = \nu \partial_z Q$ with $Q = (u^2+v^2)^{1/2}$. The linear law $u/u_\star = z/z_0$ (in the viscous sub-layer), and law-of-the-wall (\ref{eq:loglaw1}a) with $\kappa = 0.41$ and $C=5$ (i.e. the $y-$intercept at $z/z_0=1$), are also shown.}
\label{fig:EkmanBL}
\end{figure}

We consider the mean (free-stream) geostrophic velocity $\boldsymbol{U}=U \boldsymbol{1}_x$ in the bulk, which flows over a smooth or rough planar boundary. 
We introduce the local Reynolds number as
\begin{subequations}
\label{eq:Relocal}
\begin{equation}
Re_\delta = \frac{U \delta_{\nu}}{\nu} \quad \text{with} \quad \delta_{\nu}=\sqrt{\frac{2 \nu}{f}},
\tag{\theequation a,b}
\end{equation}
\end{subequations}
where $\delta_\nu$ is the laminar BL thickness, and $f=2 \Omega_s$ is the Coriolis parameter in experimental conditions.
If $Re_\delta \geq 200-300$, a turbulent Ekman BL can be established above the boundary.
When the Reynolds number is large enough, turbulent BLs are often assumed to be described by universal laws, such as the law-of-the-wall that is expected to be valid for any wall-driven turbulent flows (e.g. for channel flows).   
Such descriptions are thus cornerstones in the theory of turbulence \cite{pope2000turbulent}, as many turbulent models are calibrated to reproduce them for simple flows. 
However, there is a noticeable difference between non-rotating and rotating flows. 
The wall shear stress is expected to be along the direction of the background flow for non-rotating flows, whereas the stress is tilted from the geostrophic flow $\boldsymbol{U}$ with rotation (Figure \ref{fig:EkmanBL}a). 
Indeed, laminar and turbulent Ekman BLs display a spiral structure as a function of the distance $|z|$ from the wall and, close to the wall, the boundary shear stress is tilted from the free-stream velocity with the veering angle $\beta$.
As $Re_\delta$ increases, turbulence tends to reduce the veering angle from its laminar value of $45^{\circ}$. 
Hence, the theory has to be slightly modified to work with coordinates defined from the stress direction.

To do so, we denote by $u$ the mean shear-wise velocity (i.e. in the wall-stress direction) departing from the bulk flow $\boldsymbol{U}$, and by $v$ the (mean) cross-shear velocity.
The veering angle is thus given by $\beta = \arctan(v/u)$ at the wall.
It is then postulated that, away from the wall, the velocity decrease from the geostrophic bulk flow is independent of viscosity and is represented by the so-called velocity-defect-law. 
The latter is given from dimensional analysis by \cite{csanady1967resistance,blackadar1968asymptotic}
\begin{subequations}
\label{eq:defectsimilar}
\begin{equation}
    \frac{u-U_g}{u_\star} = G_1 \left ( \frac{f z}{u_\star}  \right ), \quad \frac{v - V_g}{u_\star} = G_2 \left (\frac{f z}{u_\star} \right ),
    \tag{\theequation a,b}
\end{equation}
\end{subequations}
where $U_g$ is the component of the geostrophic bulk flow $\boldsymbol{U}$ in the shear-wise direction (respectively $V_g$ in the cross-shear direction), $u_\star$ is the friction velocity estimated from the mean wall-stress $\tau_\star = \rho u_\star^2$, and $(G_1,G_2)$ are some universal functions.
On the contrary, near the wall, the average velocity is expected to be given by the law-of-the-wall as
\begin{subequations}
\label{eq:wallsimilar}
\begin{equation}
    \frac{u}{u_\star} = F_1 \left ( \frac{z}{z_r} \right ), \quad  \frac{v}{u_\star} = 0,
    \tag{\theequation a,b}
\end{equation}
\end{subequations}
where $z_r$ is an adjustable constant, and $F_1$ is a universal function. 
Finally, it is assumed that there is an overlap region in which the velocity law-of-the-wall can be matched with the velocity-defect law (as originally proposed for turbulent flows in channels and circular tubes \cite{millikan1938critical}). 
Admissible solutions are obtained by similarity analysis \cite{csanady1967resistance,blackadar1968asymptotic}, which shows that $(F_1,G_1)$ are logarithmic functions and $G_2$ is constant.
Notably, we obtain the usual form of the law-of-the-wall as
\begin{subequations}
\label{eq:lawofthewallrough}
\begin{equation}
\frac{u}{u_\star}=\frac{1}{\kappa} \ln \left ( \frac{z}{z_r} \right ), \quad  \frac{v}{u_\star} = 0,
\tag{\theequation a,b}
\end{equation}
\end{subequations}
where $\kappa$ is the von K\'arm\'an constant, and $z_r$ is a scale height (sometimes called the roughness length). 
Finally, the matching between equations (\ref{eq:defectsimilar}b) and (\ref{eq:wallsimilar}b) in the overlap region provides the wall stress through \footnote{Note the retraction\cite{spalart2009retraction} of the more recently proposed logarithmic law with shifted origin \cite{spalart2008direct}.}
\begin{subequations}
\label{eq:t1t3}
\begin{equation}
    \frac{U \cos(\vartheta)}{u_\star} = \frac{1}{\kappa} \ln \left ( \frac{\delta_\star}{z_r} \right ) + B, \quad  \sin(\vartheta) = A \frac{u_\star}{U}, \quad 
     \vartheta = \beta+ C_5 \frac{z_r}{\delta_\star},
     \tag{\theequation a--c}
\end{equation}
\end{subequations}
with $\delta_\star=u_\star/f$, and where $\beta$ is the veering angle between the boundary (shear) stress vector and $\boldsymbol{U}$ is the free-stream geostrophic velocity.
Note that the basic theory states that $C_5=0$ \cite{csanady1967resistance,blackadar1968asymptotic}, but the $C_5$-correction has been later proposed to have a better agreement with \textsc{dns} when $Re_\delta < 5000$ for smooth boundaries \cite{spalart1989theoretical}.

As illustrated in Figure \ref{fig:EkmanBL}~(b), turbulent Ekman BLs are characterised by different sub-layers in agreement with the aforementioned analysis.
First, there is a viscous sub-layer close to the wall in which $u_\star/U$ varies almost linearly with the distance from the wall \cite{pope2000turbulent}. 
Far enough from the wall, there is the overlap region in which $u/u_\star$ approximately obeys the law-of-the-wall, before reaching the outer layer (i.e. the last sub-layer before the bulk free-stream flow).
Between the viscous and overlap sub-layers, there is a transition zone (the buffer layer) in which the mean velocity is neither linear nor logarithmic.

\subsubsection{Adjustable constants and boundary stress}
The above analysis is often assumed to be a universal feature of wall-driven turbulence \cite{pope2000turbulent}, but there is a kind of black magic to estimate which results are independent of the turbulence mechanism.
For instance, there are no universally accepted values for $\kappa$.
Proposed values for non-rotating BLs are $\kappa = 0.36 -0.45$ \cite{george2007there}, and fitted values range between $0.36$ and $0.59$ for turbulent Ekman BLs \cite{caldwell1972laboratory}.
However, the standard value is often chosen to be $\kappa \simeq 0.41$.
Without roughness, it is generally assumed that $z_r \propto \nu/u_\star$ \cite{coleman1999similarity,shingai2004study}, such that the law-of-the-wall is generally rewritten as 
\begin{subequations}
\label{eq:loglaw1}
\begin{equation}
\frac{u}{u_\star} = \frac{1}{\kappa} \ln \left ( \frac{z}{z_0} \right ) + C, \quad z_0 = \frac{\nu}{u_\star}.
\tag{\theequation a,b}
\end{equation}
\end{subequations}
In this case, the value of $C$ is also not consensual. 
Traditional estimates for non-rotating BLs put it between $4.9$ and $5.1$, but values from $4$ to $10$ may also be acceptable \cite{george2007there}. 
For turbulent Ekman BLs, very different values $C = 0.2 - 0.4$ \cite{caldwell1972laboratory} and $C = 5-5.5$ \cite{spalart1989theoretical,sous2013friction} have been proposed in the literature.
The rough case has received much less consideration for Ekman layers \cite{howroyd1975characteristics}, but has been investigated without global rotation with various roughnesses \cite{raupach1991rough,kadivar2021review}. 
For roughness of height $h_r$, such works have shown that deviations from the smooth regime occur when $u_\star h_r/\nu>2-15$, and a fully rough regime is reached when $u_\star h_r/\nu>25-90$.
In the fully rough state, data for sand roughness suggested that $z_r \approx h_r/30$. 
Note that the higher-order $C_5-$term has not been tested yet with roughness.

In addition to the von K\'arm\'an constant $\kappa$ and $z_r$, there are also three unknown constants $(A,B,C_5)$ that appear in equations (\ref{eq:t1t3}a-c) to complete the theory.
It is natural to try estimating their values using \textsc{dns}, as it is possible to design numerical models that are very close to the theoretical assumptions of the turbulence model.
These adjustable parameters are also assumed to have the same values for rough and smooth walls \cite{spalart1989theoretical}. 
Hence, some works have attempted to obtain their values using smooth boundaries with \textsc{dns}.
To do so, it is customary \cite{braun2020turbulence} to introduce the local Reynolds number given by Equation (\ref{eq:Relocal}a) such that $\delta_\star/z_0=Re_\delta^2 (u_\star/U)^2/2$ \cite{coleman1999similarity}.
Next, equations (\ref{eq:t1t3}a-c) can be rewritten as \cite{shingai2004study,shih2023turbulent} 
\begin{subequations}
\begin{equation}
    \frac{U}{u_\star} \cos(\vartheta) + \frac{2}{\kappa} \ln \left (\frac{U}{u_\star} \right ) = \frac{2}{\kappa} \ln(Re_\delta) - \frac{\ln(2)}{\kappa} + B, \quad  \sin(\vartheta)  = A \frac{u_\star}{U}, \quad 
     \vartheta = \beta+ \frac{2 C_5}{Re_\delta^2} \left( \frac{U}{u_\star} \right)^2.
     \tag{\theequation a--c}
\end{equation}
\end{subequations}
For consistency, note that there is an inconsequential typo in the sign of $\kappa^{-1 }\ln (2)$ in \cite{deusebio2014numerical}.
Typically, it is found that $C_5 \approx -52$ \cite{coleman1990numerical,coleman1999similarity,shingai2004study,deusebio2014numerical} when $Re_\delta< 5000$, with for instance the mean values $(5.91,1.35)$ when $2000 \leq Re_\delta \leq 4000$ \cite{braun2020turbulence}.
Note that different values are obtained when $C_5$ is set to zero or not.
For instance, it has been proposed that $(6.9,-0.35)$ when $C_5=0$ or $(5.4,0.26)$ when $C_5\approx-52$ \cite{shingai2004study}, but the others values $(7.1,-0.73)$ with $C_5=0$ or $(5.1,0)$ with $C_5 \approx -52$ could also be acceptable \cite{coleman1999similarity}.
The similarity theory has also been compared with measurements of the atmospheric boundary layer \cite{hess2002evaluating,hess2002evaluating2}, showing a broad agreement, but with variations of $(A,B)$ as a function of the latitude. 
This strongly suggests that, for turbulent Ekman BLs, the different coefficients do not have universal values in the similarity theory.

Finally, we can use the above theory to estimate the drag coefficient for geophysical applications. 
The wall stress being $\rho u_\star^2 \cos (\beta)$ in the $\boldsymbol{U}$ direction, the friction coefficient between the fluid and a rigid wall can be estimated as $k_f=(u_\star/U)^2 \cos(\beta)$.
For precession-driven flows, the Ekman BL theory with either $C_5=0$ \cite{cebron2019precessing,nobili2021hysteresis} or $C_5 \approx -52$ \cite{shih2023turbulent} has been used.
Note that the simpler non-rotating limit with $\vartheta=\beta=0$ has also been employed (e.g. for Venus \cite{yoder1995venus}, sometimes together with $B=\kappa^{-1} \ln (2 \kappa)$ as in \cite{williams2001lunar}), which simplifies further to a constant $U/u_\star$ if $\delta_\star/z_0$ is imposed.
For instance, the constant value $(u_\star/U)^2=k_f=0.002$ was used to model the friction at the Moon's \textsc{cmb} \cite{yoder1981free}.

\subsubsection{Experimental investigations}
Only a few experimental works have explored the Ekman BL turbulence.
Two kinds of experiments have been designed, using either annular rotating wind tunnels \cite{caldwell1972laboratory,kreider1973ekman,howroyd1975characteristics} or water-filled tanks mounted on a turntable \cite{ferrero2005physical,sous2012tsai,sous2013friction}.
Contrary to \textsc{dns}, the comparison between experimental results and the similarity theory is more complicated.
First, there is an influence of the vortical nature of the geostrophic flow in the interior, whereas the imposed flow is rectilinear in the theory.
This effect can be quantified by introducing the apparatus Rossby number $Ro_a = U/(rf)$, where $r$ is the radius of the observation point.
Typical experimental values are $Ro_a \sim 0.1 - 1$ in \cite{caldwell1972laboratory,kreider1973ekman} and $Ro_a \sim 0.48 - 2.36$ \cite{sous2013friction}, whereas we have $Ro_a \to 0$ in the theory.
A large enough turntable should thus be employed to approach the theoretical regime.
For instance, as stated in \cite{howroyd1975characteristics}, a wind tunnel with a radius of approximately $11$~m should be used to avoid significant curvature effects. 
To weaken such effects, recent experiments \cite{ferrero2005physical,sous2012tsai,sous2013friction} have been carried out on the \textsc{coriolis} table, which is the largest turntable worldwide with a tank whose diameter is $13$~m. 
Another difficulty comes from the experimental measurements.
The friction velocity $u_\star$ has been obtained using wall-stress
sensors \cite{caldwell1972laboratory} (requiring difficult calibration and alignment procedures), by using linear regression of law-of-the-wall (\ref{eq:lawofthewallrough}) \cite{kreider1973ekman,howroyd1975characteristics}, or from the rate of change of the angular momentum during spin-up or spin-down processes \cite{sous2013friction}.
The comparison suggests that the two latter techniques are the more accurate ones \cite{caldwell1972laboratory,sous2013friction}.
For the velocity profiles, the oldest studies used hot-wire anemometry \cite{caldwell1972laboratory,kreider1973ekman,howroyd1975characteristics}, whereas the most recent ones employed \textsc{piv} techniques \cite{ferrero2005physical,sous2012tsai,sous2013friction}. 

\begin{figure}
    \centering
    \begin{tabular}{cc}
        \includegraphics[width=0.48\textwidth]{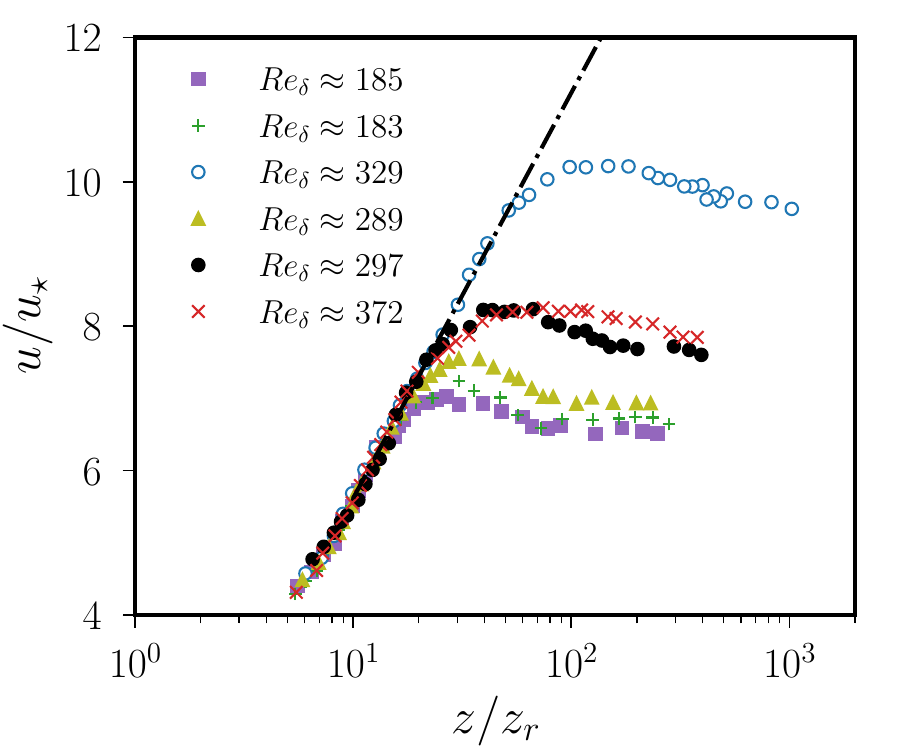} & 
        \includegraphics[width=0.48\textwidth]{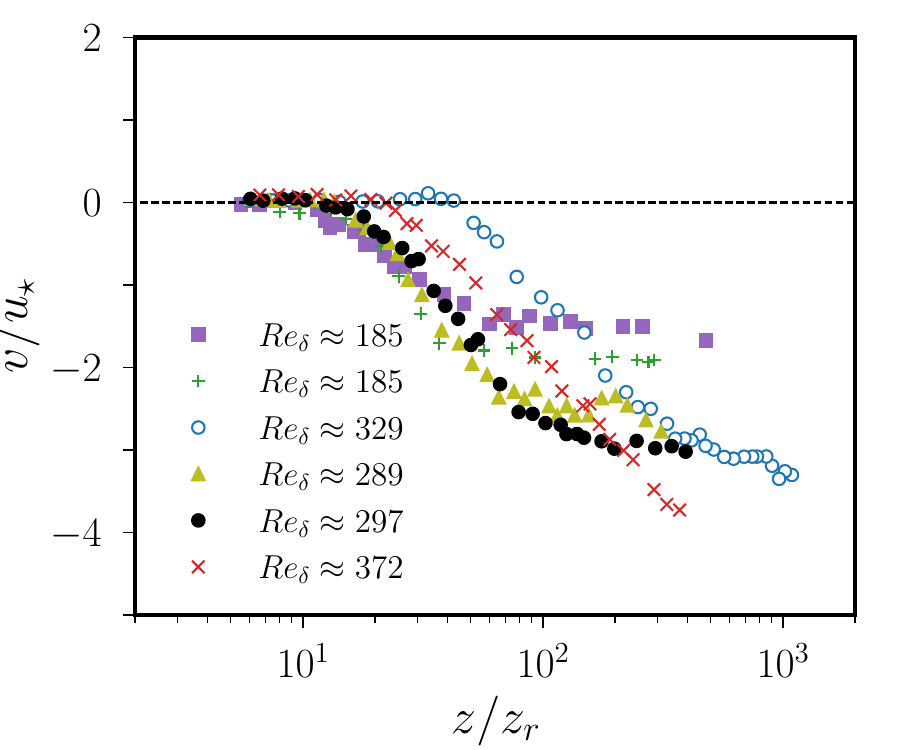} \\
        (a) & (b) \\
    \end{tabular}
    \caption{Turbulent Ekman BLs in the pioneering experiments of Kreider \cite{kreider1973ekman}. \textbf{(a)} Normalised shear-wise velocity $u/u_\star$ as a function of $z/z_r$. Dash-dot line shows law-of-the-wall (\ref{eq:lawofthewallrough}a) with $\kappa \approx 0.41$. \textbf{(b)} Normalised cross-shear velocity $v/u_\star$ as a function of $z/z_r$. Dashed line shows law-of-the-wall (\ref{eq:lawofthewallrough}b) for the cross-shear velocity.}
    \label{figu:kreiderexp}
\end{figure}

Given the intrinsic experimental difficulties, it is a tour de force that experiments have managed to confirm the similarity theory.
The first experimental results, obtained at $Re_\delta \sim 600-2000$ \cite{caldwell1972laboratory}, indeed showed a broadly good agreement with the similarity theory when $z/z_r > 10^2$.
Yet, all the theoretical predictions were not confirmed at that time.
For instance, the shear-wise velocity-defect-law (\ref{eq:defectsimilar}a) was not fully obtained in \cite{caldwell1972laboratory}, as well as a vanishing cross-shear velocity as expected from law-of-the-wall (\ref{eq:lawofthewallrough}b) in the overlap region. 
The experimental confirmation of such predictions was later given by Kreider \cite{kreider1973ekman}, as illustrated in Figure \ref{figu:kreiderexp} where we have digitised some of his results.
In particular, we observe that the measurements of the cross-shear velocity roughly vanish when $z/z_r \lesssim 20$, as expected from law-of-the-wall (\ref{eq:lawofthewallrough}b) in the overlap region.
Another striking observation is that these old measurements are already in good agreement with the similarity theory, even for the moderate values $Re_\delta \sim 150-400$.
More recent experiments \cite{ferrero2005physical,sous2013friction} did further confirm the robustness of the similarity theory for smooth boundaries with cleaner measurements, but, quantitatively, the different studies slightly disagree. 
For instance, the constants $(A,B)$ were found to be $(7.32,-2.67)$ in \cite{sous2013friction} and $(6.01,0.018)$ in \cite{caldwell1972laboratory}, which typically gives a friction about $15-20$~\% higher in \cite{caldwell1972laboratory} than reported in \cite{sous2013friction}.
Such differences have been attributed to roughness by \cite{sous2013friction}, which was not well controlled in \cite{caldwell1972laboratory}.

\begin{figure}
    \centering
    \includegraphics[width=0.95\textwidth]{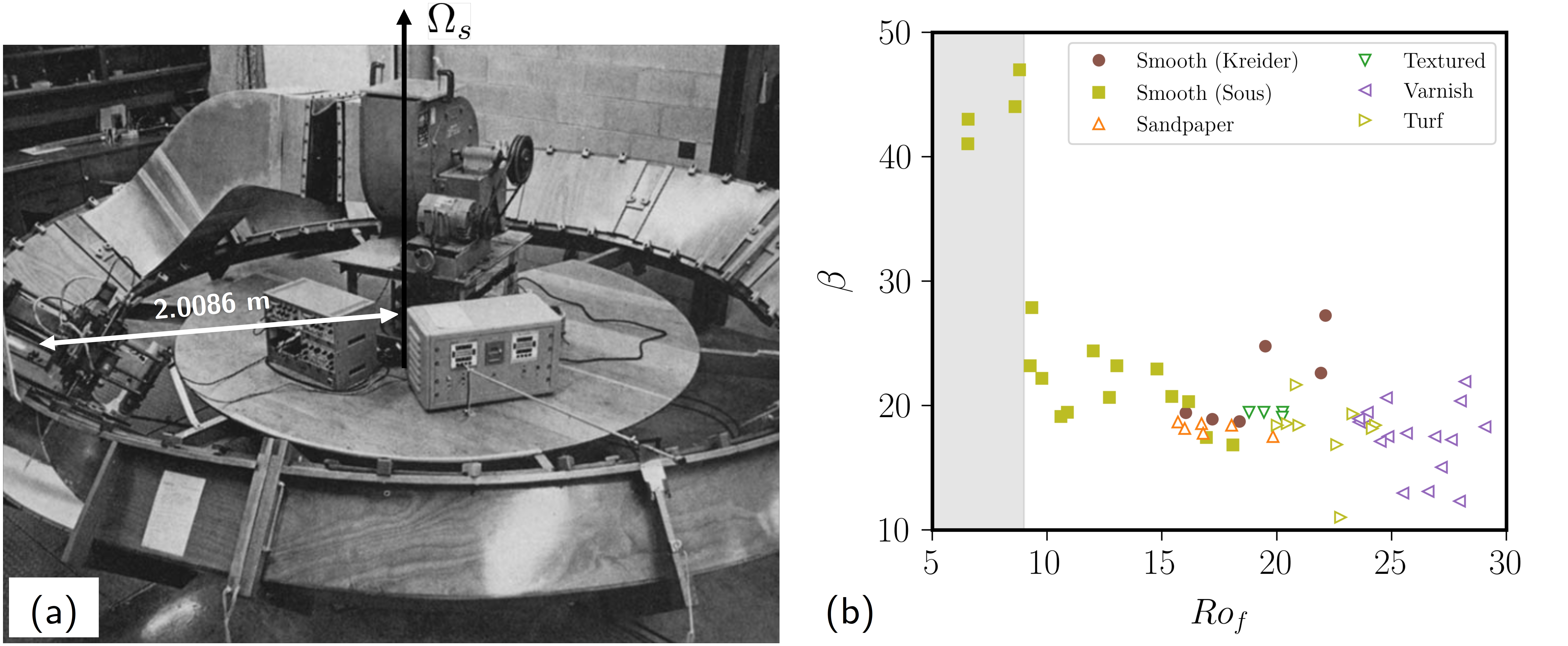}
    \caption{ \textbf{(a)} Experimental setup to investigate a turbulent Ekman BL with roughness \cite{howroyd1975characteristics}. Annular rotating wind tunnel with a duct in wood, whose floor is inclined from an angle of $45^\circ$ to the horizontal. The duct has a width of $45.72$~cm and a height of $30.48$~cm. \textbf{(b)} Veering angle $\beta$ (in degrees) as a function of the modified Rossby number $Ro_f = u_\star/\sqrt{f' \nu}$, where $f'\approx f+U/r$ is the modified Coriolis parameter \cite{sous2013friction}. Measurements for smooth \cite{kreider1973ekman,sous2013friction} rough boundaries \cite{howroyd1975characteristics} are shown. The transition from a laminar to a turbulent Ekman BL occurs at $Ro_f \simeq 9$ \cite{sous2013friction}, corresponding to a BL instability at a local (boundary layer) Reynolds number of $Re_\delta \approx 150$ \cite{caldwell1970characteristics,sous2013friction}.}
    \label{fig:BLexp}
\end{figure}

\begin{figure}
    \centering
    \begin{tabular}{cc}
        \includegraphics[width=0.47\textwidth]{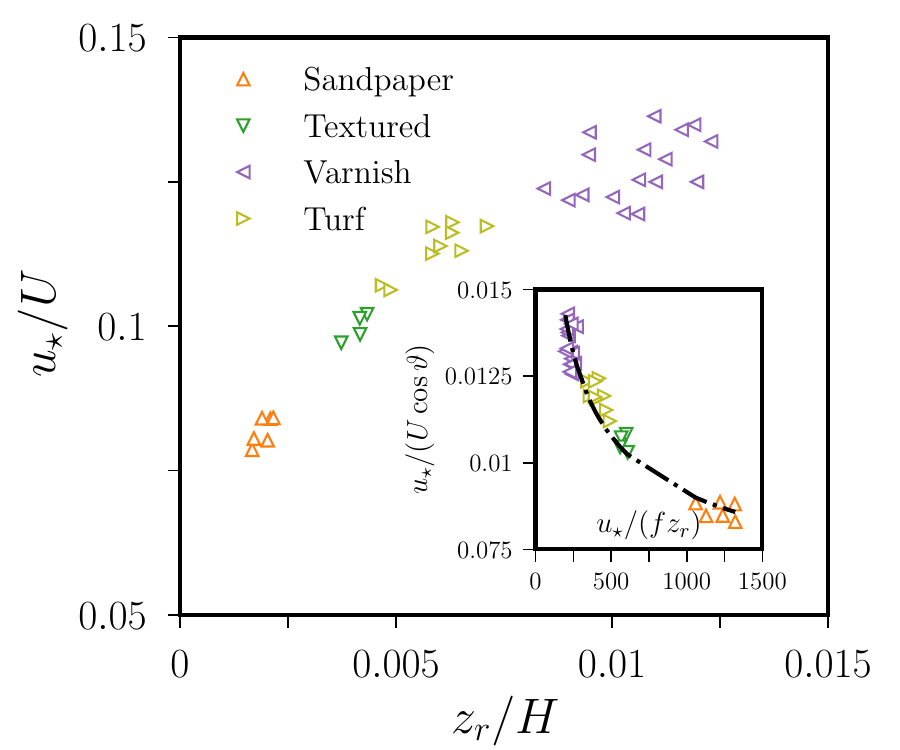} &
        \includegraphics[width=0.47\textwidth]{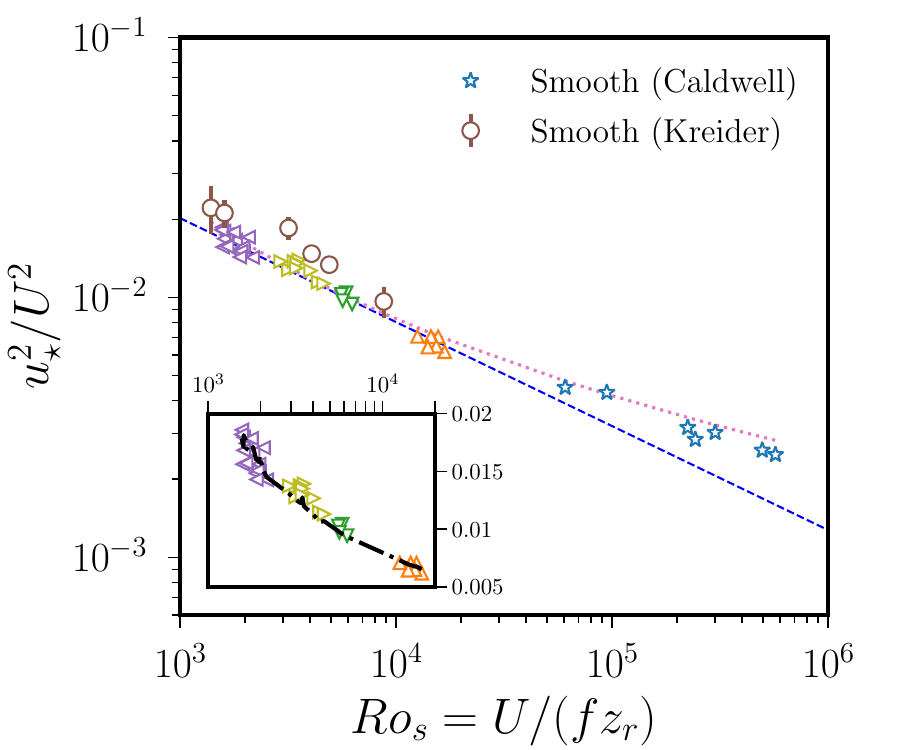} \\
        (a) & (b) \\
    \end{tabular}
    \caption{Effects of roughness of turbulent Ekman BLs from \cite{howroyd1975characteristics}. Common symbols are used in the two panels. \textbf{(a)} Dimensionless friction velocity $u_\star/U$ as a function of $z_r/H$. $H$ is the geostrophic height above which the angular deflection of the velocity vector no longer changes in magnitude. Inset shows the comparison between the rough points and law (\ref{eq:t1t3}a) with the fitted parameters $\kappa \approx 0.41$ and $B \approx -5.87$ assuming $C_5=0$. \textbf{(b)} Geostrophic drag coefficient $(u_\star/U)^2$ as a function of the surface Rossby number $Ro_s = U/(fz_r)$. Experimental points for smooth boundaries from \cite{caldwell1970characteristics,kreider1973ekman}. Dashed (blue) line shows the empirical power law $u_\star/U \propto Ro_s^{-0.2}$, which is close from the nonlinear law $u_\star/U = \alpha_1/(\log_{10}(Ro_s) - \alpha_2)$ obtained from a nonlinear eddy-viscosity model \cite{kung1968momentum,hess2002evaluating} with the fitted parameter $\alpha_1 = 0.2238$ and $\alpha_2 \approx 1.542$ (dotted pink line).
    Inset shows, using the same axes, the comparison between the rough points and law (\ref{eq:t1t3}a) with the fitted parameters $\kappa \approx 0.41$ and $B \approx -5.87$ assuming $C_5=0$.}
    \label{fig:dragcoeff}
\end{figure}

So far, there is only one study that investigated roughness effects on turbulent Ekman BLs \cite{howroyd1975characteristics}.
The experimental setup is illustrated in Figure \ref{fig:BLexp}~(a).
Various kinds of bottom roughness were employed, namely a coarse grained sandpaper, a textured paint, bare duct floor coated with clear varnish, and an artificial celanese turf (usually used for indoor-outdoor carpeting).
However, note that these experiments only allowed exploring the regime with $0.2 \leq u_\star z_r /\nu \leq 4.6$. 
A simple way to explore the effects of roughness on the Ekman layer is to look at the veering angle $\beta$ and the drag coefficient.
As shown in Figure \ref{fig:BLexp}~(b), we observe that the rough points are very close to the smooth points (as expected in this regime). 
However, the roughness height $z_r$ is found to have more influence on the friction velocity (Figure \ref{fig:dragcoeff}a).
Indeed, $u_\star/U$ increases when the dimensionless roughness height $z_r/H$ increases (where $H$ is defined as the geostrophic height above which the angular deflection of the velocity vector no longer changes in magnitude).
Moreover, all the data collapse on a master curve, suggesting that roughness has somehow universal effects.
As shown in the inset, the rough points are well predicted by law (\ref{eq:t1t3}a) with $\kappa \approx 0.41$ and $B \approx -5.87$. 
Finally, to compare the results with \cite{howroyd1975characteristics} and without \cite{caldwell1972laboratory,kreider1973ekman} roughness on a single plot, we show in Figure \ref{fig:dragcoeff}~(b) the evolution of the geostrophic drag coefficient $u_\star^2/U^2$ as a function of the surface Rossby number $Ro_s = U/(f z_r)$. 
We find that all the rough points seem to collapse on a power law of the form $u_\star^2/U^2 \propto Ro_s^{-0.4}$, whose exponent is larger than that predicted by a prior eddy-viscosity model yielding $u_\star^2/U^2 \propto Ro_s^{-0.238}$ \cite{prandtl1924windverteilung}. 
Note that the smooth and rough points also almost collapse on a single curve, which seems to be well described by a nonlinear eddy-viscosity model \cite{kung1968momentum,hess2002evaluating}. 
Moreover, we see that the geostrophic drag coefficient estimated from \cite{kreider1973ekman} is larger than that estimated with roughness \cite{howroyd1975characteristics}.
This is somehow unexpected, but this may suggest that roughness was present but not reported in the experiment \cite{kreider1973ekman}.
Future experimental studies could thus start reproducing these points to elucidate the origin of this puzzling observation, before exploring other rough regimes with larger values of $u_\star z_r/\nu$.

\subsubsection{Remaining open questions and experimental perspectives}
The similarity theory outlined above was derived for high-$Re_\delta$ turbulent Ekman BLs above planar boundaries, but experiments and \textsc{dns} showed that it is also quite robust for moderately large values $Re_\delta \sim 200-1000$. 
Yet, all these works show that the constants in the theory are not as universal as initially expected, as they can significantly vary in the different configurations.
Therefore, this calls for dedicated studies for each application.
Another challenge is to properly assess the validity of the similarity theory with roughness, to go beyond the pioneering experiments presented in \cite{howroyd1975characteristics}.
Indeed, roughness is expected in many geophysical systems (e.g. at a planet's \textsc{cmb} \cite{narteau2001smallscale,le2006dissipation}). 
In particular, next-generation experiments should investigate the different regimes as a function of the roughness height $z_r$, to properly quantify the enhancement of the friction by roughness.
Another limitation is that, in the similarity theory, the turbulent Ekman BL is forced by a steady geostrophic flow. 
However, the orbital forcings discussed in \S\ref{sec:ellipsoid} can sustain oscillatory and non-geostrophic large-scale flows (e.g. equatorial rotation with a nearly diurnal frequency in the rotating frame for precession \cite{noir2013precession}), which may lead to turbulent Ekman BLs \cite{cebron2019precessing}.
So far, a numerical study proposed the fitted values $(A,B,C_5)=(5.74,1.55,-25)$ for the precession-driven flow oscillating at $f/2$ over a smooth boundary \cite{shih2023turbulent}, which lead to smaller values of $u_\star$ and $\beta$ compared to those obtained with a steady flow. 
Therefore, it would be worth pursuing in this direction by considering the effects of oscillatory flows on turbulent Ekman BLs in future works (e.g. to assess the effects of the flow frequency). 
In particular, previous experimental works on oscillatory turbulent BLs \cite{jonsson1980new,sleath1987turbulent}, which used oscillatory-flow water tunnels, could be revisited by including global rotation.

For geophysical applications to thick-layer fluid systems, we may also question the validity of the similarity theory with curvature effects. 
It is well-known that the laminar BL solution fails to be valid at the so-called critical latitudes \cite{greenspan1968theory}, where the laminar BL thickness diverges.
Hence, we may also anticipate some issues for the similarity theory at these critical latitudes.
Indeed, a prior numerical work \cite{shih2023turbulent} showed that it is not straightforward to properly extrapolate such local laws for turbulent Ekman BLs to a global near-spherical geometry. 
Moreover, the Ekman BL could be turbulent at some latitudes and laminar at others (e.g. near the pole and the critical latitudes, respectively). 
The proper extrapolation of the similarity theory to this case, where the mass fluxes at each latitude are interconnected (e.g. by the Ekman pumping), remains to be found. 
Moreover, the turbulent theory may not be valid in the presence of meso-scale topographies, which deserves further investigation.
If the meso-cale topography has slender or non-slender bumps, we may preserve or disrupt the viscous sub-layer. 
It is also known that wall-driven turbulence is affected by the convexity of meso-scale topography \cite{mendez2018boundary}. 
Another limitation of the above turbulent theory is that some thick-layer systems may only be at the margin of turbulence. 
For instance, the expected values of the BL Reynolds number $Re_\delta \sim 100-500$ due to precession-driven flows at the Earth's \textsc{cmb} \cite{buffett2021conditions,sikdar2023differential} might be large enough to sustain wall-driven instabilities, but not to obtain a fully turbulent BL.
Since experimental works clearly show some deviations from the similarity theory when $Re_\delta$ is not large enough, the similarity theory may not give accurate enough predictions for such applications.

\begin{figure}
    \centering
    \includegraphics[width=0.85\textwidth]{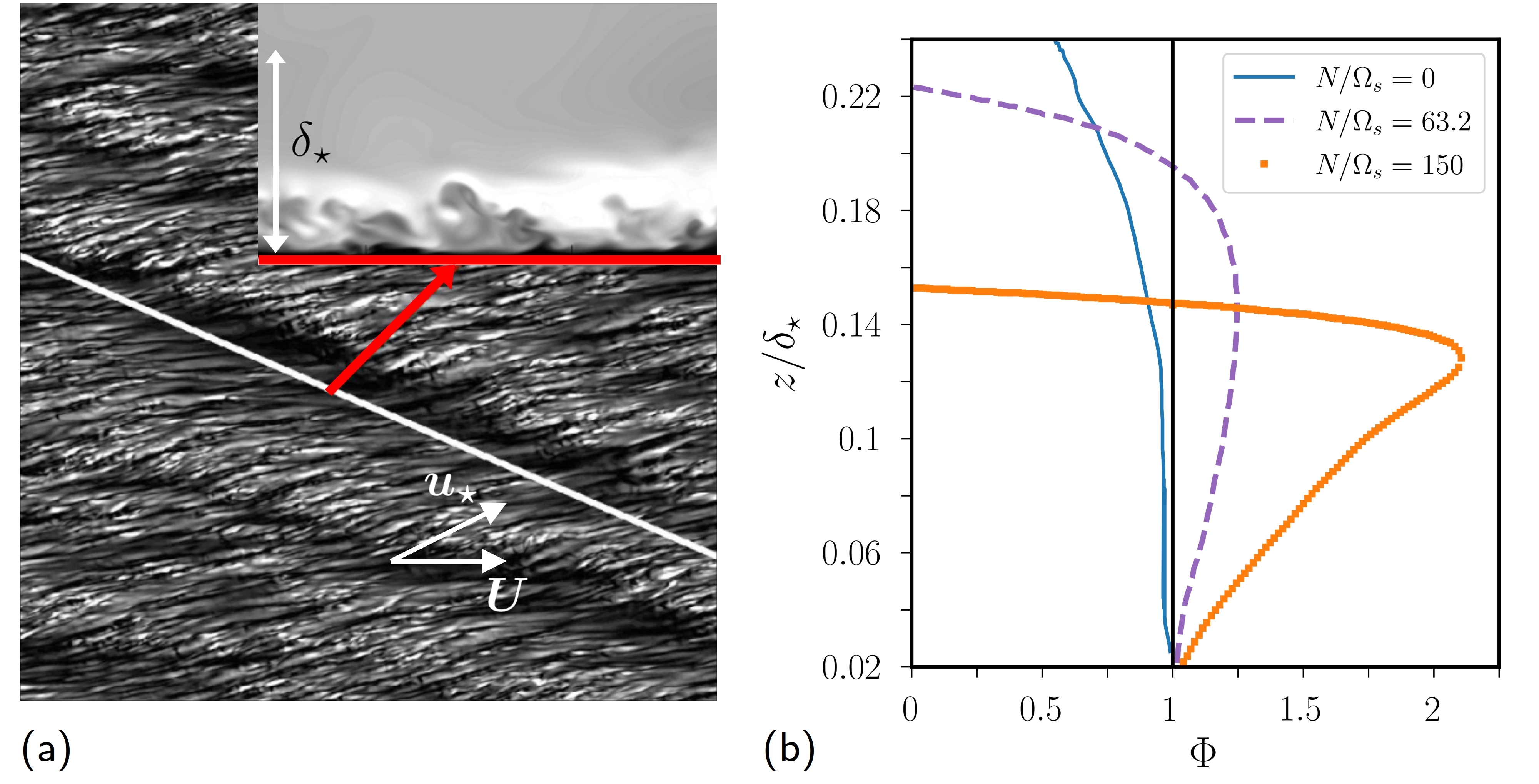}
    \caption{Turbulent Ekman BL with stratification in \textsc{dns}. \textbf{(a)} Instantaneous visualisation of wall shear stress. Colour coded from black (low) to white (high). White arrows indicate the orientation of bulk geostrophic flow $\boldsymbol{U}$ and that of $u_\star$. Tilted white line indicates a large-scale structure in the flow. Inset shows a vertical cross-section of $u/|\boldsymbol{U}|$ along the red arrow, with colour coded from black to white in the range $[0.12, 1.11]$. \textsc{dns} for a turbulent BL at $Re_\delta \simeq 500$ in the intermediate stratified regime, adapted from figures 9 and 11 in \cite{ansorge2014global}. \textbf{(b)} Deviations from law-of-the-wall (\ref{eq:loglaw1}) measured by the quantity $\Phi = (\kappa z/u_\star) \ \partial_z Q$ with $\kappa = 0.41$, as a function of the ratio $N/\Omega_s$. \textsc{dns} (using a large-eddy-simulation approximation) at $Re_\delta \simeq 5 \times 10^{7}$ and $Pr \sim 5-10$, adapted from figure 9 in \cite{taylor2008stratification}.}
    \label{fig:BLS}
\end{figure}

Note that turbulent Ekman BLs can also be affected by density variations. 
In particular, buoyancy effects deserve consideration to understand the interplay between turbulent BLs and convective flows. 
Temperature should be governed by a similar law-of-the-wall \cite{huang1995law}, but the validity of a universal law-of-the-wall for temperature was also questioned \cite{bradshaw1995law}.
The interaction between rotating convection and turbulent Ekman BLs remains to be studied, but the non-rotating case has already been considered.
It could indeed be key for elucidating the transition between the different regimes of turbulent convection (see in \S\ref{sec:convection}), or for estimating basal melting of ice shelves \cite{rosevear2021role}. 
The stably stratified regime is also of paramount importance for some geophysical applications. 
As illustrated in Figure \ref{fig:BLS}~(a), stratified Ekman BLs have some similarities with unstratified Ekman BLs (e.g. a friction velocity $u_\star$ tilted from the bulk geostrophic flow). 
However, strong departures from the law-of-the-wall can be obtained depending on the value of $N/\Omega_s$ (even for high-$Re_\delta$ BLs, see Figure \ref{fig:BLS}b).
Different turbulent regimes can be identified as a function of the strength of stratification \cite{ansorge2014global}, from weakly stratified Ekman BLs (similar to neutral ones) to strongly stratified Ekman BLs (which can exhibit global intermittency and turbulent collapse, as for non-rotating stratified BLs \cite{mahrt2014stably}). 
Except in the weakly stratified regime, the standard law-of-the-wall fails to describe the observed turbulence with stratification \cite{taylor2008stratification,ansorge2014global}.
A quantitative understanding of stratified turbulent Ekman BLs is thus considerably more difficult than without stratification. 
Other similarities theories have been proposed to go beyond the classical log-law theory (e.g. Monin-Obukhov \cite{monin1970atmospheric,obukhov1971turbulence} or Perlin \cite{perlin2005modified,taylor2008stratification} theories), but their range of validity remains to be carefully determined. 
Finally, wave motions can also be emitted by the turbulent dynamics in stratified Ekman BLs \cite{taylor2007internal,taylor2008stratification,deusebio2014numerical}, and be radiated away in the bulk. 
The vertical energy flux associated with such waves is likely negligible compared to the turbulent friction in the BL, but these waves could affect entrainment and mixing processes in the near-wall region (e.g. without global rotation \cite{linden1975deepening,xuequan1986mixing,munroe2014internal}).

To finish with, we have pointed out that turbulent Ekman BLs are far from being fully understood. 
As such, it calls for new experimental (and numerical) studies before conducting geophysical extrapolation of the results. 
Such works should explore the BL dynamics for low to medium values of $Re_\delta$ to understand the route to wall-driven turbulence, as well as high-$Re_\delta$ BLs to characterise the properties of the fully turbulent regimes and to assess the validity of the turbulence theories.

\section{Convection-driven flows with topography}
\label{sec:convection}
\begin{figure}
    \centering
    \includegraphics[width=0.75\textwidth]{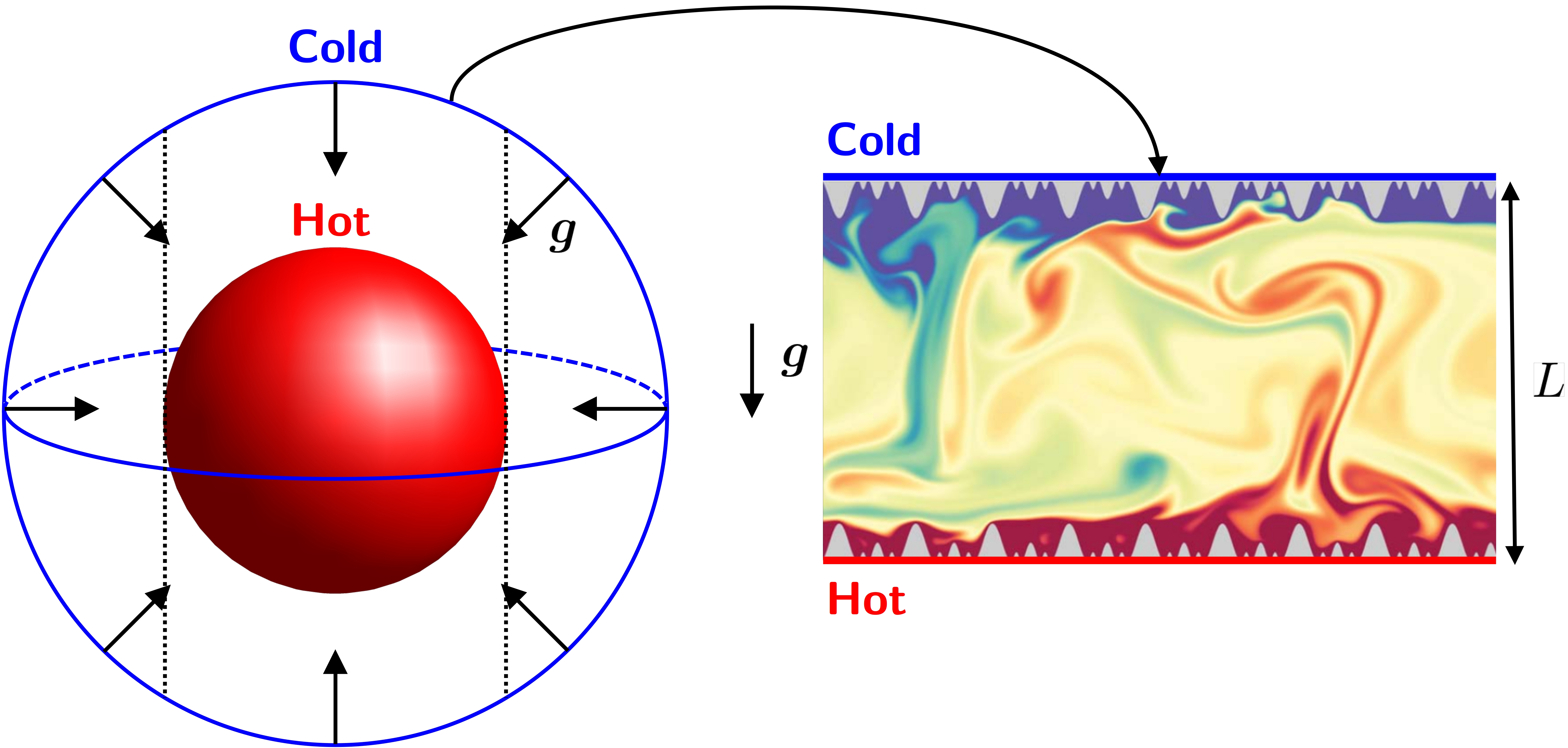}
    \caption{Thermal convection in a thick-layer envelope (e.g. with radial gravity $\boldsymbol{g}$ in a spherical shell), as modelled by Rayleigh-B\'enard convection (RBC) in experiments (with axial gravity). Visualisation of temperature variations with multiple-scale topography on top and bottom planes is also illustrated (adapted from figure 2 in \cite{zhu2019scaling}).}
    \label{fig:sketchRB}
\end{figure}

Many geophysical systems are subject to thermal convection (e.g. the Earth's liquid core \cite{jones2015tog}, subsurface oceans \cite{soderlund2024physical}, or the Sun \cite{hanasoge2016seismic}). 
It is thus natural to seek physical insight into turbulent convection using experiments.
Experimental setups have mainly considered variations around the classical configuration of Rayleigh-B\'enard convection (RBC), for which convection is driven by the temperature difference $\Delta T$ between a bottom plate and a top one at the distance $L$.
As illustrated in Figure \ref{fig:sketchRB}, RBC is well suited to experimentally model the local properties of turbulent convection of a thick-layer fluid system (even with rotation, as suggested by a recent numerical work \cite{gastine2023latitudinal}). 

In the following, we assume that convection can be described using the Oberbeck-Boussinesq (OB) approximation (i.e. with constant physical properties, except its density that varies with temperature in the buoyancy force).
Despite the restricted range of validity of the OB approximation \cite{spiegel1960boussinesq,gray1976validity}, it is often employed in theoretical works and usually offers a good (qualitative) understanding of most experimental results.
RBC is then controlled by the Rayleigh number $Ra$, the Prandtl number $Pr$, and the convective Rossby number $Ro_c$ (which is a proxy of the Rossby number based on $L$).
They are defined by
\begin{subequations}
\begin{equation}
Ra = \frac{\alpha g \Delta T L^3}{\nu \kappa}, \quad Pr = \frac{\nu}{\kappa}, \quad Ro_c = E \left ( \frac{Ra}{Pr} \right )^{1/2},
\tag{\theequation a--c}
\end{equation}
\end{subequations}
where $\alpha$ is the thermal expansion coefficient of the fluid, $g$ is the acceleration of ambient gravity, $\nu$ is the kinematic viscosity of the fluid, and $\kappa$ is the thermal diffusivity.
Convection is affected by global rotation when $Ro_c \ll 1$ and, in contrast, is barely modified by rotation when $Ro_c \geq 1$. 

One of the difficulties in convection modelling is to reach the extreme parameters characterising geophysical systems (Table \ref{table:convection}).
For a water-filled tank with $\Delta T \sim 100$~K (at maximum), it is difficult to obtain high-$Ra$ convection given the thermal expansion coefficient of water $\alpha \sim 10^{-4}$~K${}^{-1}$  (which does not efficiently convert temperature differences into density variations to explore high-$Ra$ convection).
This effect can be partly balanced by increasing the height $L$ of the tank (e.g. with $L\sim4$~m in the \textsc{tropoconvex} experiment \cite{cheng2020laboratory}), but there are then some mechanical and thermal constraints (e.g. it is difficult to ensure thermal insulation on the lateral walls).
Therefore, water-filled experiments are typically limited to $Ra \sim 10^{12}-10^{13}$, and most experiments probing more vigorous convection use other fluids with $0.6 \leq Pr \leq 7$ \cite{roche2020ultimate} (e.g. He or SF${}_6$).

A detailed account of the fundamental aspects of convective flows is available elsewhere for non-rotating \cite{roche2020ultimate,lohse2024ultimate} and rotating \cite{ecke2023turbulent} convection, so we will not discuss them here.
Instead, we will briefly discuss experimental (and numerical) works dedicated to the interplay between turbulent convection and small-wavelength topography. 
The effects of a topographic $\beta-$plane, as well as of global spherical-like geometries, are also beyond the scope of the following review (although they are directly relevant for modelling thick-layer systems). 
We refer the interested reader to \cite{cardin2015tog,potherat2024seven} for geophysically motivated descriptions of such convection experiments.  
We first discuss the interplay between small-wavelength topography and RBC (both without and with rotation), and then we outline some experimental perspectives.

\begin{table}
    \caption{Typical values of the Rayleigh number $Ra$, the Prandtl number $Pr$, the Ekman number $E$, and the convective Rossby number $Ro_c$ in natural (turbulent) systems and in laboratory experiments.}
    \begin{tabular}{lcccc}
    \hline
    {} & $Ra$ & $Pr$ & $E$ & $Ro_c$ \\
    \hline
    Earth's core \cite{jones2015tog} & $10^{20}-10^{30}$ & $10^{-2} - 1$ & $10^{-15}$ & $\ll 1$ \\
    Fully molten silicate layers \cite{solomatov2015tog,zhang2022internal} & $10^{25} - 10^{30}$ & $1-10^4$ & $10^{-13}$ & $10^{-2}-10^2$ \\
    Sun \cite{hanasoge2016seismic,schumacher2020colloquium} & $10^{19}-10^{24}$ & $10^{-7}-10^{-4}$ & $\leq 10^{-14}$ & $\geq 1$ \\
    Experiments \cite{roche2020ultimate,ecke2023turbulent} & $\leq10^{15}$ & $10^{-2}-10^3$ & $5 \times 10^{-9}-\infty$ & $10^{-4}-\infty$\\
    \hline
\end{tabular}
\label{table:convection}
\end{table}

\subsection{Non-rotating (and rotating) RBC with small-scale topography}
We first consider the limit case of non-rotating RBC (i.e. $Ro_c \to \infty$), which is appropriate for the dynamics of slowly rotating convective systems (e.g. solar convection, and possibly molten silicate layers). 
In particular, estimating the efficiency of heat transport by convection is key for understanding the heat budget of these systems and their long-term thermal evolution over billions of years. 
Convective heat transport is usually characterised by the Nusselt number $Nu$ given by $Nu = Q L/(k \Delta T)$, where $k$ is the thermal conductivity (in W.m${}^{-1}$.K${}^{-1}$), and $Q$ is the heat flux per unit of area with convection (in W.m$^{-2}$).  
There is still a long-standing controversy about the asymptotic scaling law $Nu \sim f(Ra,Pr)$ for turbulent convection in the asymptotic high-$Ra$ regime.
Indeed, classical theory predicts $Nu \sim Ra^{1/3}$ \cite{malkus1954heat}, whereas the ultimate theory states that $Nu \sim Ra^{1/2}Pr^{1/2}$ (at least for $Pr \leq 1$) when convection operates in a diffusion-free regime \cite{spiegel1963generalization,kraichnan1962turbulent}. 
Actually, in the theory, the ultimate regime of RBC is expected to be linked to the existence of turbulent BLs. 
A few studies have thus investigated the nature of the BLs in high-$Ra$ convection. 
It was first conjectured that BL turbulence would occur when the local friction Reynolds number is greater than $\sim 200$ in non-rotating RBC \cite{schumacher2016transitional}. 
Later on, some experiments \cite{ahlers2012logarithmic,ahlers2014logarithmic} and \textsc{dns} \cite{van2015logarithmic,zhu2018transition,he2024turbulent,scheel2017predicting,blass2021effect} did confirm that the viscous and thermal BLs become turbulent and can be described by similar laws-of-the-wall when the friction Reynolds number is large enough. 
Yet, some local departures from the law-of-the-wall have also been documented in the presence of large-scale plumes \cite{van2015logarithmic,zhu2018transition,he2024turbulent}.
Moreover, despite the presence of turbulent BLs, contradictory results were found for the scaling of $Nu$ with $Ra$ \cite{zhu2018transition,he2024turbulent}.
Whether very high-$Ra$ RBC with smooth plates is governed by the classical theory or falls into the ultimate regime is still thus unknown in experimental conditions, and it remains the subject of active research \cite{roche2020ultimate,lohse2024ultimate}.

\begin{figure}
    \centering
    \includegraphics[width=0.95\textwidth]{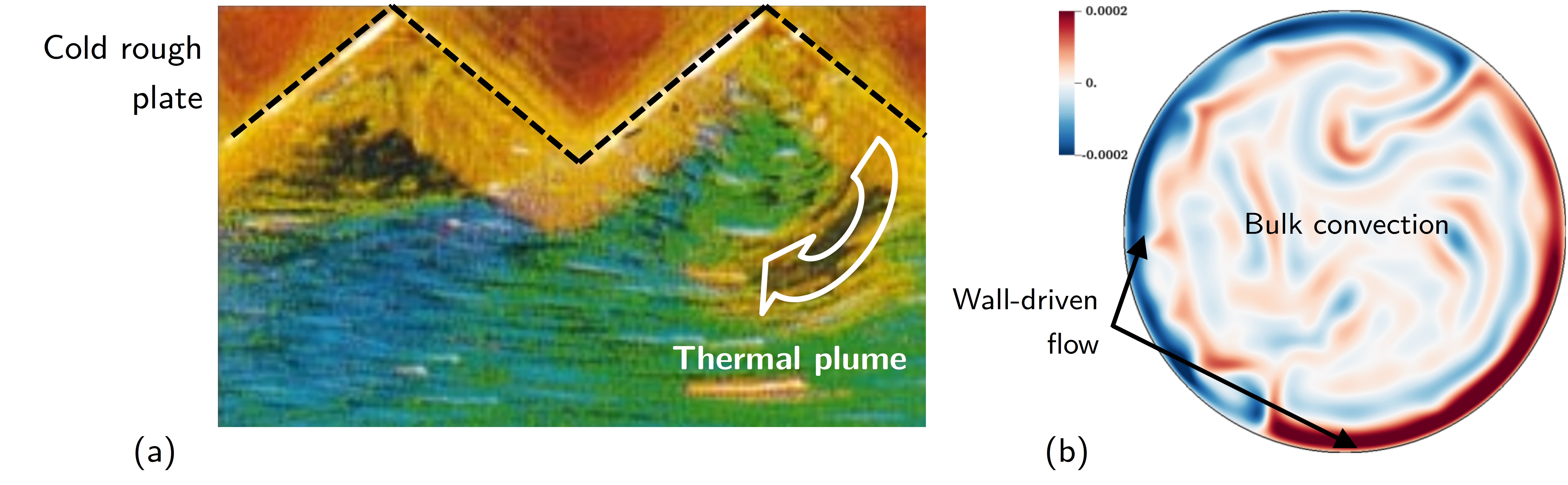}
    \caption{\textbf{(a)} Observation of topography-driven flows in an experiment of RBC with rough plates at $Ra \simeq 2.6\times10^9$ and $Pr \simeq 5.4$. Visualisation of the flow (streak image of the fluid, seeded with thermochromic liquid crystal spheres) near the upper rough (cold) plate. Wall-driven (cold) eruptions are in brown, and warmer flow structures are in green and blue. Adapted from figure 4 in \cite{du2000turbulent}. \textbf{(b)} Observation of wall-driven mode in rotating RBC. Numerical simulation in a cylinder at $Ra=5 \times 10^{10}$, $Pr=5.2$ and $E=10^{-7}$. Colour bar shows the vertical velocity in the mid-plane of the cylinder. Adapted from figure 5 in \cite{favier2020robust}.}
    \label{fig:convetion}
\end{figure}

This fundamental problem has been revisited by using topographic effects for a few years. 
Indeed, small-scale topography can favour wall-driven flows \cite{liot2016boundary,liot2017velocity}, such as thermal plumes and wall-driven turbulence (see Figure \ref{fig:convetion}a).
As such, it could possibly favour the transition to the ultimate regime \cite{toppaladoddi2017roughness}. 
Different experiments of RBC with small-scale topography on the walls have thus been conducted, using topography with different shapes and heights \cite{shen1996turbulent,du1998enhanced,ciliberto1999random,du2000turbulent,tisserand2011comparison,salort2014thermal,wei2014heat,xie2017turbulent,rusaouen2018thermal,zhu2019scaling}. 
Most studies employed either water or another fluid with $Pr \geq 1$, but a few used an air-filled experiment \cite{liot2016boundary,du2024thermal}.
These studies did find that small-scale topography can often enhance heat transport by turbulent convection, with either an exponent larger than $1/3$ or with a higher numerical prefactor.  
Scaling laws with exponents close to $1/2$ have even been experimentally reported in some studies \cite{roche2001observation,zhu2018wall}. 
This suggests that turbulent convection might reach a diffusion-free regime in the presence of wall-driven turbulence (at least for the considered values of $Ra$).
However, topography does not always enhance heat transport.
By further increasing the value of $Ra$, the $1/2$ scaling can be lost \cite{zhu2017roughness}, which can depend on the chosen topographies \cite{zhu2019scaling}.

Finally, contrary to the non-rotating case, the interaction between rotating RBC and small-scale topography has only received scant attention 
\cite{joshi2017heat,tripathi2024regimes}. 
Further work is thus required to get insight into the effects of topography on RBC.
Rotating RBC is also often impacted by the presence of (zonal) wall modes along the lateral boundaries (Figure \ref{fig:convetion}b).
Their presence is largely indifferent to the bulk convection \cite{zhang2020boundary,de2023robust}, and they may have a non-trivial topological origin \cite{favier2020robust,zhang2024non}. 
From an experimental viewpoint, these modes can affect heat flux measurements in the bulk, and reduce our ability to precisely study the regimes of bulk turbulence \cite{ecke2023turbulent}. 
A recent theoretical work \cite{terrien2023suppression} showed that such wall-driven flows can be strongly weakened by including well-chosen topographic defaults on the lateral walls, which paves the way for future experimental applications.

\subsection{Experimental perspectives}
Characterising heat transport of RBC with small-scale topography in the presence (or absence) of wall-driven turbulence is key from a fundamental viewpoint, but it could also be important for engineering \cite{garcia2012influence} and geophysical applications.
For instance, the BL Reynolds number due to convection may have the typical value $100-1000$ at the Earth's \textsc{cmb} \cite{jones2015tog}, which could be sufficient for the onset of BL instabilities and wall-driven turbulence.
Further experimental work, together with joint theoretical and numerical studies, may therefore help to make progress on this fundamental question
For rotating RBC, including topographic defaults on lateral walls seems key to weaken the strength of wall-driven flows in future experiments \cite{terrien2023suppression}.
Finally, we could naturally wonder how much the various convection scaling laws remain valid in the presence of non-Oberbeck-Boussinesq (NOB) effects (i.e. non-uniform material properties of the fluid due to temperature variations) or compressibility.
Such effects can be important in some experimental conditions (e.g. a gas near its saturated vapor curve \cite{urban2019elusive} to explore high-$Ra$ convection, or a gas subject to extreme rotation rates \cite{menaut2019experimental}), but also in some geophysical systems (e.g. for mantle convection in super-Earth \cite{ricard2023compressible}, or in gas giants).
As for incompressible RBC \cite{doering1996variational}, theory shows that the Nusselt number in non-rotating compressible (anelastic) convection remains bounded by $Ra^{1/2}$ \cite{alboussiere2024upper}.
Hence, the ultimate regime of non-rotating RBC might also exist with compressibility.
As such, the interplay between small-scale topography and compressible convection would certainly deserve consideration in the future.

Finally, exploring the effects of meso-scale topographies on rapidly rotating convection in planetary-core geometries would be worth considering.  
Indeed, the low-frequency dynamics of planetary-core convection is strongly shaped by global rotation, which tends to sustain columnar motions that are nearly invariant along the rotation axis of the fluid. 
However, prior theoretical \cite{bell1996influence,herrmann1998stationary,bassom1996localised} and experimental \cite{westerburg2003centrifugally} works in the cylindrical annulus geometry showed that this low-frequency dynamics can be modified in the presence of sinusoidal deformations on top and bottom walls (e.g. in trapping columnar motions near bumps, or in sustaining low-frequency Rossby waves).
Similar effects were found in \textsc{dns} using a quasi-geostrophic approximation, in a rotating spherical shell with a non-slender meridional ridge \cite{calkins2012effects}.
The convective flow was dominated by zonal circulation when $Ra$ is large enough and, at low values of $Ro$, blocked flows were reported (as in \cite{weeks1997transitions} for slender obstacles).
Moreover, it was observed that topography can transfer energy from large-scale (convective) flows to smaller-scale flows in the form of Rossby waves (whose amplitude scales as $Ro^{1/2}$). 
Future experiments could build upon such prior works to explore these effects, such as in the idealised annulus geometry that is amenable to experimental investigation (where Earth's gravity is replaced by centrifugal gravity \cite{potherat2024seven}).

\section{Brief discussion of rotating stratified flows}
\label{sec:stratification}
Beyond neutrally buoyant fluids and convection, understanding the dynamics of rotating and stratified flows with topography is also crucial for some geophysical applications (e.g. the wave-driven drag over a topography, wave-driven mixing processes in rotating stratified ellipsoids, or the turbulence in stratified Ekman BLs). 
The effects of density stratification have not been reviewed in the previous sections, as they deserve specific review articles.
We refer the interested reader to the two reviews in this special issue on regimes of stratified turbulence without topography \cite{cortet2024turbulence,lefauve2024cras}, and to \cite{legg2021mixing,baines2022topographic} for theoretical descriptions of topographic effects in stratified flows. 
However, rotating experiments with stratification and topography are not very common \cite{boyer2000laboratory,stewart2024stratified}. 
In brief, these experimental works have mainly explored the regime $N/\Omega_s \sim \mathcal{O}(1)$, and often considered the low-frequency dynamics (e.g. to investigate spin-up processes, or the quasi-geostrophic dynamics).
In the following, we explain why it is still very difficult to study rotating stratified flows in the laboratory.
Indeed, there are intrinsic difficulties that need to be overcome in future experimental works before answering unsolved geophysical questions.

\begin{figure}
    \centering
    \begin{tabular}{cc}
    \includegraphics[width=0.47\textwidth]{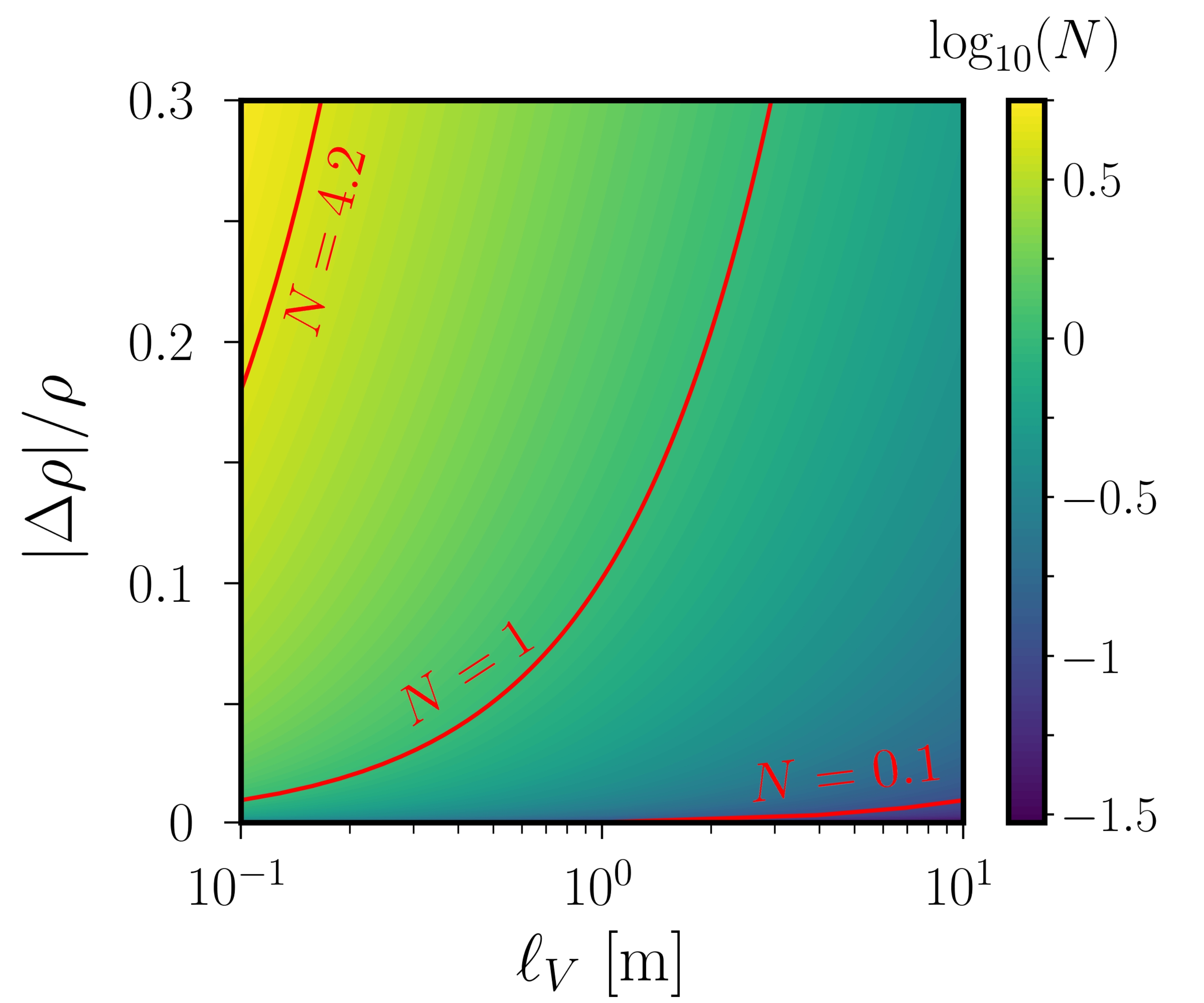} & 
    \includegraphics[width=0.47\textwidth]{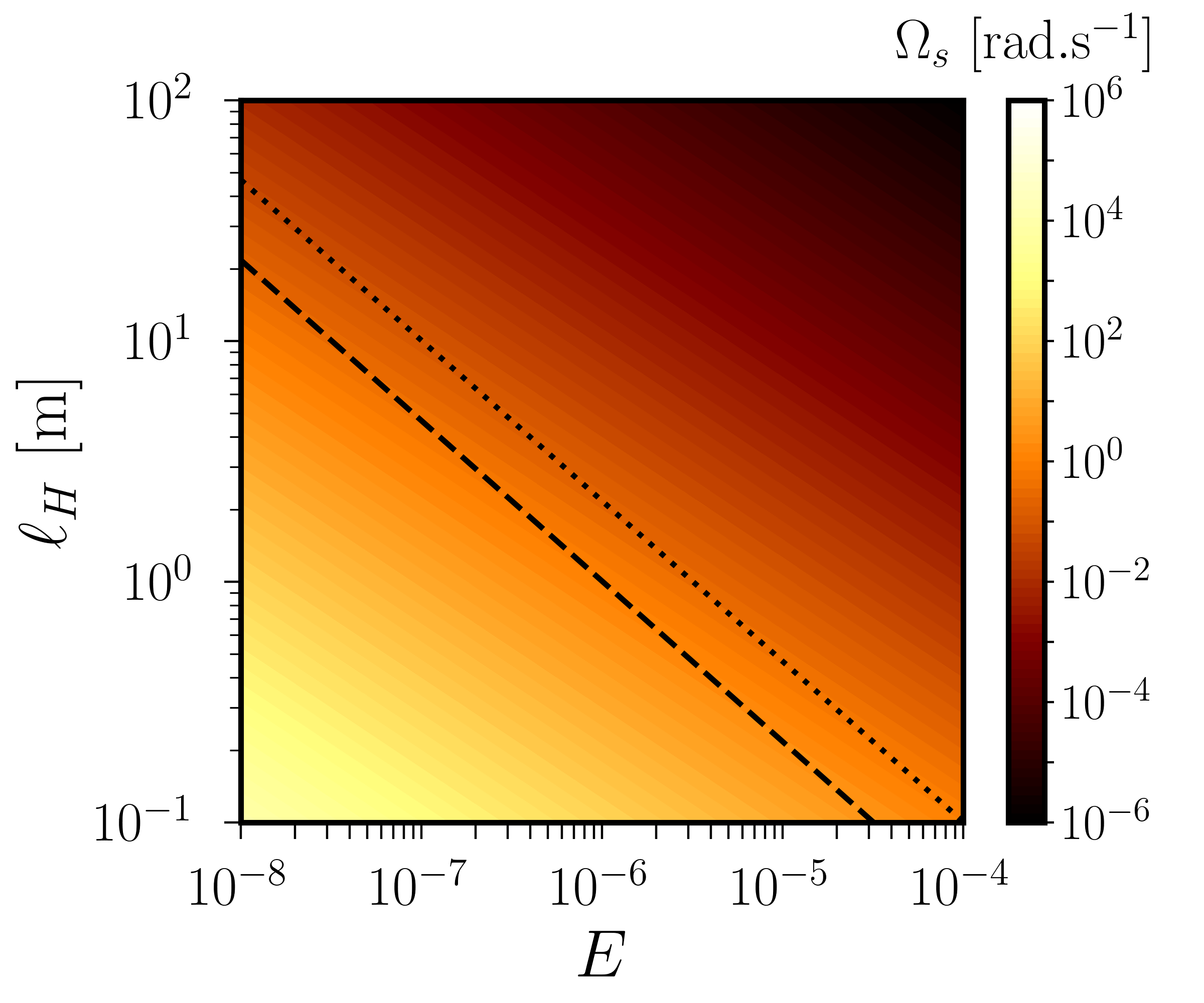} \\
    (a) & (b) \\
    \end{tabular}
    \caption{\textbf{(a)} Typical values (in logarithmic scale) of the BV frequency $N \sim \sqrt{(g/\ell_V) |\Delta \rho|/\rho}$ (in rad.s${}^{-1}$) in a water-filled experiment at room temperature, where $\ell_V$ is the height of the fluid container (in metres), $g \simeq 9.81$~m.s${}^{-2}$ is the acceleration of gravity, and $|\Delta \rho|/\rho$ is the (relative) density anomaly. The water density can be changed by varying the salt concentration (from $0$ to approximately $300$~kg.m${}^{-3}$). Red curves show the values $N \in [0.1,1,4.2]$~rad.s${}^{-1}$. \textbf{(b)} How to keep centrifugal gravity under control in a rotating experiment. Colour bar shows the rotation rate $\Omega_s$ (in rad.s${}^{-1}$) required to reach the Ekman number $E$ for a water-filled experiment with equal horizontal and vertical length scales $\ell_H \sim \ell_V$. Dashed line (respectively, dotted line) shows the length scale $\ell_H$ given by Equation (\ref{eq:negligibleCG}) such that $\Delta_g = 10^{-1}$ (respectively, $\Delta_g = 10^{-2}$).}
    \label{fig:stratified}
\end{figure}

\subsection{Accessible experimental regimes}
Since the strength of stratification is subject to strong uncertainties for geophysical applications, exploring the dynamics for a broad range of values of $N/\Omega_s$ would be desirable. 
We can estimate BV frequency (\ref{eq:BVN}) for an incompressible fluid from $N^2 \sim (g/\ell_V) |\Delta \rho|/\rho$ in laboratory conditions, where $\ell_V$ is the height of the container (in metres), $g \simeq 9.81$~m.s${}^{-2}$ is the value of Earth's gravity, and $|\Delta \rho|/\rho$ is the relative density anomaly.
For salty (NaCl) water, the density anomaly can vary from $0$ to $\sim 0.3$ (at most) by changing the salt concentration. 
As such, the values of $N$ that are accessible in a water-filled experiment are limited (Figure \ref{fig:stratified}a). 
Typically, for a metre-size tank (which is required to lower the value of $E$), we cannot impose an initial stratification with a BV frequency greater than $N \sim 1.7$~rad.s${}^{-1}$.
For a rotating table with a rotation rate $\Omega_s$ between $0.1-10$~rad.s${}^{-1}$, the maximum value would be $N/\Omega_s \sim 10-20$. 
Beyond the regime $N/\Omega_s \leq 1$, stratified regimes with $N/\Omega_s \geq 1$ could also be considered in experiments, but a first technical difficulty is to impose the initial stratification profile. 
The standard approach (known as the double-bucket method) does not offer great flexibility to impose arbitrary stratification, but other strategies have been proposed \cite{economidou2009density}. 
Another difficulty would also result from the velocity measurements. 
To probe the low-$Ro$ dynamics with $Ro \sim 10^{-3}-10^{-2}$ in a metre-size experiment, the velocity would have a typical amplitude $u_\ell \sim 10^{-4}-10^{-1}$~m.s${}^{-1}$ for an angular velocity $\Omega_s \sim 0.1 - 10$~rad.s${}^{-1}$, which could be below the detection level of the different velocimetry techniques.  
Therefore, it seems difficult to simultaneously explore low-$Ro$ and strongly stratified regimes with currently available experiments.

Rapidly rotating regimes with $N/\Omega_s \leq 1$ would also be difficult to probe because of centrifugal effects.
Indeed, if global rotation is strong enough, centrifugal gravity will deform the initial vertical stratification. 
To maintain a vertical stratification with negligible centrifugal effects, as well as small viscous ones with $E \ll 1$, a large tank with a relatively moderate rotation is required. 
This can be estimated by the ratio between centrifugal forces to Earth's gravity given by $\Delta_g =\Omega^2 \ell_H/g$ (where $\ell_H$ is the horizontal length scale of the tank, where centrifugal gravity is maximum). 
Assuming that the tank has a typical aspect ratio $\ell_H/\ell_V \sim 1$, we can introduce the Ekman number $E$ such that $\Delta_g = \nu^2/(E^2 \ell_H^3 g)$.
For a fixed value of $\Delta_g$, the maximum value of $\ell_H$ is thus given by
\begin{equation}
   \ell_H \sim \left( \frac{\nu^2}{g \Delta_g E^2} \right)^{1/3},
   \label{eq:negligibleCG}
\end{equation}
where $g\simeq9.81$~m.s$^{-2}$ is the value of Earth's gravity, and $\nu=10^{-6}$~m$^2$.s$^{-1}$ is the kinematic viscosity of water at ambient temperature.
Formula (\ref{eq:negligibleCG}) shows that when $\Delta_g$ is reduced (i.e. the more centrifugal gravity is negligible), then the horizontal length scale $\ell_H$ increases.
As shown in Figure \ref{fig:stratified}~(b), to reach a value $E=10^{-6}$ with a rotation rate $\Omega_s \sim 1$~rad.s${}^{-1}$, we must use a tank with $\ell_H \sim 1$~m if $\Delta_g=10^{-1}$ (or $\ell_V \sim 4$~m if $\delta_g=10^{-2}$). 
For a value of the Ekman number $E \sim 10^{-8}$, we would instead need a much larger tank with $\ell_H \sim 20$~ m.
Therefore, the dynamics of rotating stratified fluids with $E < 10^{-6}$ cannot be probed with a metre-size tank.
The only alternative would be to change the aspect ratio of the tank. 
While keeping $\ell_H \sim1$~m to weaken wall-driven flows, we could increase $\ell_V$ up to $4-5$~m (e.g. as in the \textsc{tropoconvex} convection experiment) but at the expense of a smaller value of the BV frequency (according to Figure \ref{fig:stratified}a).

\subsection{Specific properties of inertia-gravity waves in experimental conditions}
\begin{figure}
    \centering
    \begin{tabular}{cc}
    \includegraphics[height=0.45\textwidth]{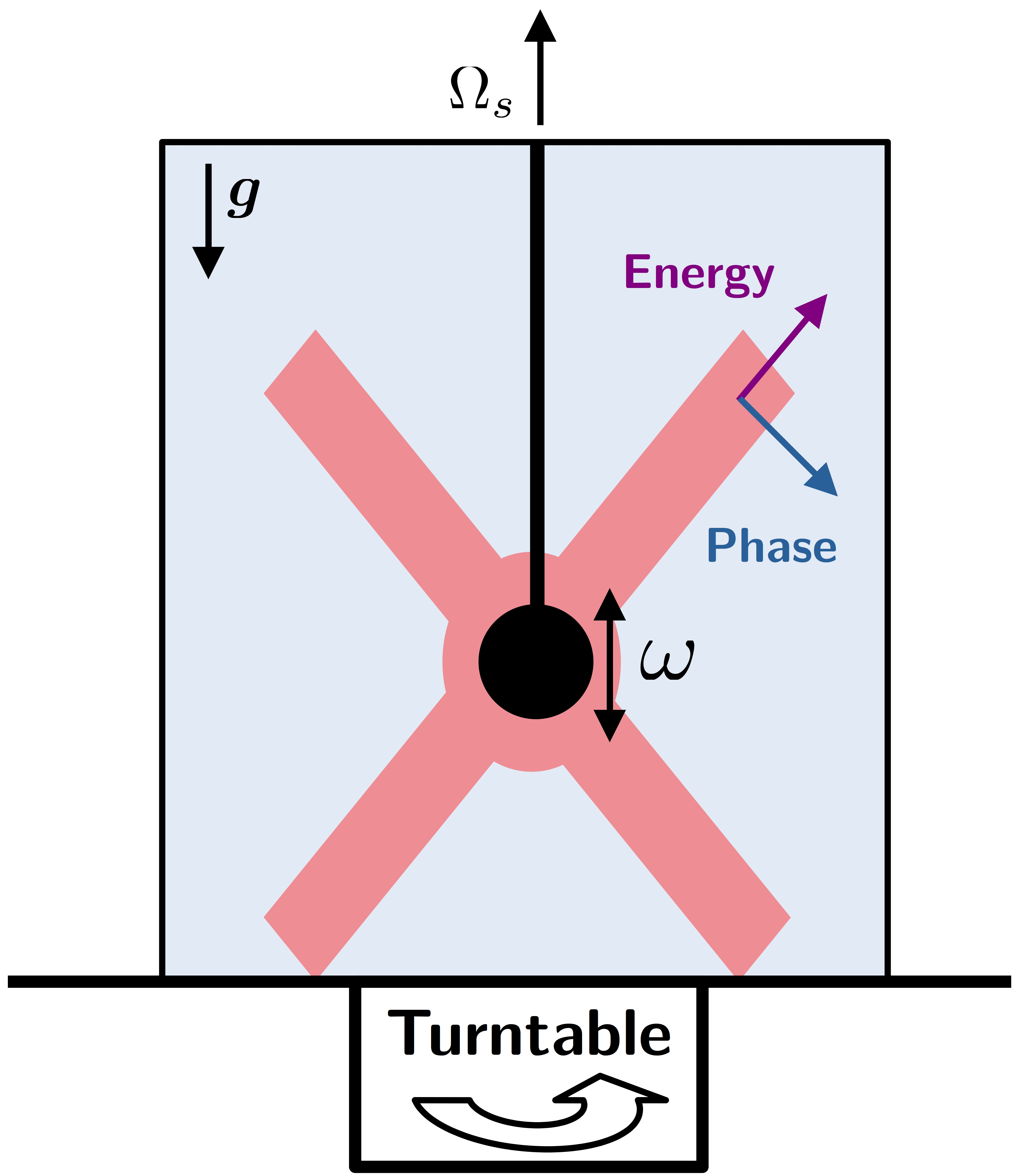} & 
    \includegraphics[height=0.45\textwidth]{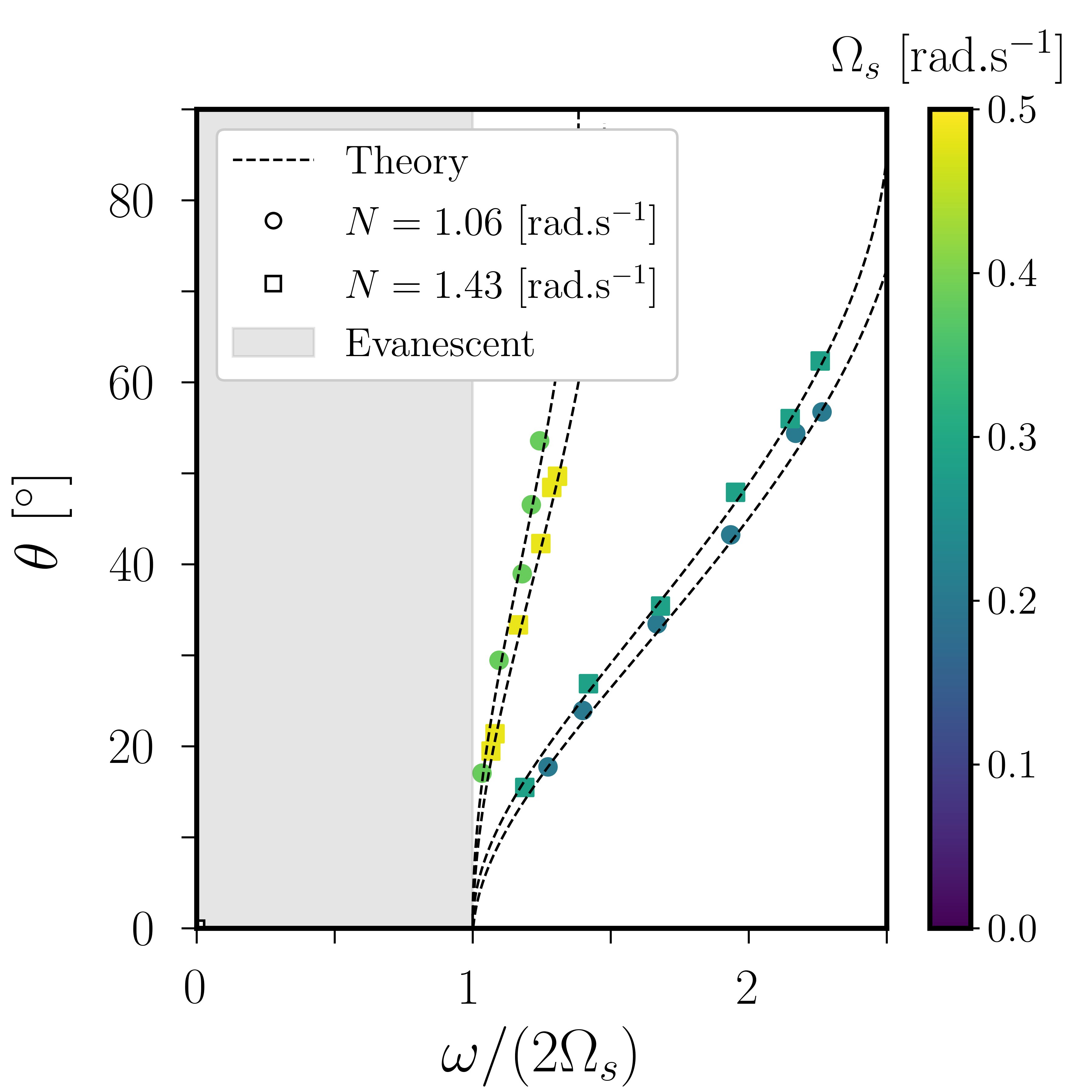} \\
    (a) & (b) \\
    \end{tabular}
    \caption{Experimental investigation of inertia-gravity waves (IGWs), for a rotating stratified fluid with aligned global rotation $\boldsymbol{\Omega}=\Omega_s \boldsymbol{1}_z$ and gravity $\boldsymbol{g} = - g \boldsymbol{1}_z$ in the regime $N/\Omega_s > 2$. Adapted from \cite{peacock2005effect}. \textbf{(a)} Sketch of the experimental setup. A solid sphere, which is embedded in the fluid, oscillates at the angular frequency $\omega$. The oscillation generates IGWs, which manifest in the form of wave beams (red regions, in which the phase and group velocities are orthogonal). \textbf{(b)} Angle $\theta$ (in degrees) between the wave vector $\boldsymbol{k}$ and gravity in dispersion relation (\ref{eq:IGWrelationdisp}). Excellent agreement is found between the experimental results and theory. No waves were reported when $|\omega| \leq 2 \Omega_s < N$.}
    \label{fig:IGW}
\end{figure}

Inertial waves and modes have proven useful to characterise the flow response of rotating fluids with large-scale ellipsoidal topographies (see \S\ref{sec:ellipsoid}), but also with small-wavelength topographies (see \S\ref{sec:smallscale}).
When global rotation and density stratification are coupled, inertia-gravity waves (IGWs) could be expected to play a similar role. 
IGWs are mixed waves that are sustained by Coriolis force and gravity. 
Let us discuss their properties in the case of a fine-tuned experiment, in which rotation is aligned with gravity and centrifugal effects are negligible.  
We consider a fluid that is stratified in density with a constant BV frequency $N>0$ along Earth's gravity $\boldsymbol{g} = -g \boldsymbol{1}_z$ (where $g$ is a constant), and rotating along the same axis at the angular velocity $\boldsymbol{\Omega} = \Omega_s \boldsymbol{1}_z$. 
In the rotating frame, IGWs satisfy the eigenvalue problem given by \cite{vidal2024igw}
\begin{subequations}
\label{eq:IGW1}
\begin{equation}
-\omega_i^2 \boldsymbol{u}_i + \mathrm{i} \omega_i (2 \boldsymbol{\Omega} \times \boldsymbol{u}_i) + N^2 (\boldsymbol{u}_i \boldsymbol{\cdot} \boldsymbol{1}_z) \boldsymbol{1}_z = -\mathrm{i} \omega_i \nabla \Phi_i, \quad \nabla \boldsymbol{\cdot} \boldsymbol{u}_i = 0,
\tag{\theequation a,b}
\end{equation}
\end{subequations}
where $\omega_i$ is the angular frequency of the wave, and $\boldsymbol{u}_i$ is the eigenvector for the wave velocity (respectively $\Phi_i$ for the pressure). 
It turns out that Equation (\ref{eq:IGW1}a) is a mixed hyperbolic-elliptic equation, whose nature changes as a function of $\omega_i$ only (i.e. the equation is either entirely hyperbolic in the fluid volume for a given value of $\omega_i$, or is entirely elliptic in the volume).
As such, it can be shown (using a plane-wave analysis) that local IGWs can only exist when $\min(2\Omega_s,N)\leq |\omega_i| \leq \max (2\Omega_s,N)$.
In this interval, $\omega_i$ satisfies the dispersion relation given by
\begin{equation}
    \omega_i^2 \boldsymbol{k}^2 = N^2 |\boldsymbol{1}_z \times \boldsymbol{k}|^2 + 4 \Omega_s^2 (\boldsymbol{1}_z \boldsymbol{\cdot} \boldsymbol{k})^2,
    \label{eq:IGWrelationdisp}
\end{equation}
where $\boldsymbol{k}$ is the (local) wave vector of the plane wave. 
As illustrated in Figure \ref{fig:IGW}, the local analysis agrees very well with experimental results \cite{peacock2005effect}. 
In the other interval $0 < |\omega_i| \leq \min(2\Omega_s,N)$, the equation is elliptic such that there are no wave motions (i.e. plane waves are evanescent, as for high-frequency disturbances without stratification \cite{nosan2021evanescent}).
However, if the entire fluid is forced in volume (not like in \cite{peacock2005effect}, where the oscillatory forcing was only local in space), then some low-frequency eigenmodes can exist in this frequency interval \cite{friedlander1982internal,vidal2024igw}.
Such $3-$D modes, which appear similar to the low-frequency Kelvin waves in oceanography, owe their existence to the presence of lateral (vertical) walls \cite{cdv2024gravito}. 

The existence of a frequency gap in experimental conditions has annoying consequences, as it hampers our ability to safely extrapolate experimental results to thick-layer systems.
Indeed, a frequency gap implies that an oscillatory forcing (e.g. a velocity flowing over a small-wavelength topography) will not be able to excite (locally) propagating waves for some frequencies in experimental conditions (but it may excite evanescent disturbances, as found without stratification \cite{nosan2021evanescent}).
However, such a frequency gap is not globally expected in a thick-layer fluid system in which gravity is radial \cite{friedlander1982gafd,dintrans1999gravito}. 
In such systems, for a given value of $\omega_i$, the fluid volume is indeed separated into regions in which waves can propagate, and other regions in which wave motions are locally evanescent.
Therefore, at different latitudes, the same oscillatory forcing could excite or not waves in thick-layer fluid systems (contrary to the experimental setup). 
The mathematical properties of IGWs are thus drastically different in experimental models and in geophysical systems, such that experimental results should be extrapolated with care to thick fluid envelopes. 

\section{Concluding remarks}
\label{sec:ccl}
\subsection{A common need for new experiments}
This paper enlightens the critical role of experimental studies in advancing our understanding of topographic effects in geophysical fluid dynamics. 
One key aspect discussed is the need to bridge the gap between numerically accessible regimes, and the often inaccessible flow regimes encountered in geophysical fluid envelopes. 
Given the inherent challenge of simulating flows over topography with \textsc{dns}, we have here emphasised the importance of experiments in complementing theoretical and numerical works. 
These problems have been extensively studied in the context of thin-layer systems (e.g. stratified oceans and atmospheres) using experiments, \textsc{dns} or theory. 
However, those results are not directly relevant for thick-layer systems, such as planetary cores and subsurface oceans. 
Those have received comparatively less attention from theoretical and experimental viewpoints, partly because of the scarcity of direct observations. 
Meanwhile, indirect observations (e.g. the accurate measurements of the Earth's and Moon rotation, or the measurements of the upcoming space missions to Jovian satellites) require a better understanding of the role of topography for mixing, angular momentum transfer, or magnetic field generation.
Density stratification is also not well-constrained in such hidden layers, such that all possible situations should be considered (including neutral regimes, weak or strong stable stratifications, or convection).

Since the most advanced \textsc{dns} are limited in the accessible range of parameters, but also in the geometries they can simulate, experiments have thus played a crucial role in making progress in this field.
For instance, they often allow us to validate asymptotic theoretical models or derive (empirical) scaling laws, which can be extrapolated to the out-of-reach range of parameters of geophysical bodies.
However, particular attention must be given to the regimes in which the experiments operate (i.e. are they already governed by the dynamics prevailing at planetary conditions?). 
An illustrative example is the debate about the expected rotating turbulence regime driven by mechanical (orbital) forcings. 
Indeed, experiments conducted at moderately small values of the Ekman and Rossby numbers likely operate in a different turbulent regime from that in geophysical conditions.

\begin{figure}
    \centering
    \includegraphics[width=0.95\textwidth]{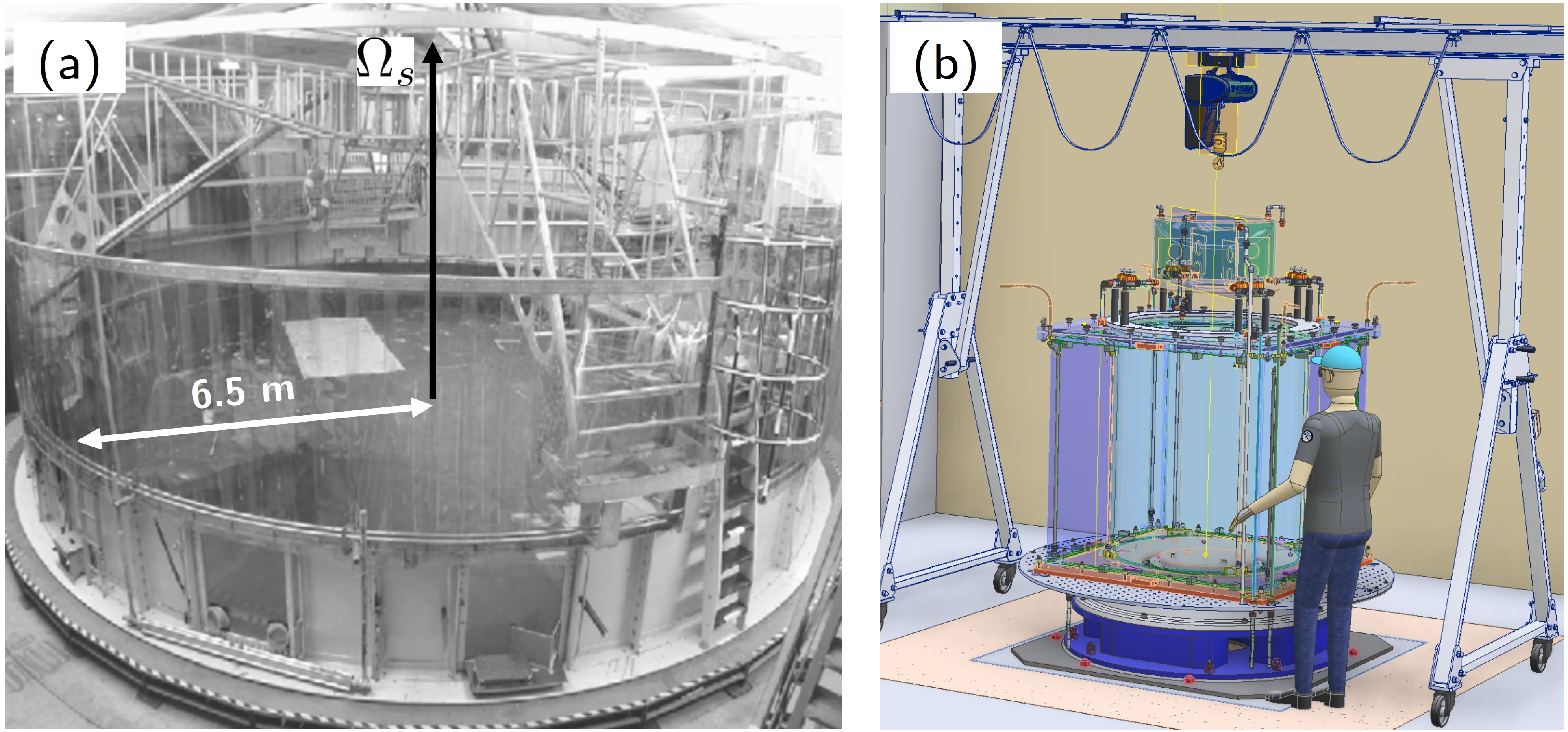}
    \caption{\textbf{(a)} \textsc{coriolis} experimental facility in \textsc{legi} laboratory (Grenoble, France). This is the largest turntable worldwide, with a diametre of $13$~m and a height of $1$~m, rotating at $6$~\textsc{rpm} at maximum. It has notably been used to investigate the dynamics of Ekman layers  \cite{aelbrecht1999experimental,ferrero2005physical,sous2012tsai,sous2013friction}. \textbf{(b)} \textsc{theia} experimental facility, currently under construction in \textsc{isterre} laboratory (Grenoble, France). It will be dedicated to the study of rotating flows with topography. The turntable will be capable of embarking tanks with a diameter of $\sim 1~$m and a height of $\sim2$~m, rotating at a maximum rotation frequency of $10$~\textsc{rpm}.}
    \label{fig:ccl}
\end{figure}

We have also identified some key points that would help to pave the way for future experiments. 
Firstly, it appears quite clear that we should consider larger experimental devices.
This is required to reduce the value of the Ekman number to probe high-$Re$ turbulence while keeping the Rossby number small enough, but also to allow measurements in BLs and in the vicinity of topographical features. 
Such measurements are indeed currently too challenging using sub-metre size experiments. 
If we target a typical value $E\sim 10^{-8}$, it would be achievable with a two-metre size experiment rotating at $60$ \textsc{rpm}, and using water at ambient temperature as a working fluid. 
Few devices of this class exist, such as the Big Sister three-metre sodium experiment in Maryland \cite{zimmerman2011bi,zimmerman2014turbulent} or the forthcoming two-metre \textsc{dresdyn} device \cite{stefani2019dresdyn}.
While the targeted low-$E$ values are accessible with such experiments, they are dedicated to specific forcing and focus on hydromagnetic flows.
Hence, they do not offer enough flexibility to address the questions discussed here. 
It may be more desirable to think of the next-generation experiments as an experimental platform that would allow investigating a broader range of problems, similar to the unique \textsc{coriolis} facility hosted in the \textsc{legi} laboratory (Figure \ref{fig:ccl}a).
This is an open tank with a diameter of $13$~m, rotating at $6$~\textsc{rpm} and embarking a fluid layer with a maximum height of one metre (resulting in an Ekman number $E\sim10^{-6}$). 
This unique instrument is well suited to model thin-layer dynamics with stratification at the cost of a moderate Ekman number. 
Yet, it is not ideal to investigate the flow dynamics in closed containers representative of thick-layer systems. 

Another benefit of large experiments would be to investigate the combined effects of stratification and rotation when $N \lesssim \Omega_s$, in contrast with the existing studies dedicated to oceanic dynamics that mainly explored the other regimes with $N \gg \Omega_s$. 
As shown in Figure \ref{fig:stratified}, it would be possible to perform experiments in the weakly stratified regimes at low Ekman numbers, and possibly in the more strongly stratified regime $1 \leq N/\Omega_s \leq 10 $. 
Finally, a larger device would also allow us to investigate the same range of Ekman numbers as we currently do with sub-metre size experiments, but using higher viscosity fluids ($10-100$ times) in order to increase the Ekman BL thickness by a factor $3$ to $10$. 
Detailed investigations of the effect of roughness in such a device would be achievable with topographic features, ranging from a few microns to several millimetres. 

\subsection{Larger experiments, a realistic prospect?}
We have argued that substantial advancement in the field of rapidly rotating geophysical fluid dynamics is contingent upon the development of new large-scale facilities.
A few laboratories have already acquired large-capacity turntables capable of embarking experiments of 1-2 m in size, such as the ones used in \cite{machicoane2018wake,lemasquerier2021zonal} or the upcoming \textsc{theia} experimental facility in Grenoble (Figure \ref{fig:ccl}b).
Even if they cannot reach low-$E$ regimes with $E \ll 10^{-8}$, these facilities pave the way for the development of even more ambitious devices. 
However, there will be some unavoidable difficulties when working with large-scale devices at small Ekman numbers. 
As anticipated from stability analyses, such rapidly rotating flows will be prone to instability for much smaller-amplitude perturbations than in the current devices.
It is therefore essential to pay close attention to the potential adverse effects, including small temperature fluctuations across the tank, the rotation of the Earth that will prevent from achieving perfect solid-body rotation \cite{boisson2012earth,triana2012precessional}, or any deviation from the intended geometry exceeding some characteristic length scales (e.g. the BL thickness).
Another challenge inherent to a unique apparatus is the repeatability and validation of the results.
It has become standard practice in the computational fluid dynamics community to conduct periodic benchmark studies on canonical problems, which allow testing and validating the different algorithms (e.g. \cite{christensen2001numerical,marti2014full} for the geodynamo).
The development of a similar approach, based on well-understood simple problems (e.g. the resonances of eigenmodes, or linear spin-up/down), would be particularly important in the context of unique devices aimed at achieving unprecedented ranges of parameters.

In our opinion, developing such new large-scale facilties should only be undertaken within the context of an open-access collaborative project (e.g. as in astronomy with the large telescopes, or in particule physics with the \textsc{cern}). 
As outlined below, it is essential that the collaborative structure in question fulfils some fundamental conditions.
\vspace{0.1cm}
\begin{paragraph}{(1) Motivations}~It should be driven by common interests, identified by a broad scientific community (e.g. the needs for a better understanding of the rapidly rotating fluid dynamics and related phenomena, which are shared among all geophysical sciences). 
\end{paragraph}
\vspace{0.05cm}
\begin{paragraph}{(2) Accessibility}~It should provide \emph{easy access} to the facility for all users (e.g. similar to astronomical observatories or supercomputers).
\end{paragraph}
\vspace{0.05cm}
\begin{paragraph}{(3) Scientific recognition}~While open access to publications is becoming a standard practice in academia, technical solutions or inconclusive results are rarely acknowledged and never rewarded. To maintain a culture of risk-taking and ambitious project development, it is essential to recognise the significant time, energy and creativity required. This may not always result in a peer-reviewed publication, but it is nevertheless a crucial aspect of academic endeavour. A collaborative platform involving recognised experts should offer an alternative to the classical publication scheme to share this unconventional outcomes, for instance in the form of open access repositories (as done by the \textsc{cern}). 
\end{paragraph}
\vspace{0.05cm}
\begin{paragraph}{(4) Scientific commitment}~
The success of such an enterprise is contingent upon the dedication of senior scientists who are past the stage of developing their careers. 
This would allow them to assume the risks inherent to these high-risk and high-gain projects. 
Moreover, it would be worth incorporating early-career and intermediate-career scientists into the management structure (to stimulate creativity and to ensure the continuity of the legacy).
\end{paragraph}
\vspace{0.05cm}
\begin{paragraph}{(5) Long-term funding and legacy}~It is not feasible to obtain financial support for a collaborative platform through the conventional grant schemes, which are constrained by temporal limitations. To achieve this, a host institution must demonstrate a long-term commitment, while national and international institutions must pledge sustainable financial support. 
Finally, given the likelihood of more rigorous financial constraints in the near future, a shared financial support would be the most optimal management strategy to secure such ambitious research activities.  
\end{paragraph}
\vspace{0.05cm}

For all these reasons, we believe that the experimental geophysical fluid dynamics community would greatly benefit from large collaborative projects rather than individual incentives. 
While already done in other fields (e.g. in astronomy or particle physics), this seems however at odds with the current research organisation that tends to reward individuals rather than consortiums.


\section*{Declaration of interests}
The authors do not work for, advise, own shares in, or receive funds from any organisation that could benefit from this article, and have declared no affiliations other than their research organisations.

\section*{D\'eclaration d'int\'er\^ets}
Les auteurs ne travaillent pas, ne conseillent pas, ne possèdent pas de parts, ne reçoivent pas de fonds d'une organisation qui pourrait tirer profit de cet article, et n'ont déclaré aucune autre affiliation que leurs organismes de recherche.

\section*{Acknowledgements}
The authors thank S. Fauve and M. Le Bars for the opportunity to write this review article. DC thanks the French Academy of Science and Electricit\'e de France for granting their `Amp\'ere Prize' to the ISTerre `Geodynamo' team. JV acknowledges Y. Colin de Verdi\`ere for fruitful discussions about waves in rotating and stratified fluids. DC and RM thank D. Jault for discussions on topographic effects in planetary cores. DC thanks M. Solazzo for his valuable help and his engineering work on rotating experiments with topographic effects. The authors would like to dedicate the paper to Stijn Vantieghem, who made major contributions in our field but passed away in 2023 at the age of 39.


\def\bysame{\leavevmode ---------\thinspace}
\makeatletter\if@francais\providecommand{\og}{<<~}\providecommand{\fg}{~>>}
\else\gdef\og{``}\gdef\fg{''}\fi\makeatother
\def\cdrandname{\&}
\providecommand\cdrnumero{no.~}
\providecommand{\cdredsname}{eds.}
\providecommand{\cdredname}{ed.}
\providecommand{\cdrchapname}{chap.}
\providecommand{\cdrmastersthesisname}{Memoir}
\providecommand{\cdrphdthesisname}{PhD Thesis}

\end{document}